\documentclass[aps,prx,reprint,superscriptaddress,noeprint,longbibliography]{revtex4-1}

\usepackage[normalem]{ulem}
\usepackage{amsmath}
\usepackage{amsbsy}
\usepackage{hhline}
\usepackage{multirow}
\usepackage{amssymb} 
\usepackage{graphicx}
\usepackage{float}
\usepackage{upgreek}
\usepackage[caption=false]{subfig}
\usepackage{hyperref}
\hypersetup{
	colorlinks=true,       
	linkcolor=blue,          
	citecolor=green,       
	filecolor=magenta,      
	urlcolor=blue          
}

\usepackage[dvipsnames]{xcolor}

\newcommand{\bx}{{\mathbf x}}
\newcommand{\bq}{{\mathbf q}}
\newcommand{\ra}{{\rm a}}
\newcommand{\rb}{{\rm b}}

\DeclareMathOperator{\Tr}{Tr}

\begin{document}
	\title{Non-equilibrium fixed points of coupled Ising models}	

	\author{Jeremy T. Young}
	\email[Corresponding author: ]{jjttyy27@umd.edu}
	\affiliation{Joint Quantum Institute, NIST/University of Maryland, College Park, Maryland 20742 USA}
	
	\author{Alexey V. Gorshkov}
	\affiliation{Joint Quantum Institute, NIST/University of Maryland, College Park, Maryland 20742 USA}
	\affiliation{Joint Center for Quantum Information and Computer Science, NIST/University of Maryland, College Park, Maryland 20742 USA}
	\author{Michael Foss-Feig}
	\affiliation{United States Army Research Laboratory, Adelphi, Maryland 20783, USA}
	
	\author{Mohammad F. Maghrebi}
	\affiliation{Department of Physics and Astronomy, Michigan State University, East Lansing, Michigan 48824 USA}
	
	\date{\today}
	
	\begin{abstract}
		Driven-dissipative systems are expected to give rise to non-equilibrium phenomena that are absent in their equilibrium counterparts. However, phase transitions in these systems generically exhibit an effectively classical equilibrium behavior in spite of their non-equilibrium origin. In this paper, we show that multicritical points in such  systems lead to a rich and genuinely non-equilibrium behavior. Specifically, we investigate a driven-dissipative model of interacting bosons that possesses two distinct phase transitions: one from a high- to a low-density phase---reminiscent of a liquid-gas transition---and another to an antiferromagnetic phase. Each phase transition is described by the Ising universality class characterized by an (emergent or microscopic) $\mathbb{Z}_2$ symmetry. 
		They, however, coalesce at a multicritical point, giving rise to a non-equilibrium model of coupled Ising-like order parameters described by a $\mathbb{Z}_2 \times \mathbb{Z}_2$ symmetry.  
		Using a dynamical renormalization-group approach, we show that a pair of non-equilibrium fixed points (NEFPs) emerge that govern the long-distance critical behavior of the system.
		We elucidate various exotic features of these NEFPs. In particular, we show that a generic continuous scale invariance at criticality is reduced to a discrete scale invariance. This further results in complex-valued critical exponents and spiraling phase boundaries, and it is also accompanied by a complex Liouvillian gap even close to the phase transition.
		As direct evidence of the non-equilibrium nature of the NEFPs, we show that the fluctuation-dissipation relation is violated at all scales, leading to 
		an effective temperature  
		that becomes ``hotter'' and ``hotter'' at longer and longer wavelengths.
		Finally, we argue that this non-equilibrium behavior 
		can be observed in 
		cavity arrays with cross-Kerr nonlinearities.
	\end{abstract}
	
	
	\maketitle
	
	\section{Introduction}
	
	The increasing control over synthetic quantum systems has provided new avenues into studying many-body physics that are not accessible in conventional condensed matter systems. 
	In particular, driven-dissipative systems, defined by the competition between a coherent drive and dissipation due to the coupling to the environment, have emerged as a versatile setting to investigate non-equilibrium physics \cite{Diehl2008}. 
	They are very naturally realized by a variety of emerging quantum simulation platforms ranging from exciton-polariton fluids \cite{Deng199,Kasprzak06,Littlewood06,Byrnes14,Rodriguez16,Rodriguez2017}, to trapped ions \cite{Bohnet16,Schindler13}, to Rydberg gases \cite{Peyronel12,Firstenberg13,Carr13,Malossi2014}, to circuit-QED platforms \cite{Houck2012,Fitzpatrick2017}.
	At long times, these systems approach a non-equilibrium steady state due to the interplay of drive and dissipation. The steady states can potentially harbor novel phases and exhibit exotic dynamics.  Situated far from equilibrium, understanding the properties of these steady states
	requires methods beyond those suitable in or near equilibrium. The quest to realize and characterize macroscopic phases of these non-equilibrium systems has sparked a flurry of theoretical and experimental investigations. 
	
	Given their non-equilibrium dynamics, driven-dissipative systems are expected to exhibit universal, critical properties different from their equilibrium counterparts. In spite of this, it has become increasingly clear that 
	an effective temperature, and thus an effectively  classical equilibrium behavior, emerges in
	a large class of driven-dissipative phase transitions \cite{mitra06,wouters06}.
	In particular, 
	the equilibrium Ising universality class and, more generally, the model A dynamics of the Hohenberg-Halperin classification---describing the critical behavior of a non-conserved order parameter in or near equilibrium---have emerged in a variety of driven-dissipative phase transitions; 
	these include bosonic or photonic Bose-Hubbard systems 
    \cite{LeBoite2013,Weimer2015,FossFeig2017,Vicentini2018}, various driven-dissipative spin models near an Ising \cite{Carr13,Overbeck2017,Marcuzzi2014,Maghrebi2016,Owen2017,Chan2015,Wilson2016}, antiferromagnetic \cite{Lee2011,Lee2013,Jin2013,Jin2014,Hoening2014,Maghrebi2016,Owen2017,Wilson2016,Chan2015}, or limit-cycle phase \cite{Lee2011,Owen2017,Chan2015,Wilson2016}, as well as
	driven-dissipative condensates consisting of polaritons \cite{Sieberer2013,Tauber2014a}.
	A possible exception is a two-dimensional driven-dissipative condensate, where it has been argued that the non-equilibrium Kardar-Parisi-Zhang universality class governs the long-wavelength dynamics \cite{Altman2015, Sieberer2016}. 
	While existing experiments are consistent with an effective thermal behavior  \cite{Roumpos2012,Nitsche2014}, the Kardar-Parisi-Zhang dynamics is expected to emerge under different experimental conditions. 
	In general, an important goal of the field is to identify whether generic driven-dissipative systems can escape the pull of an effective equilibrium behavior and instead give rise to new non-equilibrium universality classes. 
	In particular, it has proved difficult to identify non-equilibrium universal behavior which is genuinely of a quantum nature; see Refs.~\cite{Marino2016,Torre2010,Cheung2017} for recent proposals to achieve non-equilibrium quantum criticality and Ref.~\cite{Rota2019} for numerical evidence of an equilibrium quantum critical point in a driven-dissipative system.
	
	An effective equilibrium behavior 
	is not special to driven-dissipative quantum systems. In \textit{driven-diffusive} classical systems too, where the drive, as well as the dynamics, is entirely classical, effective equilibrium seems to be remarkably robust. For instance, 
	an Ising-type dynamics governing a non-conserved order parameter 
	is 
	argued to be stable against all dynamical, non-equilibrium perturbations \cite{Bassler1994}. More generally, the universal dynamics of various models in the Hohenberg-Halperin classification \cite{Hohenberg1977} are shown to be robust against a class of
	non-equilibrium perturbations which violate detailed balance \cite{Garrido1989,Wang1988,Marques1989,Marques1990,Tome1991,DeOliveira1993,Achahbar1996,Godoy2002,Risler04,Risler05};
	truly non-equilibrium behavior emerges under more constrained dynamics, for example, in the presence of a conserved order parameter in an anisotropic medium \cite{Grinstein90,Garrido90,Cheng91,Tauber1997,Tauber2002,Katz1984,Zia2010}. 
	In much of the previous work, 
	situations have been considered where the phase transition is governed by 
	a single order parameter. Due to the restriction that this places on the dynamics, a description based on an effective Hamiltonian often becomes available, hence the emergence of an effective equilibrium behavior.

\begin{figure}
	\includegraphics[scale=.58]{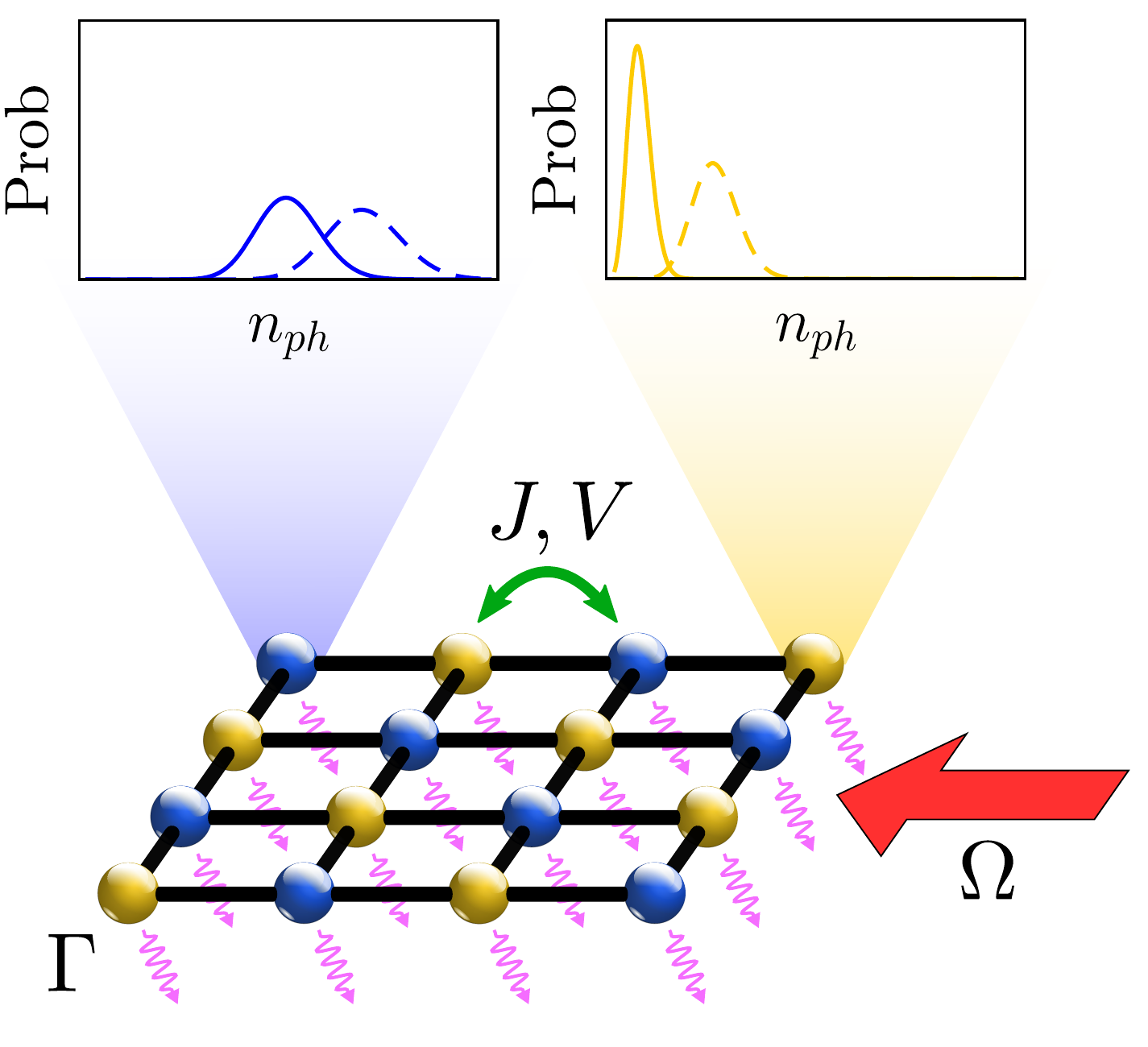}
	\caption{Schematic illustration of the physical setup. The contrast of the two checkerboard (light yellow and dark blue) sublattices defines the antiferromagnetic order parameter. On each sublattice, there is a low-population (low $n_{ph}$; solid curve) and a high-population (high $n_{ph}$; dashed curve) steady state, corresponding to the bistability order parameter. The large  arrow indicates the drive, while the wavy arrows represent the dissipation. $J$ and $V$ denote the hopping and nearest-neighbor interactions, respectively. \label{schematic}}	
	\end{figure}

	In this work, we consider a driven-dissipative model that gives rise to multicritical points defined by the joint transition of \textit{two} order parameters. 
	In particular, we investigate the interplay of two phase transitions, each of which has been studied extensively in driven-dissipative settings: One is the many-body analog to optical bistability; in the other, a sublattice symmetry is spontaneously broken, leading to an antiferromagnetic ordering. A schematic illustration of this combination is shown in Fig.~\ref{schematic}.
	With two order parameters,
	the non-equilibrium dynamics is much less constrained than that of equilibrium and
	an immediate identification of an effective Hamiltonian is no longer possible.
	Remarkably, we show that a new, genuinely non-equilibrium universal behavior emerges at the multicritical point, giving rise to exotic critical behavior and dynamics. 
	Our proposal to observe non-equilibrium critical behavior relies on tuning the system parameters (such as drive and detuning, which are easy to control) to a multicritical point.
	In fact, the driven-dissipative setting of our model can be experimentally realized using cross-Kerr nonlinearities in cavity arrays \cite{Jin2013,Jin2014,Kounalakis2018}.

	In order to determine the critical behavior,
	we will employ the Keldysh-Schwinger and functional integral formalism suited for the non-equilibrium setting of driven-dissipative systems \cite{Torre2010,Gopalakrishnan2010,Torre2012,Torre2013,Sieberer2013,Sieberer2014, Tauber2014a,Altman2015,Sieberer2016,Maghrebi2016,Marino2016,FossFeig2017,Maghrebi2017}. 
	While the presence of two order parameters prevents an immediate Hamiltonian description, the long-wavelength universal behavior---and whether or not the macroscopic behavior escapes an equilibrium fixed point---is determined
	by investigating how the parameters evolve under a dynamical version of renormalization-group (RG) techniques \cite{Tauber2014}. 
	
	The remainder of this paper is organized as follows. In Sec.~\ref{sum}, we present the main results of this paper and a summary of the non-equilibrium critical properties emerging in our driven-dissipative model. In Sec.~\ref{model}, we discuss the phase diagram of the model and identify the multicritical points where two distinct phase transitions meet. 
	In Sec.~\ref{RGsec}, we present the RG analysis and show that a pair of new classical non-equilibrium fixed points (NEFPs) emerge that exhibit a variety of novel critical behaviors. In Sec.~\ref{expt}, we discuss an experimental setting based on cavity arrays to realize the multicritical points of our model. 
	Finally, in Sec.~\ref{out}, we conclude our paper with a discussion of possible future directions which are motivated by the results of our work. In the Appendices, we present technical details omitted from the main text.
	
	\section{Main Results}
	
	\label{sum}
	In this section, we present the main results of this paper. 
	We consider a driven-dissipative model which displays two distinct phase transitions, each of which arises generically in various settings. 
	The first one is a many-body version of bistability where two stable solutions arise with a low or high population of photons (or excitation of spins). 
	In the thermodynamic limit, 
	the bistable region is reduced to a line of first-order phase transitions that terminates at a critical point, reminiscent of a liquid-gas phase transition.
	The second type of phase transition is 
	one to an antiferromagnetic phase where 
	the population takes different values on the two sublattices (a or b) of the system. 
	We shall consider a model where these phase transitions coalesce at a multicritical point and investigate the exotic dynamics that arise due to the interplay of the respective order parameters.
	These features are provided, for example, in 
	a driven-dissipative model of weakly interacting bosons with
	nearest-neighbor density-density interactions on a $d$-dimensional cubic lattice. The coherent dynamics of the model is governed by the Hamiltonian 
	\begin{multline}
	\label{modeq1}
	H = \sum_i -\Delta a_i^\dagger a_i + \Omega(a_i^\dagger + a_i) \\
	+ \sum_{\langle ij \rangle} -J (a_i^\dagger a_j + a_j^\dagger a_i) + V a_i^\dagger a_i a_j^\dagger a_j,
	\end{multline}
	where $\Delta = \omega_D - \omega_0$ is the detuning of the drive ($\omega_0$ is the frequency of the bosons and $\omega_D$ the drive frequency), $\Omega$ the drive strength, $J$ the hopping strength, and $V$ the strength of the nearest-neighbor interactions.
	The incoherent dynamics is due to loss of bosons, characterized by the Lindblad operators $L_i = \sqrt{\Gamma} a_i$, where $\Gamma$ defines the loss rate. 
	The (mixed) state of the system $\rho$ evolves under the quantum master equation 
	\begin{equation}
	\label{modeq2}
	\dot{\rho} = - i[H,\rho] + \sum_i L_i \rho L_i^\dagger - \frac{1}{2} \{\rho, L_i^\dagger L_i\},
	\end{equation}
	until it approaches a non-equilibrium steady state at long times where $\dot \rho=0$.
	The interplay of the coherent drive (the linear term in the Hamiltonian) and dissipation together with the interactions tends to give rise to bistability, while the nearest-neighbor interactions 
	can lead to an antiferromagnetic phase. 
	We stress that our general results should hold beyond the specific model considered here; for example, the addition of on-site interactions or density-dependent hopping terms to our model also gives rise to multicritical points whose universal properties should be independent of the microscopic model considered. 
	More generally, the relevant features of our bosonic model also arise in a variety of driven-dissipative systems 
	including spin models 
	\cite{Lee2011,Jin2013,Jin2014,Chan2015,Maghrebi2016,FossFeig2017,Marcuzzi2014}. We have chosen this particular model as a minimal driven-dissipative setting that gives rise to multicritical points, although our general conclusions should apply to a large class of models.

	Each phase transition in our model is characterized by an Ising-like order parameter (low/high density in bistability and sublattice a/b in the antiferromagnetic transition). 
	The simple structure of the order parameter, together with the incoherent nature of the dynamics, puts a strong constraint on the universal properties of the phase transition. Thus, it may be expected that each phase transition alone is described by the Ising universality class that also governs the Ising-type transitions in equilibrium. It can be argued, on more formal grounds, that this is indeed the case. 
	Associating the order parameter $\phi_1$ with bistability and $\phi_2$ with antiferromagnetic ordering, their long-wavelength properties in the steady state are governed by a thermal distribution but with respect to 
	the effective (classical) Hamiltonians 
	\begin{subequations}
		\begin{equation}
		{\cal H}_1 = \int_\bx \frac{D_1}{2} |\nabla \phi_1|^2 - h \phi_1 + \frac{r_1}{2} \phi_1^2 + \frac{g_1}{4} \phi_1^4 ,
		\end{equation}
		\begin{equation}
		{\cal H}_2 = \int_\bx \frac{D_2}{2}|\nabla \phi_2|^2 + \frac{r_2}{2} \phi_2^2+ \frac{g_2}{4} \phi_2^4 ,
		\end{equation}
	\end{subequations}
	with $D_i$ characterizing the stiffness, $g_i$ the interaction strength, $r_i$ the distance from the critical point (which will shift once fluctuations are taken into account), 
	and $h$ an effective magnetic field. Note that due to sublattice symmetry, there is no magnetic field associated with the antiferromagnetic phase. 
	Furthermore, the incoherent nature of the model leads to stochastic Langevin-type dynamics of the order parameters as \cite{Tauber2014}
	\begin{equation}
	\label{stochlang}
	\zeta_i \partial_t \phi_i = - \frac{\delta {\cal H}_i}{\delta \phi_i} +  \xi_i,
	\end{equation}
	where $\zeta_i$ is a ``friction'' coefficient and $\xi_i$ describes Gaussian white noise with correlations
	\begin{equation}
	\label{noise1}
	\langle \xi_i(t,\mathbf{x}) \xi_j(t', \mathbf{x}') \rangle = 2 \zeta_i T_i\delta_{ij} \delta(t-t') \delta(\mathbf{x}-\mathbf{x}'),
	\end{equation}
	where $T_i$ is the effective temperature of the system.
	Near equilibrium, the ``friction'' coefficients $\zeta_i$ control the rate at which the system relaxes to a thermal state via dissipating energy and thus is a purely dynamical quantity. The noise itself is related to the dissipation (i.e., friction) through temperature in what is known as the Einstein relation, which itself is a consequence of the fluctuation-dissipation theorem \cite{Tauber2014}. 
	In the non-equilibrium context of driven-dissipative models, where there is no intrinsic temperature, the ratio of the noise level to the dissipation can be used to define an effective temperature at long wavelengths. Even a non-equilibrium system that is effectively (i.e., at large scales) governed by the Hamiltonian dynamics [as in Eq.~(\ref{stochlang})] 
	is effectively in thermal equilibrium. This condition is 
	often satisfied for a single Ising-like order parameter \cite{Tauber1997}, although notable examples exist where this is not the case \cite{Grinstein90,Garrido90,Cheng91,Tauber1997,Tauber2002,Katz1984,Zia2010}. Notice that, with the appropriate scaling of the fields $\phi_i$, the effective temperatures can be set to $T_i=1$. Therefore, as long as the two order parameters are not coupled, their distribution in the steady state is given by $e^{-{\cal H}_1-{\cal H}_2}$.

	\begin{figure}
\includegraphics[scale=.45]{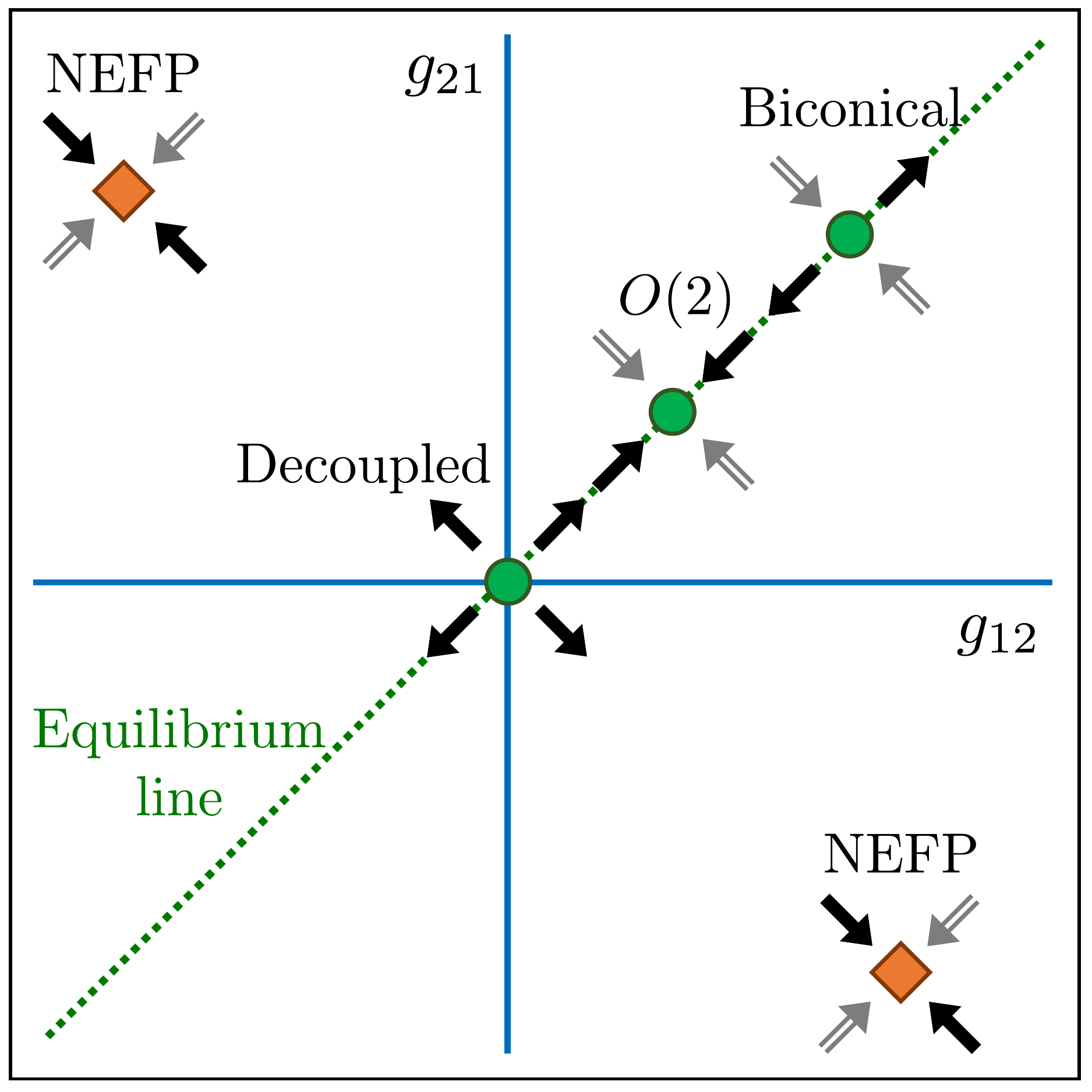}
\caption{A schematic RG flow diagram projected to the $g_{12}$-$g_{21}$ plane. (The full RG flow requires a five-dimensional space, which precludes a more complete sketch of the RG flow in this two-dimensional space; see Sec.~\ref{RGsec}.)
In addition to the equilibrium fixed points where $g_{12}=g_{21}$ (green circles), a pair of stable NEFPs (orange diamonds) emerge in the sector defined by the opposite signs of $g_{12}$ and $g_{21}$. These new fixed points exhibit exotic critical behavior reflecting their truly non-equilibrium nature.
Filled (black) arrows represent the stability while partial (gray) arrows indicate the expected stability of the various fixed points in different directions. 
Stability is known to lowest order in $\epsilon = 4-d$ only in directions which preserve the ratio $g_{12}/g_{21}$. The RG flow cannot cross the $g_{12},g_{21}$ axes, which is represented by double lines.
The stable equilibrium fixed point is characterized by an emergent $O(2)$ symmetry, while the unstable equilibrium fixed points include a biconical fixed point as well as various decoupled fixed points (which all lie at the origin in this diagram) corresponding to combinations of Ising and Gaussian fixed points.
\label{flow}}
\end{figure}	
	
	The situation is entirely different in the vicinity of multicritical points where the two order parameters are generally coupled. 
	Given the underlying symmetries, the dynamics can always be brought to the form 
	\begin{subequations}
		\begin{equation}
		\zeta_1 \partial_t \phi_1 = - \frac{\delta {\cal H}_1}{\delta\phi_1} - g_{12} \phi_1 \phi_2^2 + \xi_1,
		\end{equation}
		\begin{equation}
		\zeta_2 \partial_t \phi_2 = - \frac{\delta {\cal H}_2}{\delta\phi_2} - g_{21} \phi_2 \phi_1^2 + \xi_2.
		\end{equation}
	\end{subequations}
	Notice that the new terms that couple the two fields respect the underlying Ising symmetry of both order parameters. 
	The noise correlations are given by Eq.~(\ref{noise1}); we shall again exploit our freedom in scaling the fields to set $T_1=T_2=1$. 
	With the two order parameters coupled,
	the condition for an effective equilibrium description becomes much more restrictive. A thermal description requires the entire dynamics to be described by a single Hamiltonian. This only occurs when $g_{12}=g_{21}$,
	leading to the effective Hamiltonian 
	\begin{equation}
	{\cal H} = {\cal H}_1 + {\cal H}_2 + \frac{g_{12}}{2} \int_\bx \phi_1^2 \phi_2^2,
	\end{equation}
	in which case the steady-state distribution is given by $e^{-{\cal H}}$.
	However, this will not generally be the case, so 
	we must consider how various parameters flow under RG. While the microscopic (though coarse-grained) dynamics is not immediately described by a thermal state, it could very well be the case that the RG flow pulls the system into a thermal fixed point where $g^*_{12}=g^*_{21}$ at macroscopic scales. Indeed, we show that this is the case roughly when the microscopic values of the coupling constants are both positive, i.e., when $g_{12}$,  $g_{21}>0$. It is rather remarkable that 
	equilibrium restores itself under RG, showcasing another instance in which equilibrium proves to be a robust fixed point even when the system is driven far from equilibrium. However, this is not the end of the story: We show that, when the microscopic couplings have opposite sign ($g_{12}g_{21}<0$), 
	a pair of two NEFPs emerge where $g^*_{12} = -g^*_{21}$. (Both equilibrium fixed points and NEFPs are shown in Fig.~\ref{flow}.) Furthermore, we will argue that for the model under consideration, the critical behavior will be governed by one of the NEFPs. These fixed points give rise to a new non-equilibrium universality class  
	exhibiting a variety of exotic features that generically do not, or even cannot, arise in any equilibrium setting. A summary of the most interesting features, including the critical behavior, of the new NEFPs is illustrated in Fig.~\ref{figsum}. In the following subsections, we discuss these features in detail.
	
	\subsection{Scaling phenomena}
	
	In the vicinity of an RG fixed point governing a phase transition, the system exhibits universal scaling behavior characterized by critical exponents, regardless of the microscopic model. The scaling behavior of the correlation and response functions at a NEFP or in its vicinity 
	takes, respectively, the form 
		\begin{subequations}\label{scaling fns}
			\begin{equation}
			\begin{aligned}
			 C(\bq,\omega) & \equiv \mathcal{F} \langle \{a^\dagger(\mathbf{x}, t), a(0, 0)\} \rangle_c/2 \\
			 & \propto |\mathbf{q}|^{-2 + \eta - z} \tilde{C}\left(\frac{\omega}{|\mathbf{q}|^z},\frac{r}{|\mathbf{q}|^{1/\nu'}},P\left(\frac{\log|\mathbf{q}|}{\nu''} \right)\right),
			 \end{aligned}
			\end{equation}
			\begin{equation}
			\begin{aligned}
			\chi(\bq,\omega) &\equiv i \mathcal{F} \Theta(t) \langle [a^\dagger(\mathbf{x}, t), a(0, 0)] \rangle \\
			& \propto |\mathbf{q}|^{-2 + \eta'} \tilde{\chi}\left(\frac{\omega}{|\mathbf{q}|^z},\frac{r}{|\mathbf{q}|^{1/\nu'}},P\left(\frac{\log|\mathbf{q}|}{\nu''}\right)\right),
			\end{aligned}
			\end{equation}
		\end{subequations}
	where $\mathcal{F}$ denotes the Fourier transform in both space ($\mathbf{x}$) and time ($t$) with $\mathbf{q}$ the momentum and $\omega$ the frequency, the curly brackets denote the anti-commutator, and the subscript $c$ indicates the connected part of the correlation function. Here, $r = \sqrt{r_1^2+r_2^2}$ is the distance from the multicritical point, $P$ is a $2 \pi$-periodic function, and the functions $\tilde{C}$ and $\tilde{\chi}$ are
	scaling functions. 
	While in principle the scaling behavior could be different for the two order parameters ($\phi_1$ and $\phi_2$), we argue, based on a systematic RG analysis, that a stronger notion of scaling emerges where the critical (static and dynamic) behavior and exponents characterizing the two order parameters become identical. Therefore,
	we can express either the correlation or the response function via a single scaling function (and not one for each order parameter) with the same set of exponents. 
	The exponents $\eta$ and $\eta'$ define the anomalous dimensions corresponding to correlation and response functions, respectively, and $z$ is the dynamical critical exponent characterizing the relative scaling of time with respect to spatial coordinates. The correlation length $\xi$ is described by the critical exponent $\nu'$ via $\xi \propto r^{-\nu'}$. Typically, it is the exponent $\nu$, associated with the scaling behavior of $r_1$ and $r_2$, that describes the scaling of the correlation length. However, the latter exponent becomes complex-valued at the NEFPs, $\nu^{-1} = {\nu'}^{-1} + i\, {\nu''}^{-1}$, with the real part determining the scaling of the correlation length and the imaginary part leading to a discrete scale invariance, as we shall discuss shortly. 
	Altogether, there are five independent critical exponents of interest: $\nu', \nu'', \eta, \eta', z$.

\begin{figure}
		\centering
		\def\arraystretch{1.35}
		\begin{tabular}{|c|c|}
			\hline
			Effective equilibrium & Non-equilibrium fixed points \\
			\hhline{|==|}
			\multicolumn{2}{|c|}{(a) Scale invariance} \\
			\hline
			Continuous scale invariance & Discrete scale invariance  \\
			\hline
			\includegraphics[scale=.3]{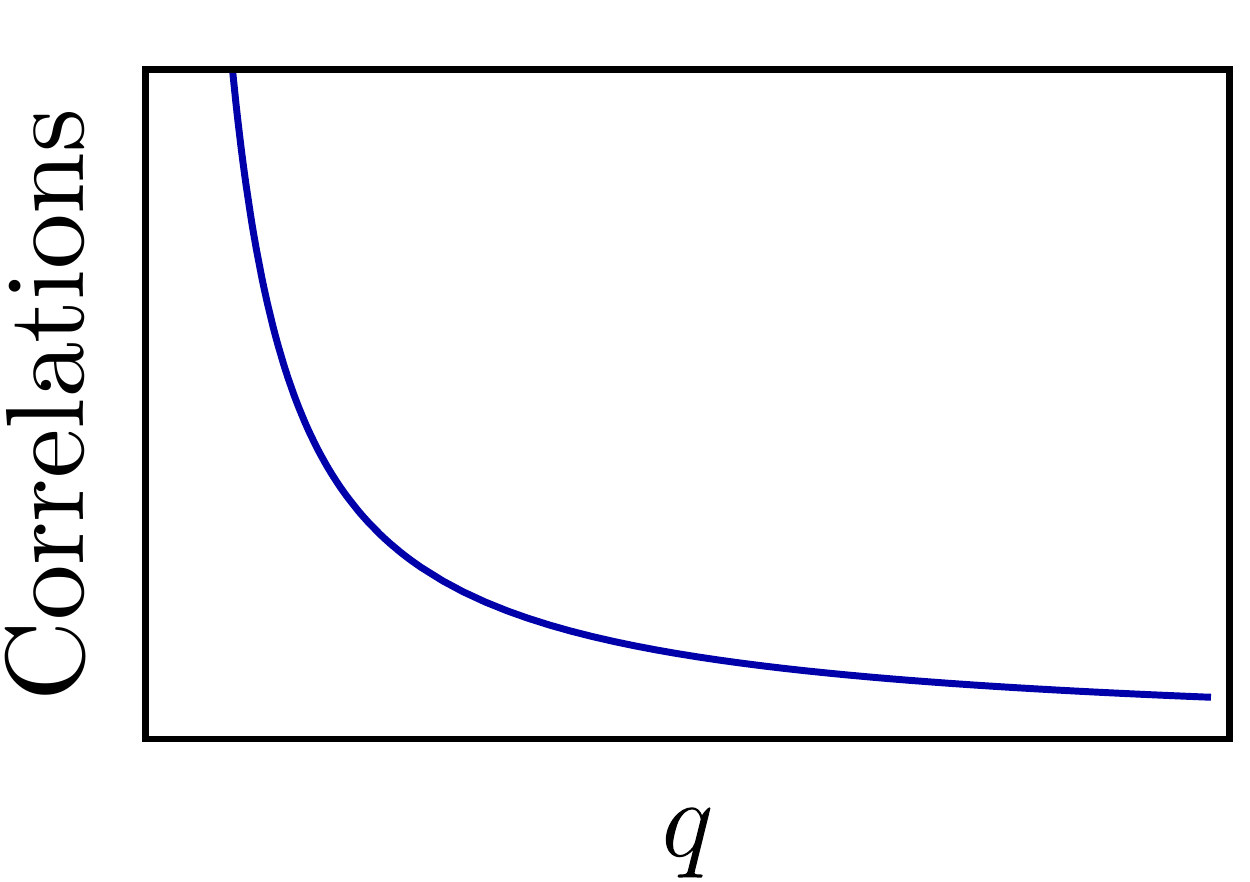} & 
			\includegraphics[scale=.3]{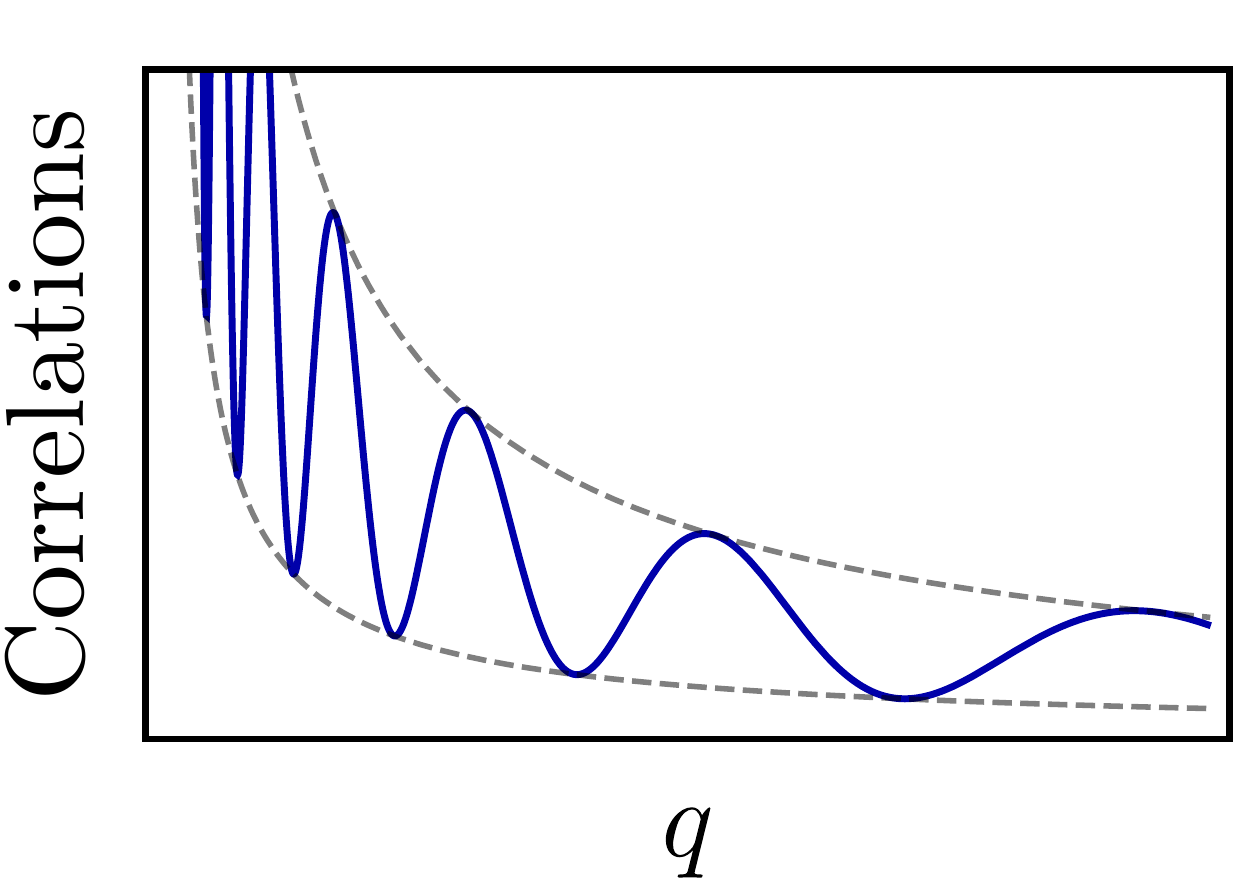} \\
			\hline
			\multicolumn{2}{|c|}{(b) Effective temperature behavior} \\
			\hline 
			\includegraphics[scale=.3,trim= 0 0 0 0]{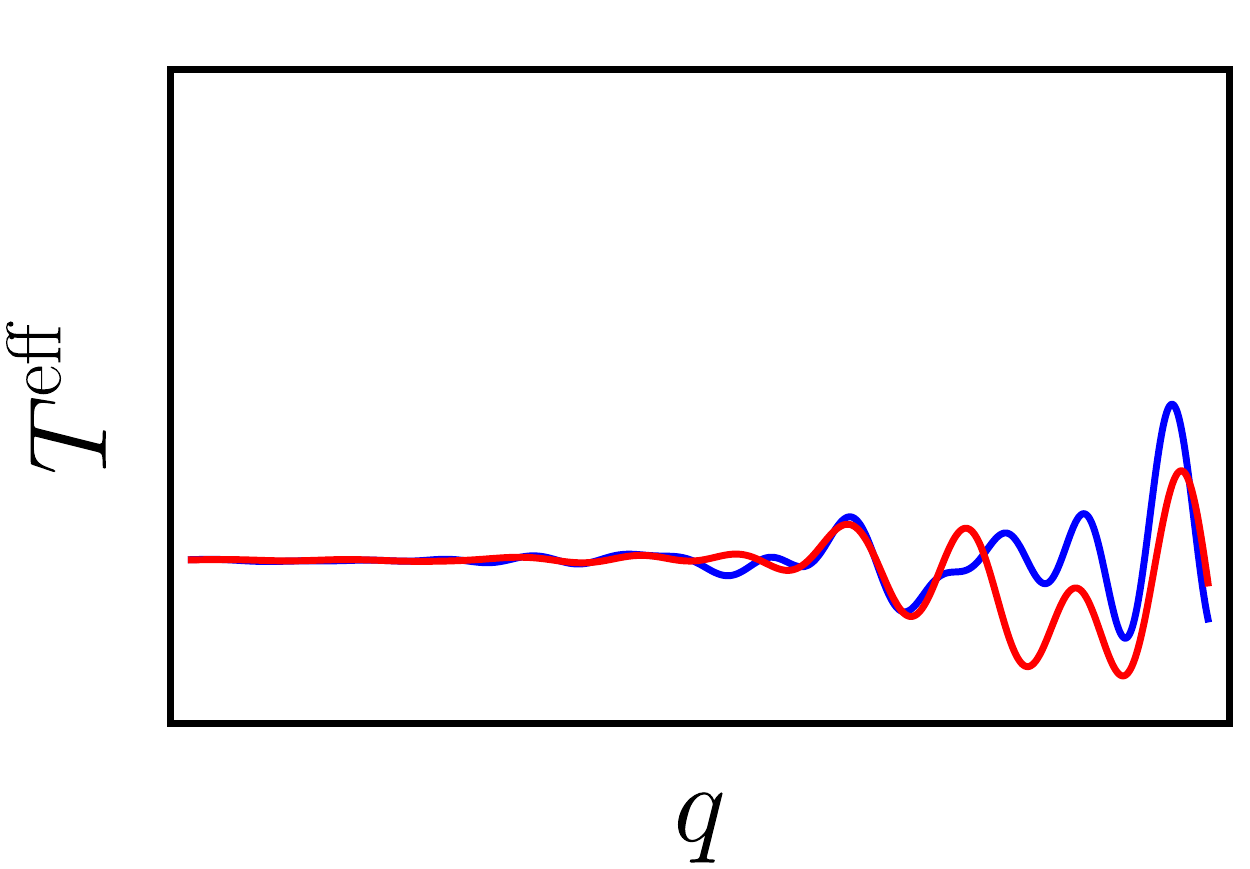}&
			\includegraphics[scale=.3,trim= 0 0 0 0]{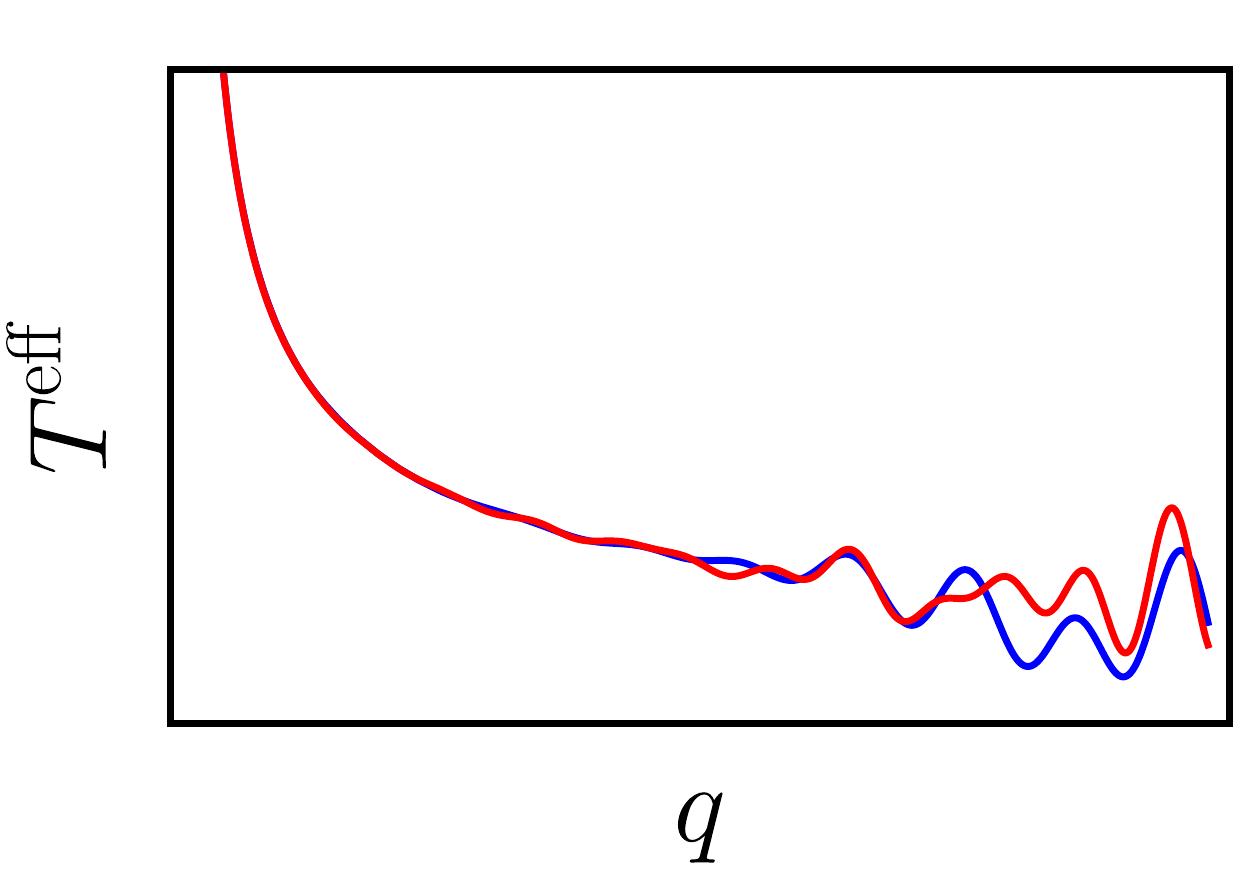} \\
			\hline
			\multicolumn{2}{|c|}{(c) Liouvillian gap closure} \\
			\hline 
			\includegraphics[scale=.11,trim= 0 -7 0 -7]{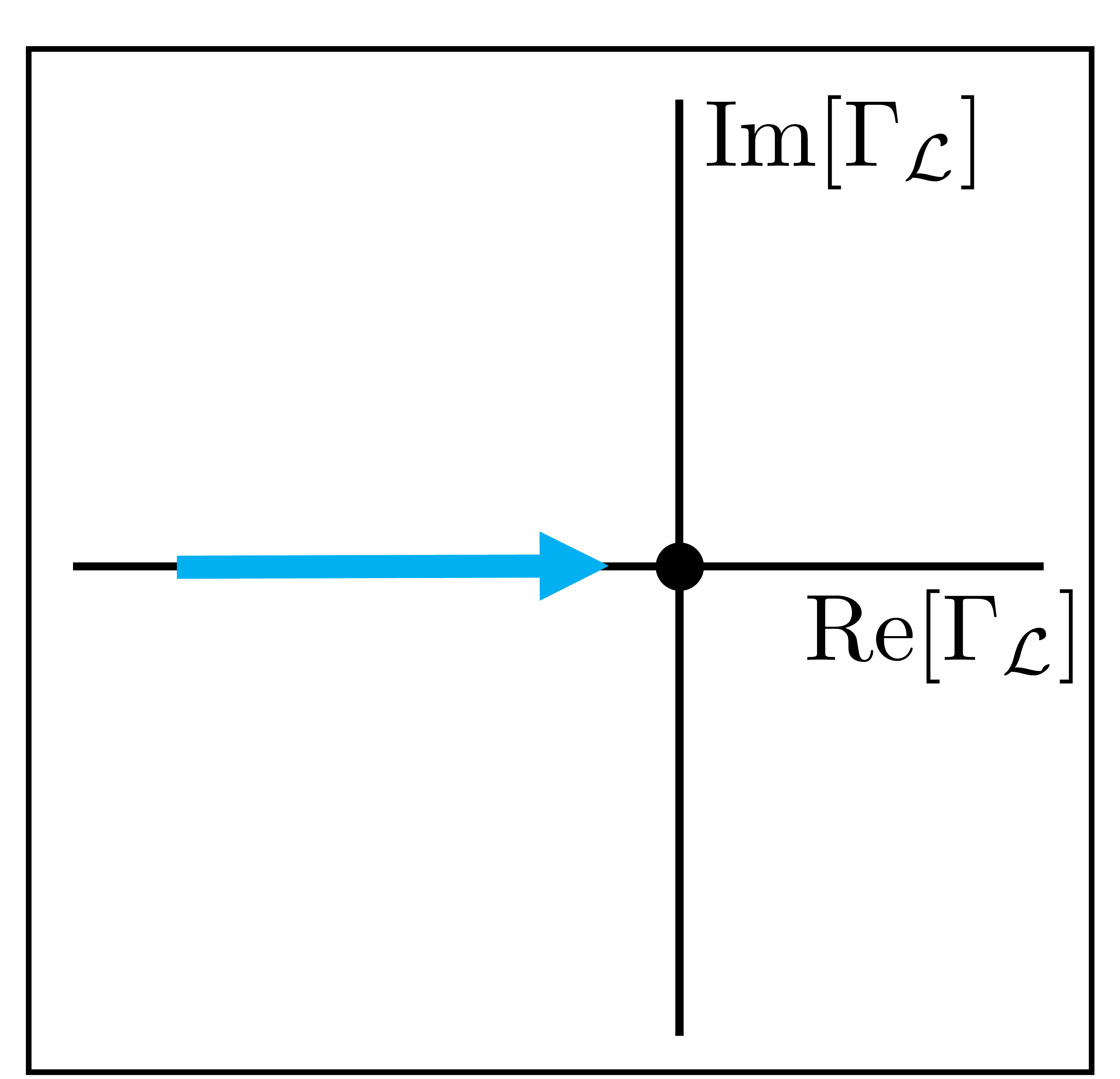} & 					\includegraphics[scale=.11,trim= 0 -7 0 -7]{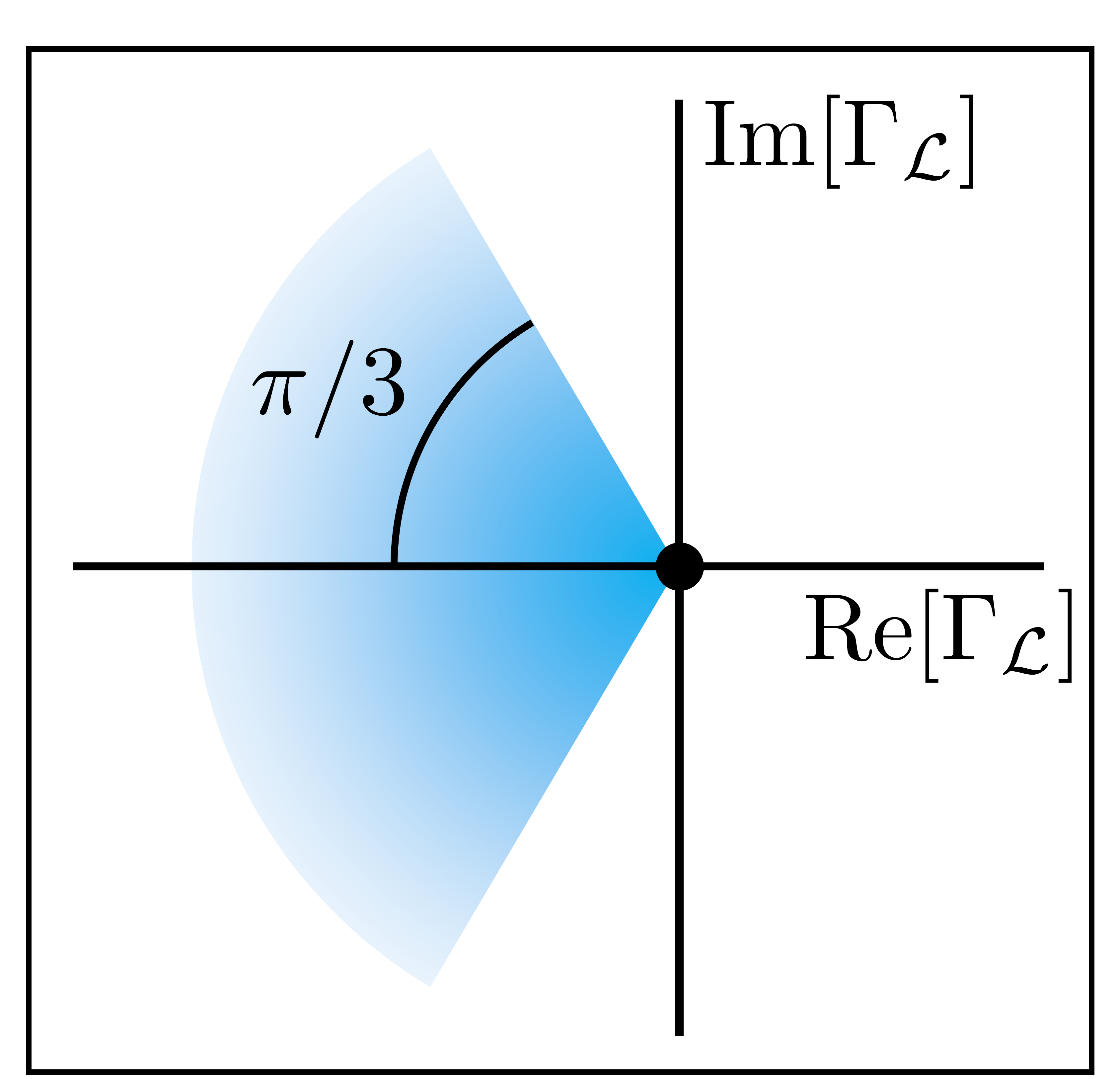} \\
			\hline 
			\multicolumn{2}{|c|}{(d) Critical exponents} \\
			\hline
			$\text{Im} [\nu^{-1} ] = 0$ & $\nu^{-1} = 2 - \left(\frac{1}{2} \pm \frac{i}{2 \sqrt{3}} \right) \epsilon$  \\
			$\eta >0$ & $\eta =  \frac{\epsilon^2}{36}(1 - 12 \log(4/3))<0$ \\
			$\eta' = \eta$ & $\eta' =  \frac{\epsilon^2}{36}$ \\
			$z - 2 = \eta' ( 6 \log(4/3) - 1)$ & $z - 2 = \eta' ( 6 \log(4/3) - 1)$ \\
			\hline
		\end{tabular} 
		\caption{Summary of the main features of the NEFPs contrasted with effective equilibrium fixed points. (a) Schematic correlation functions. A generic continuous scale invariance characteristic of criticality is reduced to a discrete scale invariance at the NEFP. 
			(b) Effective temperatures $T_i$ representing the two order parameters as a function of the length scale ($q^{-1}$). The two temperatures become identical at long length scales, but while they approach a constant at the equilibrium fixed point, they diverge at large scales at the NEFPs. (c) Gap closure upon approaching the critical point. $\Gamma_{\mathcal{L}}$ denotes the Liouvillian gap with the real part describing the relaxation rate (a.k.a. the dissipative gap) and the imaginary part characterizing the ``coherence gap''.
For the equilibrium fixed point, the gap can close only along the real line, indicated by the arrow. In contrast, the gap for the NEFP can take complex values and close along any path lying in the shaded region, making a maximum angle of $\pi/3$ with the real line.
			(d) Critical exponents to lowest nontrivial order in $\epsilon = 4-d$. The exponent $\nu$, typically associated with the divergence of the correlation length, becomes complex-valued at the NEFPs with its imaginary part characterizing the discrete scale invariance [cf.~part (a)]. $\eta$ and $\eta'$ are anomalous dimensions characterizing fluctuations and dissipation with $\eta\ne \eta'$ at the NEFPs indicating the violation of the fluctuation-dissipation theorem. $z$ is the dynamical critical exponent.
			\label{figsum}}
	\end{figure}

\begin{figure}
		\centering
			\includegraphics[scale=.64]{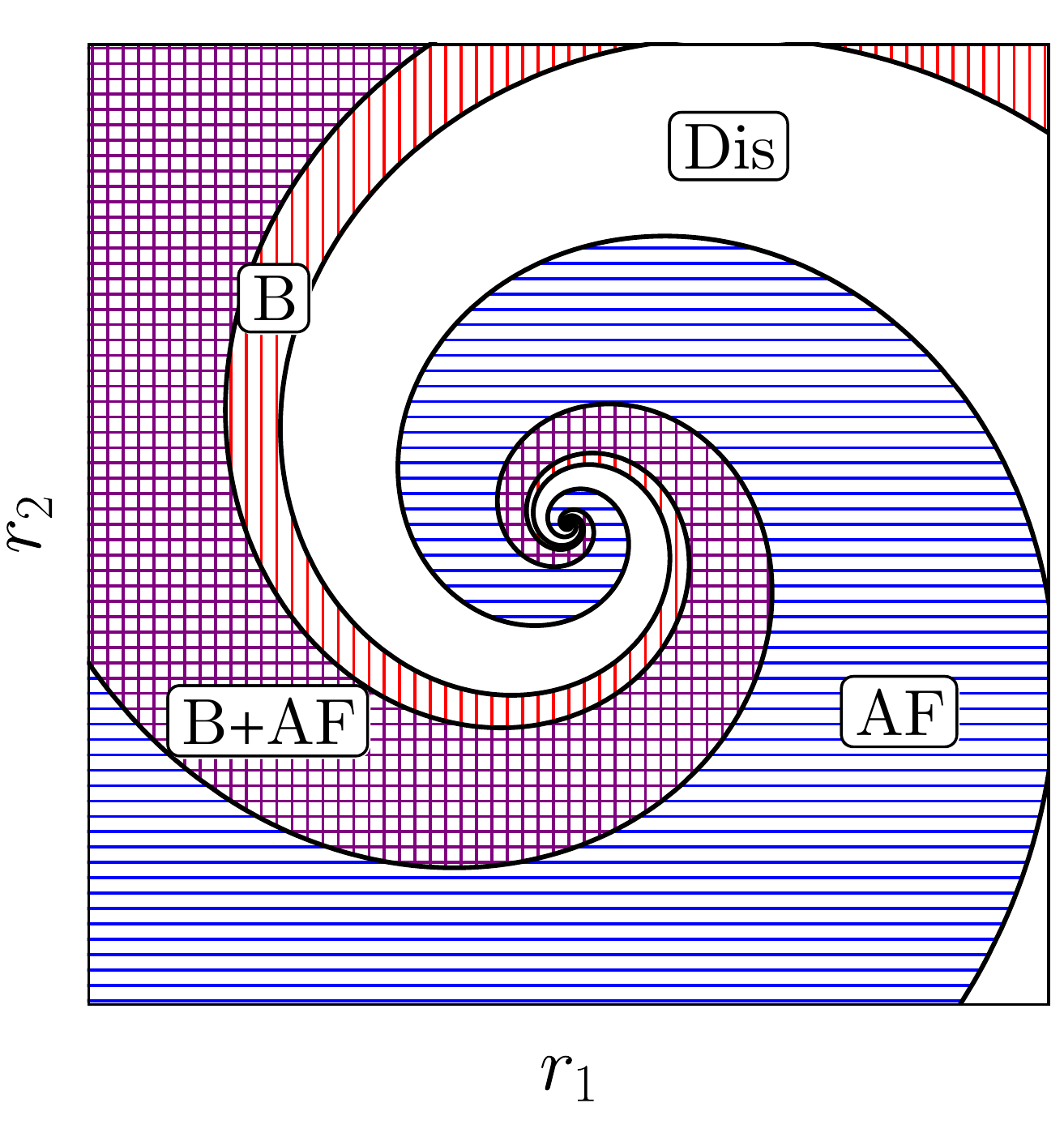}
		\caption{Phase diagram associated with the NEFPs of the non-equilibrium Ising model of two coupled fields for $h=0$. The white region indicates the disordered phase, B (red vertical shading) corresponds to the phase where the bistability order parameter undergoes spontaneous symmetry breaking, AF (blue horizontal shading) denotes antiferromagnetic ordering, and B$ + $AF (purple square shading) corresponds to the phase where both order parameters are nonzero. The solid black lines denote second-order phase transitions. The NEFP phase diagram exhibits logarithmic spirals in the phase boundaries. The other NEFP is described by an analogous diagram 
		upon switching the roles of the two order parameters (B $\leftrightarrow$ AF, $r_1 \leftrightarrow r_2$).  \label{phasediags} } 
	\end{figure}

	Critical points are generically associated with a continuous scale invariance where the system becomes self-similar upon an arbitrary rescaling of the momentum and frequency. However, due to the ``log-periodic'' function in the scaling functions, the correlations are self-similar upon the rescaling 
	$\mathbf{q} \to b_* \, \mathbf{q}$ and $\omega \to b_*^{z} \,\omega$
	for a particular scaling factor 
	\begin{equation}
	b_*=e^{2 \pi\nu''},
	\end{equation}
	or any integer powers thereof. Rather than a typical continuous scale invariance, this behavior is indicative of a \textit{discrete} scale invariance, which is reminiscent of fractals, shapes that are self-similar under particular choices of scaling \cite{Sornette1998}. A schematic depiction of the correlation functions with discrete scale invariance is shown in Fig.~\ref{figsum}(a). 
	Additionally, since the origin of the discrete scale invariance is the scaling behavior of $r_i$ that characterize the distance from the critical point, the phase boundaries themselves also exhibit a form of discrete scale invariance in $r_i$; see Fig.~\ref{phasediags} and the discussion in the next subsection titled ``Phase diagram''. 
	
	The critical exponents $\eta$ and $\eta'$ characterize the anomalous dimensions corresponding to fluctuations and dissipation, respectively. In an equilibrium setting, the fluctuation-dissipation theorem dictates that the correlation and response functions are related as \cite{Tauber2014}
	\begin{equation}
	C(\mathbf{q}, \omega) = \frac{2 T}{\omega} \text{Im} \chi (\mathbf{q},\omega).
	\end{equation}
	(We have assumed the classical limit of the fluctuation-dissipation theorem at low frequencies and at a finite temperature.) 
	In an equilibrium setting, the temperature $T$ is just a constant set by an external bath and thus is scale invariant. Therefore, the overall scaling behavior of the correlation and response functions is identical apart from the dynamical scaling (due to $\omega^{-1}$ on the rhs of the above equation) set by the critical exponent $z$.
	This in turn puts a constraint on the critical exponents as $\eta = \eta'$ for effectively equilibrium phase transitions. However, we find $\eta \neq \eta'$ at the NEFPs, indicating the violation of the fluctuation-dissipation theorem and resulting in a new exponent characterizing the non-equilibrium nature of the fixed point.
	This in turn results in an effective temperature 
	that remains scale-dependent at all scales.
	Inspired by the fluctuation-dissipation theorem, we define an effective ``temperature'' as
	\begin{equation}
	\mbox{Env}[ C(\mathbf{q}, \omega)] = \frac{2 T^{\rm eff}(\mathbf{q},\omega)}{\omega} \mbox{Env} [\text{Im}\, [\chi (\mathbf{q},\omega)]].
	\end{equation}
	To factor out the log-periodic nature of the correlation and response functions, we have made a convenient choice by postulating a fluctuation-dissipation relation between the envelope (Env) functions of the correlation and response functions. This relation can be defined via either the upper or lower envelope functions.
We can then identify the scaling behavior of the effective temperature at the NEFP, which we find to be $T^{\rm eff} \sim |\mathbf{q}|^{\eta-\eta'}$ at long wavelengths and fixed $\omega/|\mathbf{q}|^z$. Interestingly, we find that $\eta'>\eta$, so the system gets ``hotter'' and ``hotter'' at longer and longer scales. 
Of course, the divergence of the effective temperature at long wavelengths does not imply an infinitely energetic state; rather, it reflects the fact that, at longer wavelengths, the correlation function is increasingly larger compared to the response function than one would expect in an equilibrium setting based on the fluctuation-dissipation theorem.
	This behavior is illustrated in Fig.~\ref{figsum}(b) individually for the two effective temperatures corresponding to the two order parameters. At long wavelengths, these effective temperatures become identical to each other and to $T^{\rm eff}(\bq,\omega)$ defined above.
	Finally, the values of the critical exponents at the NEFPs are provided to the lowest nontrivial order in $\epsilon = 4-d$
	in Fig.~\ref{figsum}(d).
	
	\subsection{Phase diagram}

	The critical point described by the new fixed points is a tetracritical point. In the vicinity of the tetracritical point (with $h=0$), there are four different phases where none, one, or both of the order parameters undergo a continuous phase transition. 
	A particularly exotic feature of the phase diagram is that it exhibits spiraling phase boundaries. This leads to the discrete scale invariance of the phase diagram itself, a property that follows from the same feature of the scaling functions in Eq.~(\ref{scaling fns}). In contrast, depending on the microscopic model, the equilibrium fixed points can give rise to either a bicritical point---in which case there will not be a phase where both order parameters undergo a continuous phase transition---or a tetracritical point; neither of these will exhibit spiraling phase boundaries. 
	 Note that since the $\mathbb{Z}_2$ symmetry associated with the bistability transition ($\phi_1 \to - \phi_1$) is only an emergent one (when $h=0$), the full phase diagram (including $h\ne 0$) can be better described as a three-dimensional plot that also includes the first-order phase transitions characteristic of bistability; see Fig.~\ref{tetradiag}. 
	The contrast between the equilibrium fixed points and NEFPs can further provide a route to experimentally identify the new fixed points. 
	An overview of the properties of bicritical and tetracritical points in equilibrium systems can be found in Refs.~\cite{Fisher1965,Nelson1974,Bruce1975,Kosterlitz1976,Folk2008a,Folk2008b,Eichhorn2013}.

\subsection{Spectral properties}

\begin{figure}
\centering
\includegraphics[scale=.16]{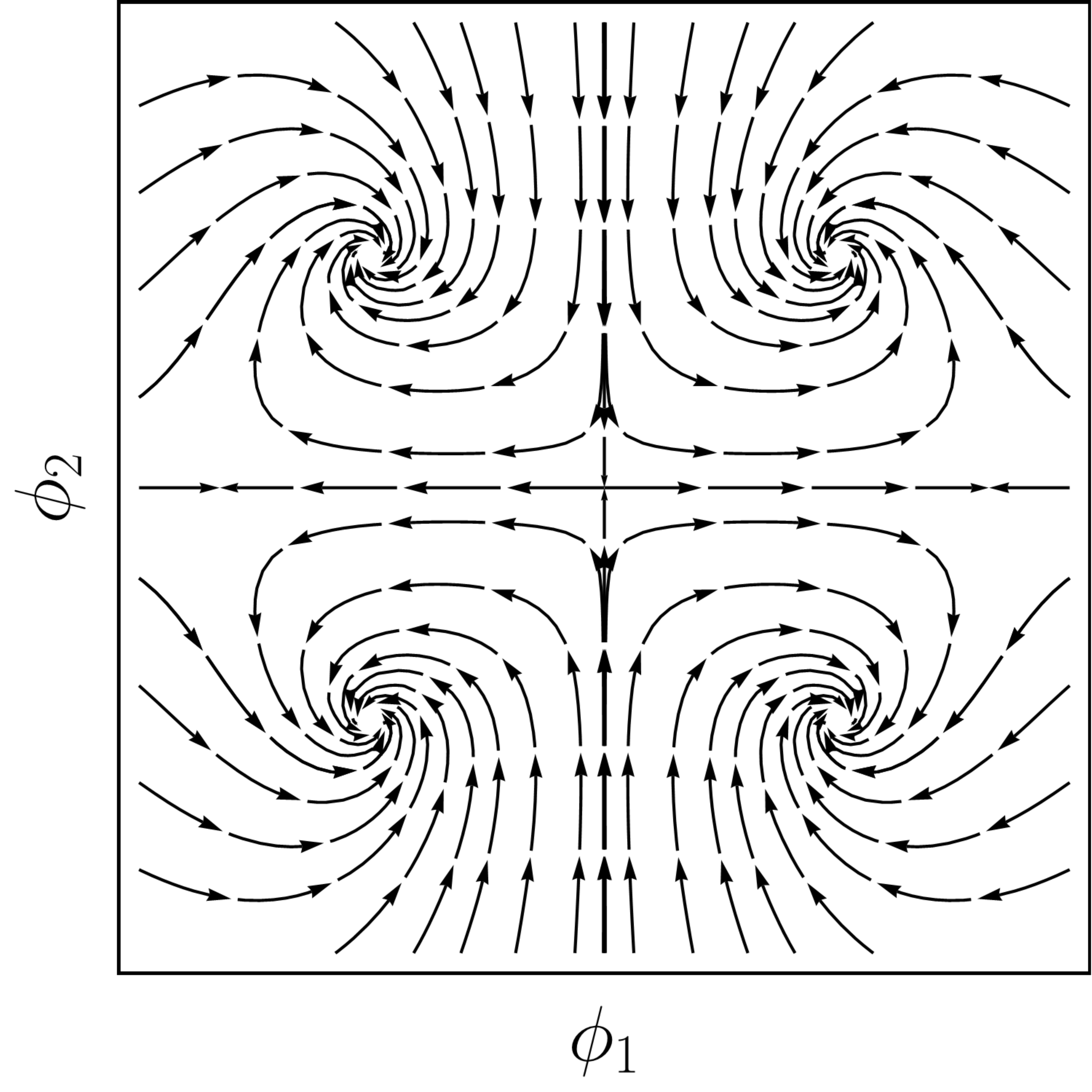}
\caption{Mean field dynamics near the NEFP within the doubly ordered phase with $|M_1|=|M_2|$. The arrows denote how the fields $\phi_i$ evolve in time, with four possible steady states. At each steady state, there is a dissipative relaxation process as well as a ``coherent'' rotation, resulting in a spiraling relaxation to the steady state. Two of the steady states spiral clockwise while two spiral counterclockwise. \label{mfspiral}}
\end{figure}

The NEFP can be further distinguished by its particular dynamics that governs the relaxation of the system to the steady state. In the non-equilibrium setting of our model, the dynamics is described by the Liouvillian $\mathcal{L}$ via [cf.~Eq.~(\ref{modeq2})]
\begin{equation}
\partial_t \rho = \mathcal{L} [ \rho],
\end{equation}
rather than a Hamiltonian. However, in analogy with the ground state 
that is described by the smallest eigenvalue of the Hamiltonian, the steady state(s) is given by the 0 eigenvalue(s) of the Liouvillian; all the other eigenvalues of the Liouvillian have a negative real part characterizing the decay into the steady state. Furthermore, the spectral gap of the Hamiltonian is naturally generalized to the eigenvalue of the Liouvillian with the smallest (in magnitude) nonzero real part. We denote this eigenvalue by $\Gamma_{\mathcal{L}}$.
For a continuous phase transition, just like the spectral gap, the closing of the Liouvillian gap results in the divergence of a time scale associated with a slow or soft mode of the dynamics. The fashion that the latter gap closes reveals characteristic information about the nature of the phase transition.
In equilibrium phase transitions at finite temperature, this gap becomes real (i.e., purely dissipative) as the critical point is approached. Even when the microscopic dynamics is far from equilibrium, the Liouvillian gap may (and typically does) become real, leading to effectively thermal equilibrium. In contrast, the dynamics near the NEFPs can close away from the real axis.  This indeed occurs in the doubly-ordered phase; let $M_i = \langle \phi_i \rangle \neq 0$ define the nonzero order parameters and redefine the fields as $\phi_i \to \phi_i + M_i$. We then find the linearized equations of motion as
\begin{subequations}
\begin{equation}
\zeta_1 \partial_t \phi_1 = -2 g_1 M_1^2 \phi_1 -2 g_{12} M_1 M_2 \phi_2 ,
\end{equation}
\begin{equation}
\zeta_2 \partial_t \phi_2 = -2 g_2 M_2^2 \phi_2 - 2 g_{21} M_1 M_2 \phi_1 ,
\end{equation}
\end{subequations}
where, at the NEFPs, $g_{12}^*=-g_{21}^*$ and $g_1^* = g_2^*$ while we can choose $\zeta^*_1/\zeta^*_2=1$; noise, gradient, and higher-order terms have been dropped. Due to the opposite signs of $g_{12}$ and $g_{21}$ in the two equations, we find a spiral relaxation to the steady state. This relaxation in turn is characterized by a complex Liouvillian gap---defined by a conjugate pair of complex eigenvalues---which exhibits both a dissipative (real) and a ``coherent'' (imaginary) part depending on the values of $M_1$ and $M_2$. We find that when $|M_1| = |M_2|$, the angle of this complex gap relative to the real line achieves its maximum value of $\pi/3$. This is illustrated in Fig.~\ref{figsum}(c). The corresponding mean-field relaxational dynamics is illustrated in Fig.~\ref{mfspiral}. 

\section{Model}

\label{model}

The representative model we have focused on is a driven-dissipative system of weakly interacting bosons defined in Eqs.~(\ref{modeq1},\ref{modeq2}). In order to understand how this model gives rise to bistability and antiferromagnetic ordering, we  begin this section with a detailed discussion of mean field theory and  corrections, or fluctuations, on top of the mean field solutions. Along the way, we will identify the soft modes of the dynamics that ultimately describe the critical behavior of the multicritical point. Finally, we conclude this section by presenting a mapping of our non-equilibrium model to a model of coupled Ising-like order parameters with a $\mathbb{Z}_2 \times \mathbb{Z}_2$ symmetry, corresponding to the sublattice symmetry as well as the emergent Ising symmetry due to bistability.

\subsection{Mean field theory}

In order to analyze the phase diagram of our model, we begin with a mean-field analysis, in which we assume different sites are uncorrelated; that is, for any two operators $A_i$ and $B_j$ on neighboring sites, we have ~$\langle A_i B_j \rangle = \langle A_i \rangle \langle B_j \rangle$ \cite{Rokhsar1991,Diehl2010}. Additionally, we assume that individual sites are described by coherent states. While the latter assumption follows from the former in our model, this will generally not be the case, for example, when on-site Hubbard interactions are present. 
However, a systematic path-integral formalism (adopted in subsequent sections) beyond mean field theory is perfectly suited to analyzing the latter type of interaction.
Finally, in anticipation of the antiferromagnetic phase transition, we separate the system into two sublattices a/b and assume each to be described by a single coherent state. 

Following these assumptions and using the fact that $\partial_t \langle O \rangle = \Tr(\dot{\rho} O)$ for an arbitrary operator $O$, the resulting mean field equations of motion are given by
\begin{subequations}\label{mean field}
\begin{equation}
i \dot{\psi}_\ra = (- \Delta - i \Gamma/2 )\psi_\ra  - \mathfrak{z} J \psi_\rb + \mathfrak{z} V |\psi_\rb|^2 \psi_\ra + \Omega,
\end{equation}
\begin{equation}
i \dot{\psi}_\rb = (- \Delta - i \Gamma/2 )\psi_\rb  - \mathfrak{z} J \psi_\ra + \mathfrak{z} V |\psi_\ra|^2 \psi_\rb + \Omega,
\end{equation}
\end{subequations}
where $\psi_i$ corresponds to the coherent state $\langle a \rangle$ on sublattice $i\in\{{\ra},{\rb}\}$ and $\mathfrak{z}$ is the coordination number; from here on, we absorb $\mathfrak{z}$ in the microscopic parameters via $\mathfrak{z} J \to J$ and $\mathfrak{z}V \to V$. It is clear from these equations that the density-density interaction behaves as an effective detuning that depends on the density of the other sublattice. This results in physics that is similar to Rydberg excitations in stationary atoms, in which the presence or absence of a Rydberg excitation on one site can either prevent (blockade) \cite{Jaksch2000,Lukin2001,Urban2009} or facilitate (antiblockade) \cite{Ates2007,Amthor2010} a Rydberg excitation on a neighboring site by shifting it away or towards the effective resonance. 

Setting $\psi_\ra = \psi_\rb$, one can immediately see that the mean-field equations become identical to those describing bistability; cf.~Ref.~\cite{FossFeig2017} where the nonlinearity due to the Hubbard interaction should be replaced by the density-density interactions in this context. 
The emergence of bistability can be understood in simple terms: Away from resonance, there is a low population on each site. 
However, once a sufficient number of sites are highly excited, they begin to facilitate the excitation of neighboring sites, resulting in a high-population steady state. This process occurs when the shift in detuning due to interactions is comparable to the detuning. 
This condition is satisfied approximately when
\(
\frac{\Omega^2}{\Gamma^2/4+(\Delta+J)^2} V \approx \Delta+J,
\)
where $J$ behaves like an effective detuning while the product of the interaction strength $V$ and the non-interacting steady-state population [$\Omega^2/(\Gamma^2/4+(\Delta+J)^2)$] 
gives the interaction-induced shift of the detuning. For $\Gamma \gtrsim \Delta+J$, this reasoning becomes blurred 
as the drive is effectively always on resonance due to the larger linewidth. As a result, a finite region of bistability emerges with low- and high-population steady states.
Beyond mean field theory, the bistable region is replaced by a line of first-order phase transitions that terminates at a critical point.

The presence of antiferromagnetic ordering in this system can be understood by inspecting the role of the density-density interactions. Since the interaction affects neighboring sites only, the blockade effects occur between sublattices but not within each sublattice. For example, if one sublattice has a high population, it can prevent further excitations in the other sublattice. 
Similar to the case of bistability, the phase boundary occurs approximately when the shift in detuning due to interactions takes the system out of resonance. This approximately occurs when
\(
|\frac{\Omega^2}{\Gamma^2/4+(\Delta+J)^2} V - \Delta- J|\gtrsim \Gamma,
\)
i.e., when one sublattice is effectively more than a linewidth out of resonance due to interactions.  
Unlike bistability, this process does not break down as $\Gamma$ and $\Omega$ are increased. As the decay $\Gamma$ is increased, the drive strength $\Omega$ can be further increased so that the interaction-induced shift in the detuning compensates for the increase of the linewidth.

\begin{figure*}
\subfloat[]{
\includegraphics[scale=.1]{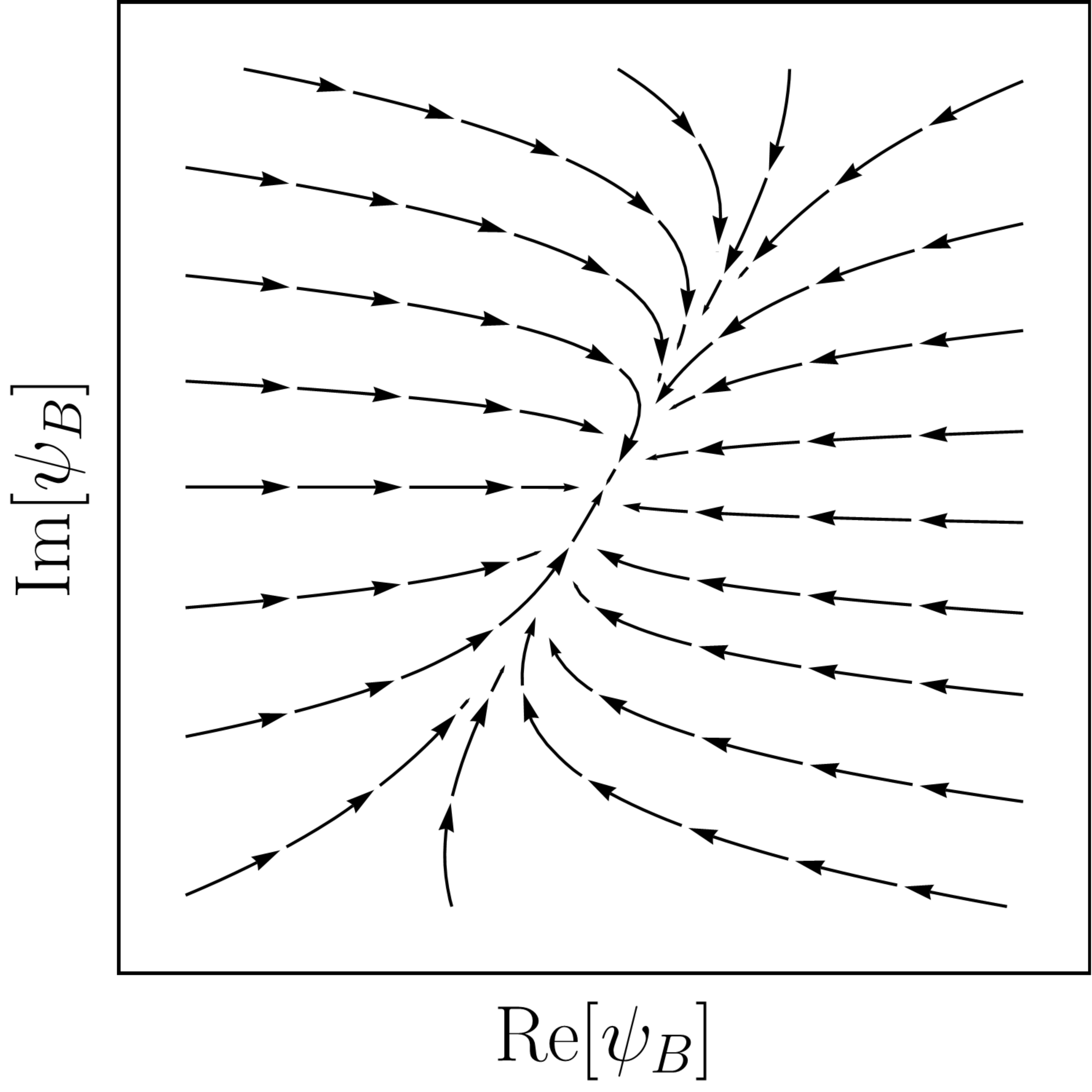}
}
\hspace{.5cm}
\subfloat[]{
\includegraphics[scale=.1]{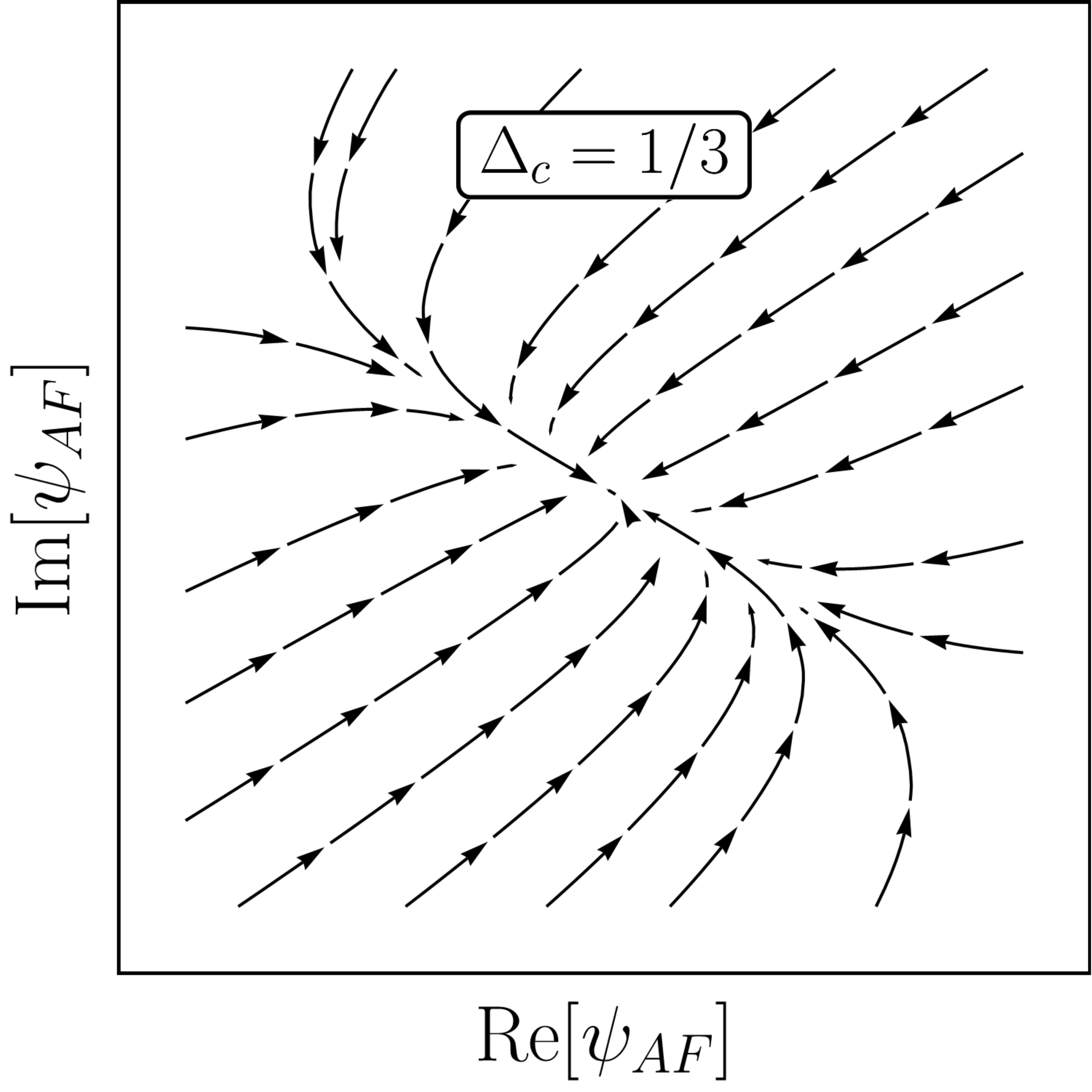}
}
\hspace{.5cm}
\subfloat[]{
\includegraphics[scale=.1]{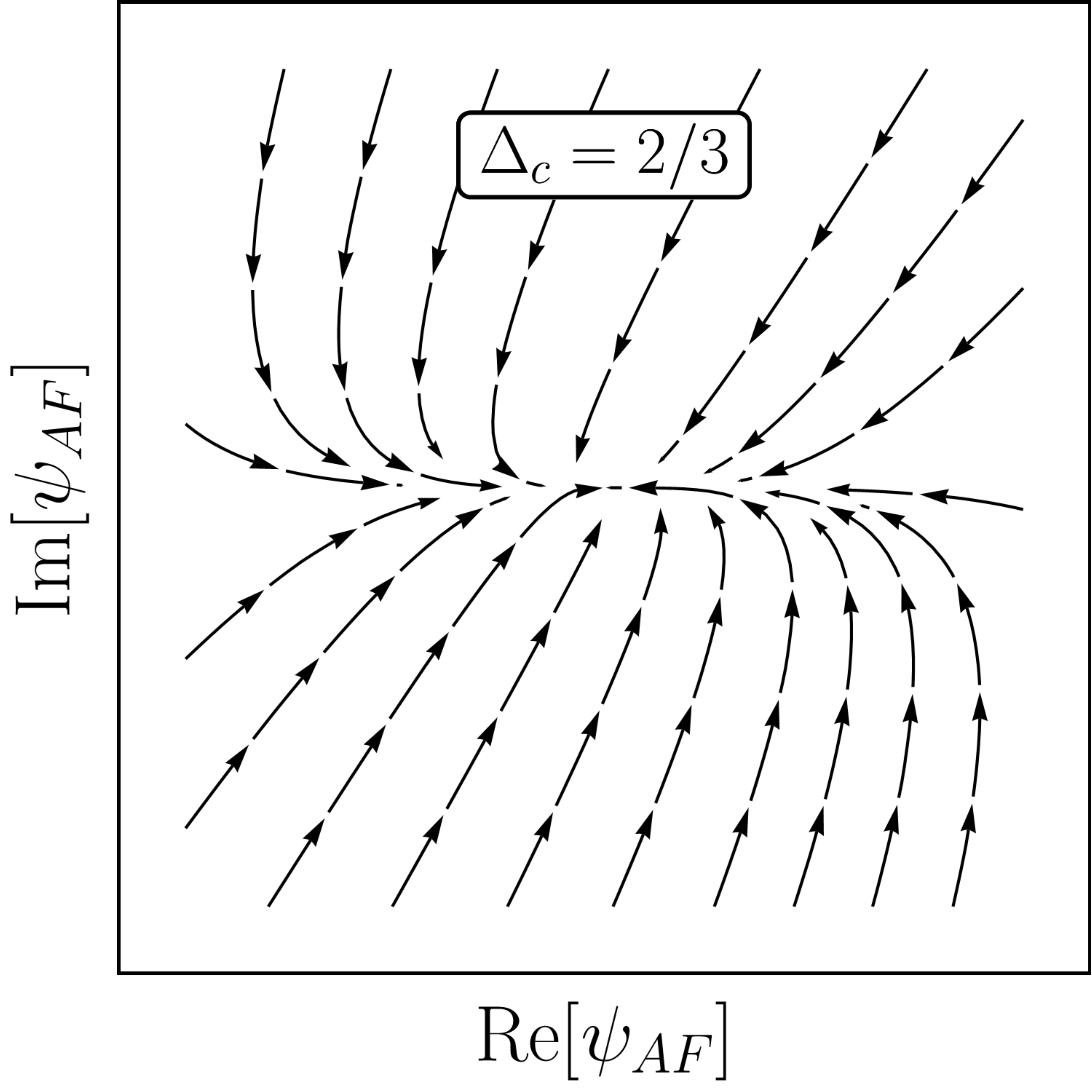}
}
\caption{Dynamics of gapped (fast or massive) and soft (slow or massless) modes with arrows indicating the linear (in $\phi_i, \phi_i'$) relaxation of the two modes. Near the critical point, the gapped mode quickly decays to a straight line defined by the slow direction of the soft mode. (a) Relaxation of the field $\psi_\text{B}$ with the soft and gapped modes lying along the angles $\pi/3$ and 0, respectively. (b) Relaxation of the field $\psi_\text{AF}$ for $\Delta_c = 1/3$ with the soft and gapped modes lying along the angles $-\pi/6$ and $\pi/6$ respectively. (c) Relaxation of the field $\psi_\text{AF}$ for $\Delta_c = 2/3$ with the soft and gapped modes lying along the angles $0$ and $\pi/3$ respectively. (We have adopted units where $\Delta+J=1$.)\label{modes}}
\end{figure*}

In order to better understand the mean-field structure of the  model, it is convenient to introduce a new set of fields corresponding to the two order parameters as
\begin{subequations}\label{BAF}
\begin{equation}
\psi_\text{B} = \frac{\psi_\ra + \psi_\rb}{2},
\end{equation}
\begin{equation}
\psi_\text{AF} = \frac{\psi_\ra - \psi_\rb}{2}.
\end{equation}
\end{subequations}
The field $\psi_\text{B}$ captures the effects of bistability while $\psi_\text{AF}$ describes the  antiferromagnetic ordering. The mean-field equations can in turn be cast in terms of these fields as
\begin{subequations}
\begin{equation}
i \dot{\psi}_\text{B} = (- \Delta - J - i \Gamma/2) \psi_\text{B} + V(\psi_\text{B}^2-\psi_\text{AF}^2) \psi_\text{B}^* + \Omega,
\end{equation}
\begin{equation}
i \dot{\psi}_\text{AF} = (- \Delta + J - i \Gamma/2) \psi_\text{AF} + V(\psi_\text{AF}^2-\psi_\text{B}^2) \psi_\text{AF}^*.
\end{equation}
\end{subequations}
At the multicritical point, $\psi_\text{AF} = 0$ and the equation governing the steady-state value of $\psi_\text{B}$ is no different than if we had not considered antiferromagnetic ordering. Thus, the critical values of $\Delta+J$, $V$, $\Omega$ as well as the steady-state value of $\psi_\text{B}$ are determined according to the critical point associated with bistability only. This leaves a single free parameter in the equation of motion for $\psi_\text{AF}$: $\Delta-J$. By properly tuning the latter parameter, the antiferromagnetic phase boundary can be manipulated so that it intersects the critical point associated with bistability. Working in units of $\Delta +  J = 1$, two multicritical points occur at
\begin{equation}\label{Critical point}
\begin{gathered}
(\Delta_c, J_c) = \left(\frac{1}{3},\frac{2}{3}\right) \quad \mbox{or} \quad\left(\frac{2}{3}, \frac{1}{3}\right), \\
\hskip -1cm\mbox{and}\quad \Gamma_c = \sqrt{4/3}, \ \ \Omega_c = (2/3)^{3/2}/\sqrt{V},
\end{gathered}
\end{equation}
as well as $\Psi_c = \sqrt{2/3V} e^{-i \pi/3}$ as the steady-state value of $\psi_\text{B}$ at the critical point (by virtue of symmetry, $\psi_\text{AF}=0$ there).

The two fields $\psi_{B/AF}$ are complex-valued, thus comprising four real (scalar) fields. However, given the Ising nature of each transition, we must anticipate that two scalar fields would be sufficient to describe the critical behavior of both types of ordering. Indeed, we find that, at the multicritical point, two massless fields emerge---defined by  appropriate components of the original fields---corresponding to the soft (or slow) modes $\phi_i$, while the other components $\phi_i'$ remain massive and are therefore noncritical (or fast). We then adiabatically eliminate the two noncritical modes by setting $\dot{\phi}_i' = 0$ and solving for $\phi_i'$ in terms of $\phi_i$. Upon substituting our solutions for the massive fields into $\dot{\phi}_i$, we find an effective description in terms of the soft modes. We closely follow Refs.~\cite{Maghrebi2016,FossFeig2017} to identify these modes. For the bistability order parameter, we can identify
\begin{equation}
\psi_\text{B} = \Psi_c+e^{i \pi/3} \phi_1+\phi_1' ,
\end{equation}
with  the real fields $\phi_1$ and $\phi_1'$ characterizing the slow and fast modes, respectively.
A similar identification has been made in Refs.~\cite{Maghrebi2016,FossFeig2017}; see also Refs.~\cite{Drummond05,Dechoum16} for a similar reasoning, although the slow and fast modes identified there make an angle of $\pi/2$.
For the antiferromagnetic field, the massless and massive components 
depend on the choice of the multicritical point in Eq.~(\ref{Critical point}) as 
\begin{subequations}
\begin{align}
\Delta_c=1/3:&& \psi_\text{AF} = \frac{1}{\sqrt{3}}\left( e^{- i \pi/6} \phi_2+  e^{i \pi/6}\phi_2'\right),\\
\Delta_c = 2/3:&&\psi_\text{AF} = \frac{1}{\sqrt{3}}\left(\phi_2 + e^{i \pi/3} \phi_2'\right).
\end{align}
\end{subequations}
Again, the unprimed fields are massless while the primed fields are massive. The slow and fast modes of the fields are illustrated pictorially in Fig.~\ref{modes}.

Next we adiabatically eliminate the massive modes to find an effective description in terms of the soft modes. Including the gradient terms---describing the coupling between neighboring sites---as well as the noise terms due to the coupling to the environment, we find the Langevin equations 
\begin{subequations}
\begin{multline}
\zeta_1 \dot{\phi_1} = h - r_1 \phi_1 + D_1 \nabla^2 \phi_1 + \xi_1 \\
+ A_{20} \phi_1^2 + A_{02} \phi_2^2 + A_{12} \phi_1 \phi_2^2+ A_{30} \phi_1^3,
\end{multline}
\begin{multline}
\zeta_2 \dot{\phi_2} = -r_2 \phi_2 + D_2 \nabla^2 \phi_2 + \xi_2\\
 + B_{11} \phi_1 \phi_2 + B_{21} \phi_1^2 \phi_2 + B_{03} \phi_2^3,
\end{multline}
\end{subequations}
with Gaussian noise
\begin{equation}
	\langle \xi_i(t,\mathbf{x}) \xi_j(t', \mathbf{x}') \rangle = 2 \zeta_i T_i\delta_{ij} \delta(t-t') \delta(\mathbf{x}-\mathbf{x}').
\end{equation}
Higher-order terms that are irrelevant in the sense of RG have been neglected. We have expressed the noise coefficients in a convenient notation that mimics the dissipative dynamics in thermal equilibrium, in spite of the underlying non-equilibrium dynamics. 
Finally, the details of the adiabatic elimination together with the explicit values of all the coefficients ($h$, $r$s, $D$s, $A$s, $B$s, $\zeta$s, and $T$s) in terms of microscopic parameters of the model are provided in 
Appendix \ref{CH}. 

It turns out that, at the level of mean field analysis, $A_{20} =0$ in the vicinity of the multicritical point. The resulting mean-field dynamics (neglecting the gradient and noise terms) of the two soft modes is then described by a cusp-Hopf bifurcation; a detailed analysis of this type of bifurcation can be found in Ref.~\cite{Harlim2007}. However, since $A_{20}$ is not protected by any symmetries, the corresponding term can be generated in the course of RG and become of the order of the other quadratic terms. While we focus on the multicritical points, further details about the full mean field phase diagram of our model and slight variations on it can be found in Refs.~\cite{Jin2013,Jin2014}.

\subsection{Non-equilibrium Ising model for two fields}

Before proceeding with our perturbative RG analysis, it is important to identify what are known as redundant operators. These are terms in the action which are generated under suitably local symmetry-preserving transformations of the fields. Since such a transformation should not change the long-distance behavior of the system, this redundancy can be used to simplify our analysis. This is an important step for perturbative RG and to identify the upper critical dimension; see Ref.~\cite{Wegner1974} for a discussion of redundant operators in an equilibrium setting. 

As a simple illustrative example, consider the generic Hamiltonian of a scalar field $\phi$ in the absence of a ${\mathbb Z}_2$ symmetry:  
\begin{equation}\label{free energy}
\mathcal{H} = \int d^d\bx \left[(\nabla \phi)^2 - h \phi+ r \phi^2 + u_3 \phi^3 +u \phi^4\right].
\end{equation}
Shifting the field by a constant as $\phi \to \phi + \phi_0$, the Hamiltonian is given by the same expression (up to an unimportant additive constant) with possibly different coefficients. This underscores a redundancy in Hamiltonians that describe the same physical system. The change of the Hamiltonian $\Delta \mathcal{H}$ (or rather the integrand) due to a constant shift of the field defines a redundant operator.
In particular, the cubic term transforms as
\begin{equation}
u_3 \to u_3 + 4 u \phi_0 .
\end{equation}
By choosing the value of $\phi_0$ properly, the $\phi^3$ term can be dropped from the Hamiltonian while shifting the coefficients of the terms $\phi$ and $\phi^2$ \cite{Cardy2000}. 

Similar to the above example, we should first identify the redundant operators in the non-equilibrium setting of the two coupled scalar fields $\phi_1$ and $\phi_2$.
In this case, we allow for a more general, nonlinear transformation which is suitably local and retains the underlying symmetries. 
We find that the set of redundant operators in our model is sufficient to remove all the quadratic terms in the Langevin equation (or equivalently, the cubic terms in the action, similar to the Hamiltonian in the above example); the details of this analysis are presented in Appendix \ref{Redundant}.
In particular, we find that, under this transformation, the new value for the ratio $A_{12}/B_{21}$ is given by $2 A_{02}/B_{11}$; therefore, the relative sign of the quadratic terms (prior to the transformation) determines the relative sign of the cubic terms in the final equations of motion. In the model that we have considered here, the two quadratic terms have opposite signs (see Appendix \ref{CH}). This fact will be important in determining the fixed point of the RG flow. In fact, we show that the above sign difference leads the system to one of the NEFPs.

With the above considerations, the Langevin equations can finally be brought into a canonical form as 
\begin{subequations}
\label{langevin_eqs}
\begin{equation}
\zeta_1 \partial_t \phi_1 = D_1 \nabla^2 \phi_1 + h - r_1 \phi_1 - g_1 \phi_1^3 - g_{12} \phi_1 \phi_2^2 + \xi_1,
\end{equation}
\begin{equation}
\zeta_2 \partial_t \phi_2 = D_2 \nabla^2 \phi_2 - r_2 \phi_2 - g_2 \phi_2^3 - g_{21} \phi_2 \phi_1^2 + \xi_2,
\end{equation}
\end{subequations}
with Gaussian noise
\begin{equation}
	\label{noise}
	\langle \xi_i(t,\mathbf{x}) \xi_j(t', \mathbf{x}') \rangle = 2 \zeta_i T_i\delta_{ij} \delta(t-t') \delta(\mathbf{x}-\mathbf{x}').
\end{equation}
Therefore, the dynamics exhibits a $\mathbb{Z}_2 \times \mathbb{Z}_2$ symmetry when $h=0$, corresponding to the emergent symmetry $\phi_1 \to - \phi_1$ in addition to the sublattice symmetry $\phi_2 \to - \phi_2$. Such emergent symmetry has previously been identified in the bistability transition \cite{Marcuzzi2014,FossFeig2017,Maghrebi2016}; our analysis shows that such symmetry emerges even in the vicinity of a multicritical point where bistability and antiferromagnetic transitions coalesce. We must point out that, even in the absence of the latter symmetry, the sublattice symmetry alone prevents any mixing of the gradient and mass terms between the two fields as well as the noise terms, a property that should hold to all orders of perturbation theory.

\section{Renormalization group analysis}

In this section, we derive the perturbative RG equations to the two-loop order (for reasons that will be explained shortly), identify the fixed points, and characterize the critical exponents that determine the scaling properties of correlations near the multicritical point. 
\label{RGsec}

\subsection{RG equations}

The Langevin-type equations can be cast in terms of the response-function formalism.
This allows us to study our non-equilibrium model by extending the standard techniques of the RG analysis to a dynamical setting;
see, for example, Ref.~\cite{Tauber2014} for more details. The non-equilibrium partition function is defined by $Z=\int {\cal D}[\phi_i,i \tilde\phi_i]e^{-{\mathcal{A}[\tilde{\phi}_i,\phi_i] }}$, where the functional integral measure and the ``action'' $\cal A$ involve both fields $\phi_i$ with $i=1,2$ and their corresponding ``response'' fields ${\tilde \phi}_i$. In the language of Keldysh field theory, $\phi$ corresponds to the classical field while $\tilde{\phi}/2i$ corresponds to the quantum field. The statistical weight of $\phi_i(t,\bx)$ can be obtained by integrating out both response fields as $P[\phi_i] = \int \mathcal{D}[i \tilde{\phi}_i] e^{- \mathcal{A}[\tilde{\phi}_i,\phi_i]}$.
While the partition function $Z=1$ by construction, the expectation value of any quantity---the fields themselves or their correlations---
can be determined by computing a weighted average in the partition function. 
For our model defined by Eqs.~(\ref{langevin_eqs},\ref{noise}), we write the action as the sum of quadratic and nonlinear (beyond quadratic) terms 
\begin{subequations}\label{MSR}
\begin{equation}
\mathcal{A}[\tilde{\phi}_i,\phi_i] = \mathcal{A}_0[\tilde{\phi}_i,\phi_i] + \mathcal{A}_{int}[\tilde{\phi}_i,\phi_i],
\end{equation}
with the quadratic action given by
\begin{equation}
\mathcal{A}_0[\tilde{\phi}_i,\phi_i] =  \int_{t,\mathbf{x}} -h \tilde{\phi}_1 + \sum_i  \tilde{\phi}_i (\zeta_i \partial_t  - D_i \nabla^2  + r_i )\phi_i - \zeta_i T_i \tilde{\phi}_i^2,
\end{equation}
and the nonlinear interaction terms 
\begin{equation}
\mathcal{A}_{int}[\tilde{\phi}_i,\phi_i] = \int_{t,\mathbf{x}} g_1 \phi_1^3 \tilde{\phi_1} + g_2 \phi_2^3 \tilde{\phi_2} + g_{12} \phi_1 \phi_2^2 \tilde{\phi}_1 + g_{21} \phi_2 \phi_1^2 \tilde{\phi}_2.
\end{equation}
\end{subequations}
Our goal is to determine the RG flow of various parameters in the action and, specifically, of the coefficients $g$ of the interaction terms. 

We begin by considering the subspace defined by $g_{12} g_{21} = 0$ when either $g_{12}=0$ or $g_{21}=0$. 
This subspace is special in that it is closed under renormalization to all orders. The reason is that when $g_{12}=0$ or $g_{21} = 0$, one of the two fields is not affected by the other at the microscopic level, a property that should hold at all scales. 
This can also be understood perturbatively in a diagrammatic scheme: If, say, $g_{12} = 0$, then all diagrams that could generate $g_{12}$ involve a causality violation; hence  it should remain zero to all orders. An important consequence of this fact 
is that the relative sign of $g_{12}$ and $g_{21}$ cannot change, as this would require passing through the closed subspace. 

Before performing the RG analysis, we first
use our freedom in rescaling the fields to cast the action in a more convenient form. In Sec. \ref{sum}, we used this freedom to set both temperatures to unity; here, for the convenience of the RG analysis, we make a different choice. 
Note that rescaling $\phi_2 \to c \phi_2$ and $\tilde{\phi}_2 \to \tilde{\phi_2}/c$ maps $g_2 \to c^2 g_2$, $g_{12} \to c^2 g_{12}$, and $T_2 \to T_2/c^2$. 
Exploiting this freedom, 
we can set the rescaled value of $g_{12}$ to be identical to $g_{21}$ up to a sign. In doing so, we have effectively shifted the renormalization of $g_{12}/g_{21}$ onto $T_1/T_2$, 
simplifying the RG analysis later.
Note, however, since $g_{12}$ is rescaled by a factor $c^2$, this transformation cannot change the relative sign of $g_{12}$ and $g_{21}$. This is indeed consistent with the closure of the $g_{12}g_{21}=0$ subspace discussed above. Having rescaled the fields appropriately, 
we can write the action as (the quadratic action is repeated for completeness)
\begin{subequations}
\begin{equation}
\mathcal{A}_0[\tilde{\phi}_i,\phi_i] = \int_{t,\mathbf{x}} -h \tilde{\phi}_1 + \sum_i  \tilde{\phi}_i ( \zeta_i \partial_t  - D_i \nabla^2  + r_i)\phi_i - \zeta_i T_i \tilde{\phi}_i^2,
\end{equation}
\begin{equation}
\label{Aint}
\mathcal{A}_{int}[\tilde{\phi}_i,\phi_i] = \int_{t,\mathbf{x}} u_1 \phi_1^3 \tilde{\phi_1} + u_2 \phi_2^3 \tilde{\phi_2} + u_{12} \phi_1 \phi_2(\phi_2 \tilde{\phi}_1 + \sigma \phi_1 \tilde{\phi}_2),
\end{equation}
\end{subequations}
where $\sigma = \pm 1$ indicates the relative sign of $g_{12}$ and $g_{21}$ and the coefficients $u_1$, $u_2$, and $u_{12}$ define the rescaled values of the interaction strengths (in an abuse of notation, we use the same notation for the other rescaled parameters of the model as well as the rescaled fields). 

Let us first briefly consider $\sigma=1$, in which case the action can be written in a suggestive form as 
\begin{equation}\label{H form}
\mathcal{A}[\tilde{\phi}_i,\phi_i] = \int_{t,\mathbf{x}} \sum_i \tilde{\phi}_i \left(\zeta_i \partial_t \phi_i +  \frac{\delta {\cal H}}{\delta \phi_i} \right) - \zeta_i T_i \tilde{\phi}_i^2,
\end{equation}
where the function $\cal H$ is given by 
\begin{equation}\label{H}
    {\cal H}=\int_\bx \sum_{i=1,2}\left(\frac{D_i}{2} |\nabla \phi_i|^2 + \frac{r_i}{2} \phi_i^2+ \frac{u_i}{4} \phi_i^4 \right) - h \phi_1 +\frac{u_{12}}{2}\phi_1^2\phi_2^2.
\end{equation}
Put in this form, Eq.~(\ref{H form}) bears close resemblance to an equilibrium setting where the dynamics is governed by a Hamiltonian (in this case, $\cal H$). However, with each field at a different temperature,
their coupled dynamics does not generally satisfy fluctuation-dissipation relations, and thus an (effective) equilibrium behavior cannot be established, at least at the microscopic level. (Note that unlike Sec.~\ref{sum}, we have already used the scaling freedom in redefining the interaction parameters which in turn fixes the ratio $T_1/T_2$.) One then should resort to an RG analysis to determine whether or not effective equilibrium is restored at long wavelengths, that is, if $T_1/T_2\to 1$ under RG. 
We shall see shortly that equilibrium proves to be a robust fixed point even when $T_1/T_2\ne 1$ at the microscopic level. 

In contrast, a Hamiltonian dynamics [similar to Eqs.~(\ref{H form},\ref{H})] is not possible when $\sigma=-1$ since a term proportional to $\phi_1^2\phi_2^2$ in the Hamiltonian leads to equations of motion that couple the two fields with the same coefficient and hence the same sign.
Therefore, in this case, the dynamics cannot flow to an equilibrium fixed point even when $T_1=T_2$, with the exception of a decoupled fixed point where $u_{12}=0$ (or $g_{12}=g_{21}=0$). Indeed, we shall argue that a pair of genuinely non-equilibrium fixed points emerge in this case. 

At a technical level, an RG analysis would be complicated as we need to consider diagrams up to two loops. This is because at one loop, no renormalization occurs for the temperatures (due to causality) as well as the diffusion constants and friction terms (owing to their momentum and frequency dependence). This is while the interaction terms ($u_{1}$, $u_2$, and $u_{12}$) are all renormalized already at one loop.
This observation---besides aesthetic reasons---has motivated the representation adopted here; 
in the original description in terms of $g_{12}$ and $g_{21}$, the ratio $g_{12}/g_{21}$ would not be renormalized at one loop. 

To perform the RG analysis, we first define the renormalized parameters as
\begin{equation}
\begin{array}{ccc}
D_{i_R} = Z_{D_i} D_i, & &r_{i_R} = Z_{r_i} r_i \mu^{-2}, \\ \\
u_{i_R} = Z_{u_i} u_i A_d \mu^{-\epsilon}, & &u_{12_R} = Z_{u_{12}} u_{12} A_d  \mu^{- \epsilon},\\ \\
\zeta_{i_R} = Z_{\zeta_i} \zeta_i, & & T_{i_R} = Z_{T_i} T_i ,
\end{array}
\end{equation}
where $A_d = \Gamma(3-d/2)/(2^{d-1} \pi^{d/2}) $ is a geometrical factor, $\Gamma(x)$ is the Euler's Gamma function, $\mu$ is an arbitrary small momentum scale (compared to the lattice spacing), and $\epsilon = 4 - d$ defines the small parameter of the epsilon expansion. The effect of renormalization is captured in the $Z$ factors that contain the divergences according to the minimal subtraction procedure. We determine these factors perturbatively to the lowest nontrivial order in $\epsilon$ or loops (the details are provided in Appendix \ref{RG}). The lowest-order corrections to $Z_{r}$ and $Z_{u}$ occur at one loop ($\sim\epsilon$), while those of $Z_\zeta, Z_T, Z_D$ appear at two loops ($\sim\epsilon^2$). These perturbative corrections, while having some similarities with their equilibrium counterparts, are more complicated due to their non-equilibrium nature.

Using the above $Z$ factors, we determine the RG flow and beta functions via
\begin{subequations}
\begin{equation}
\label{gammaflow}
\gamma_p = \mu \partial_\mu \ln(p_R/p),
\end{equation}
\begin{equation}
\beta_{u_a} = \mu \partial_\mu u_{a_R},
\end{equation}
\end{subequations}
where $p \in \{r_i,\zeta_i,D_i,T_i\}$ and $u_a\in\{u_1,u_2,u_{12}\}$.
These functions describe the flow of various parameters in the action under the change of the momentum scale $\mu$. In particular, the beta functions identify the fixed points of the interaction coefficients via $\beta_{u_a} = 0$.
At any such fixed point, 
the scaling behavior of the remaining parameters is governed by power laws whose exponents depend on $\gamma_p$. 
Here, we report the beta functions for the interaction parameters $u_a$ (the details are provided in Appendix \ref{RG}):
\begin{subequations}
\label{beta1}%
\begin{equation}
\beta_{u_1} = u_{1_R} \left(-\epsilon + 9 \frac{T_{1_R}}{\zeta_{1_R}^2 \tilde{D}_{1_R}^2} u_{1_R}\right) + \sigma \frac{T_{2_R}}{\zeta_{2_R}^2 \tilde{D}_{2_R}^2} u_{12_R}^2,
\end{equation}
\begin{equation}
\beta_{u_2} = u_{2_R} \left(-\epsilon + 9 \frac{T_{2_R}}{\zeta_{2_R}^2 \tilde{D}_{2_R}^2} u_{2_R}\right) +  \sigma \frac{T_{1_R}}{\zeta_{1_R}^2 \tilde{D}_{1_R}^2}  u_{12_R}^2,
\end{equation}
\begin{multline}
\beta_{u_{12}} = u_{12_R} \left(-\epsilon + 3 \frac{T_{1_R}}{\zeta_{1_R}^2 \tilde{D}_{1_R}^2}  u_{1_R} + 3 \frac{T_{2_R}}{\zeta_{2_R}^2 \tilde{D}_{2_R}^2} u_{2_R} \right. \\
\left. + \frac{4}{\zeta_{1_R} \zeta_{2_R}} \frac{T_1 \tilde{D}_{2_R} + \sigma T_{2_R} \tilde{D}_{1_R}}{\tilde{D}_{1_R}\tilde{D}_{2_R}(\tilde{D}_{1_R}+\tilde{D}_{2_R})} u_{12_R} \right),
\end{multline}
\end{subequations}
\noindent
where we have introduced $\tilde{D}_{i_R} \equiv D_{i_R}/\zeta_{i_R}$. These equations exhibit a number of important features. 
First, for $u_{12_R} = 0$, we can absorb a factor of $T_{i_R}/D_{i_R}^2$ into $u_{i_R}$, leaving the two beta functions for $u_i$ independent of $T_i, D_i, \zeta_i$. We thus immediately recover a pair of uncoupled equilibrium Ising phase transitions, as one should expect. 
Second, under equilibrium conditions where $\sigma = 1$ and $T_{1_R} = T_{2_R}\equiv T_R$, we recover the standard beta functions in equilibrium. In a similar fashion, we can absorb the factors of $T_{R}/D_{i_R}^2$ into $u_{i_R}$ and $T_{R}/(D_{1_R} D_{2_R})$ into $u_{12_R}$, again leaving the beta functions dependent only on the coupling coefficients. This observation underscores the important fact that, in equilibrium, static properties are entirely  decoupled from the dynamics. 
On the other hand, in the setting of our non-equilibrium model, statics and dynamics are inherently intertwined. Indeed, no redefinition of the coupling terms can lead to beta functions that would be independent of $T_{i_R}$ and $\tilde{D}_{i_R}$. 
This is not the case for $\zeta_{i_R}$ as they can always be absorbed in other parameters; for example, we can still absorb $1/\zeta_{i_R}^2$ into $u_{i_R}$ and $1/(\zeta_{1_R} \zeta_{2_R})$ into $u_{12_R}$ in the beta functions. This  reflects the fact that, through an appropriate rescaling of the fields, one can always rescale $\zeta_i$ arbitrarily without changing $T_i,\tilde{D}_i$, or the overall structure of the action. 

To set up the full RG equations, let us define the parameters
\begin{subequations}\label{Eq. parameters redefined}
\begin{equation}
v \equiv \frac{T_2}{T_1}, \quad w \equiv \frac{\tilde{D}_2}{\tilde{D}_1},
\end{equation}
\begin{equation}
\tilde{u}_{i} \equiv \frac{T_i}{D_i^2} u_{i},\quad \tilde{u}_{12} \equiv \frac{T_1}{ D_1 D_2} u_{12}.
\end{equation}
\end{subequations}
With these definitions, the beta functions for the new interaction parameters $\tilde{u}_a$ depend only on the renormalized parameters $v_R$, and $w_R$. To obtain the full RG equations, we further need to determine the RG evolution of the latter parameters. 
As we shall see, their RG equations are also closed in the (five) parameters defined in Eq.~(\ref{Eq. parameters redefined}). To see why, first notice that 
there are ten marginal parameters in the original action at the upper critical dimension ($\zeta_i,D_i,T_i,g_i,g_{12/21}$) which can define the basin of attraction for the RG flow.
Since all four fields and time can be rescaled relative to an overall momentum scale, this leaves a total of five parameters needed to define the fixed point. The remaining parameters ($r_i,h$) define relevant directions of the RG flow and thus must be tuned to their critical values.
In order to determine the RG equations for the parameters $v$ and $w$, 
we use the identity
\begin{equation}
\label{betaratio}
\beta_{p/q} = \frac{p}{q} (\gamma_p - \gamma_q).
\end{equation}
We now report the full set of beta functions of the parameters of our model (with $r_i$ and $h$ set to zero at the fixed point) 
\begin{widetext}
\begin{subequations}
\label{beta}
\begin{equation}
\beta_{\tilde{u}_1} = \tilde{u}_{1_R} \left[-\epsilon + 9 \tilde{u}_{1_R}\right] + \sigma v_R \tilde{u}_{12_R}^2,
\end{equation}
\begin{equation}
\beta_{\tilde{u}_2} = \tilde{u}_{2_R} \left[-\epsilon + 9 \tilde{u}_{2_R}\right] + \sigma v_R \tilde{u}_{12_R}^2,
\end{equation}
\begin{equation}
\beta_{\tilde{u}_{12}} = \tilde{u}_{12_R} \left[-\epsilon + 4 \frac{\sigma v_R + w_R}{ 1 + w_R} \tilde{u}_{12_R} + 3 \tilde{u}_{1_R} + 3 \tilde{u}_{2_R}\right],
\end{equation}
\begin{equation} 
\label{betav}
\beta_v = - v_R \tilde{u}_{12_R}^2 F(w_R) [v_R - \sigma][v_R +\sigma F(w_R^{-1})/F(w_R)],
\end{equation}
\begin{equation}
\label{betaw}
\beta_{w} = - w_R \left[ C \times \left(\tilde{u}_{1_R}^2 - \tilde{u}_{2_R}^2 \right) + \tilde{u}_{12_R}^2 (v_R^2 G(w_R) - G(w_R^{-1})) + 2 \sigma v_R \tilde{u}_{12_R}^2  (H(w_R) - H(w_R^{-1})) \right],
\end{equation}
\end{subequations}
\end{widetext}
where we have defined $C = 9\log(4/3) - 3/2 $ and the functions
\begin{subequations}
\begin{equation}
F(w) = -\frac{2}{w} \log \left( \frac{2+2w}{2+w} \right),
\end{equation}
\begin{equation}
G(w) = \log\left(\frac{(1+w)^2}{w(2+w)} \right) - \frac{1}{2 + 3 w + w^2},
\end{equation}
\begin{equation}
H(w) = \frac{1}{w} \log \left( \frac{2+2 w}{2+w} \right) - \frac{3 w + w^2}{8 + 12 w + 4 w^2}.
\end{equation}
\end{subequations}
The functions $F,G,H$ always appear in the RG equations in pairs, with one taking $w_R$  and the other $w_R^{-1}$ as an argument. This is because the diagrams that contribute to the beta functions come in pairs, corresponding to one from the renormalization of the terms involving $\phi_1$ only and the other from those that involve $\phi_2$ only. Similarly, under the mapping $\tilde{u}_{1_R} \leftrightarrow \tilde{u}_{2_R}$, $\tilde{u}_{12_R} \to \sigma v_R \tilde{u}_{12_R}, v_R \to v_R^{-1}$ and $w_R \to w_R^{-1}$, the beta functions are left unchanged. This reflects the fact that we can switch the role of $\phi_1$ and $\phi_2$ without changing the physics. As a result, if either $\sigma =-1$, $v_R \neq 1$, or $w_R \neq 1$ at a given fixed point, there will always be a second fixed point paired with it. 

The above equations determine the full RG equations of our non-equilibrium model, but it is instructive to first consider the RG equations under equilibrium conditions where the temperatures are equal, i.e., $v_R=1$, and  
$\sigma= 1$. We then immediately find that the temperature ratio does not flow, $\beta_v = 0$; hence the two temperatures remain identical at all scales. 
Furthermore, the temperature itself---and not just the ratio---remains scale invariant ($\gamma_T = 0$), indicating (effective) thermal equilibrium. 
Finally, as remarked earlier, the RG equations for the interaction terms 
become independent of $w_R$ under equilibrium conditions, highlighting once again the fact that, in equilibrium, the statics is decoupled from the dynamics.

There are two distinct scenarios with respect to the 
beta function $\beta_w$.
The first scenario is that the beta function vanishes 
when $\gamma_{\tilde{D}_{1}} = \gamma_{\tilde{D}_{2}}$.
Since the dynamical critical exponents are related to the flow of $\tilde{D}$ as 
$z_i = 2 + \gamma_{\tilde{D}_i}$,
we find that 
$z_1=z_2$ under this scenario. 
This means that both fields are governed by the same dynamical critical exponent, giving rise to a ``strong dynamic scaling''.
The second scenario occurs when  $\gamma_{\tilde{D}_1} \neq \gamma_{\tilde{D}_2}$, which would lead to the fixed point $w_R=0$ or $w_R=\infty$ depending on the sign of $\gamma_{\tilde{D}_1}-\gamma_{\tilde{D}_2}$. This behavior is then described by a ``weak dynamic scaling''
where the two fields exhibit different dynamical scaling properties and exponents \cite{Halperin1969,DeDominicis1977,Dohm1977}; see also \cite{Tauber2014}.
Similarly, 
one can consider the beta function $\beta_v$ characterizing the RG flow of the ratio of the temperatures. In this case too, there are two scenarios: Either the beta function vanishes for a fixed temperature ratio or rather, depending on the sign of $\gamma_{T_1}-\gamma_{T_2}$, the RG flow leads to either $v_R = 0$ or $v_R=\infty$, which both correspond to the $g_{12} g_{21} = 0$ subspace. However, this subspace does not appear to be amenable to perturbative RG. 
In this sector, we find that $w$ flows to either 0 or $\infty$, indicating weak dynamic scaling where the two fields are governed by distinct dynamical universality classes. 
However, in both cases, the fixed-point values of the coupling terms diverge, resulting in a nonperturbative regime that is not accessible within the perturbative RG analysis. This indicates that an alternative approach from our present analysis should be considered in this scenario. In this work, we shall restrict ourselves to the case where $v_R$ and $w_R$ are both finite and nonzero.

\subsection{Fixed points of RG flow}
With the RG beta functions, we can now identify the resultant fixed points. In the $\sigma = 1$ sector, the only fixed points of the RG equations are those where $w_R^*=1$, exhibiting a strong dynamic scaling, as well as $v_R^*=1$, indicating that the two temperatures become identical at the fixed point. Indeed, aside from the case of $u_{12_R}$ = 0, the only possible fixed-point value of $v_R$ at this order is 1. This can be seen by noting that the only other root of Eq.~(\ref{betav}) is $-F(w_R^{-1})/F(w_R)$, which is always negative and thus unphysical. Similarly, noting that the beta functions for $u_{1_R}, u_{2_R}$ are identical at this order, all coupled fixed points in this sector will satisfy $u_{1_R} = u_{2_R}$. In light of this, we immediately identify $w_R=1$ as the only possible solution of Eq.~(\ref{betaw}). Remarkably, an effective equilibrium behavior emerges in this sector despite the underlying non-equilibrium nature of the dynamics. 
In particular, we recover the familiar equilibrium $O(2)$ and biconical fixed points as well as various decoupled fixed points involving combinations of Gaussian and Ising fixed points. However, there are no additional NEFPs in this sector (possibly with the exception of a kind of weak dynamical scaling in the $g_{12}g_{21}=0$ subspace). Note that the emergent equilibrium is not achieved by a simple rescaling of the terms in the action to mimic an effective Hamiltonian but is truly the result of a nontrivial two-loop RG analysis.

In the $\sigma = -1$ sector, any nontrivial fixed point 
is truly non-equilibrium as it cannot be described by effective Hamiltonian dynamics that defines equilibrium. Therefore, we should first determine if there exists any nontrivial fixed point in this sector or, alternatively, if the RG evolution flows to a trivial (decoupled) fixed point. Interestingly enough, the former is the case; we find a pair of genuinely non-equilibrium fixed points as
\begin{equation}
\label{fixedvals}
\begin{gathered}
v_R^* = 1, \ w_R^* = 1,\\
u_{1_R}^* = \frac{\epsilon}{6}, \ u_{2_R}^* = \frac{\epsilon}{6}, \ u_{12_R}^* = \pm \frac{\epsilon}{2 \sqrt{3}}.
\end{gathered}
\end{equation}
These fixed points also exhibit a strong dynamic scaling since $w_R^*=1$, so the two fields are governed by the same dynamical scaling. Furthermore, we find $v_R^*=1$, implying that the two temperatures are equal, which might suggest an equilibrium behavior; however, the latter temperatures only characterize the strength of the noise (more precisely, $\gamma_i T_i$ defines the noise) while a true equilibrium description (and a genuine notion of temperature) requires Hamiltonian dynamics [similar to Eq.~\eqref{H form}], which is inherently impossible in this sector. 

While we have identified a new pair of NEFPs, this does not guarantee that they would govern the critical behavior near the multicritical point. If these fixed points are unstable under RG, further fine-tuning would be necessary to access them. Even if they are stable, depending on the initial microscopic parameters, the system could still flow to an equilibrium fixed point under renormalization. Nevertheless, we shall argue that the multicritical point is indeed governed by the new NEFPs. 

To determine the stability of the fixed points, we need to consider the stability matrix
\begin{equation}
\Lambda_{a b} = \frac{\partial \beta_a}{\partial {s_b}_R},
\end{equation}
where $s_b$ denotes the set of parameters that enter the RG beta functions.
A fixed point is stable if all of the eigenvalues of $\Lambda$ are positive. Although we have determined the lowest-order corrections to all five parameters, we can only determine these eigenvalues up to ${\cal O}(\epsilon)$. This is because in order to fully determine $\Lambda$ to ${\cal O}(\epsilon^2)$, we need to consider the two-loop corrections to the coupling terms $u$ and the three-loop corrections to $v,w$. However, when a fixed point possesses a higher symmetry than the underlying field theory, then it is possible to determine some of the ${\cal O}(\epsilon^2)$ eigenvalues without including higher-order corrections. This is a consequence of the fact that a symmetry-preserving perturbation will not generate a symmetry-violating term, so $\Lambda$ finds a block-triangular form  and the two sectors can be diagonalized separately. In 
the equilibrium limit of our model (in the sector $\sigma=1$ when $v=1$), a similar situation occurs with respect to statics and dynamics, where perturbations in the dynamics ($w$) cannot affect the behavior of the statics ($u$).
This makes it possible to inspect the stability of $w$ up to ${\cal O}(\epsilon^2)$ at the same order of the RG calculations. 
In the full non-equilibrium model, the statics is not decoupled from the dynamics, so the stability in $w$ cannot be determined using such an approach. However, for the equilibrium fixed points, equilibrium plays a role similar to a higher symmetry because equilibrium perturbations do not generate non-equilibrium terms. As a result, it is possible to determine the stability in $v$ for the two coupled equilibrium fixed points, and we find both to be stable in $v$. However, for all of the coupled fixed points, $w$ remains marginal.
In short, to the lowest order in our perturbative expansion, the system flows to the NEFP (equilibrium fixed point) in the $\sigma=-1$ ($\sigma=1$) sector.
While, in principle, non-perturbative effects or higher-order terms in $\epsilon$ could modify this behavior, this is a generic feature of perturbative RG and not specific to our model.
A qualitative sketch of the expected RG flow is illustrated in Fig.~\ref{flow} in terms of the original $g_{12}, g_{21}$ couplings.

Finally, we remark that in the case of the original microscopic model, $A_{02}$ and $B_{11}$ have opposite signs, which thus carries over to the relative sign of $g_{12}$ and $g_{21}$. Thus it is plausible to expect a critical behavior governed by the NEFPs. 

\subsection{Universal scaling behavior}

Any fixed point---equilibrium or not---exhibits critical behavior and exponents  characterizing correlations and dynamics among other properties of the system. 
In particular, we consider the anomalous dimensions $\eta$ and $\eta'$ of the original and response fields, the dynamical critical exponent $z$, as well as the exponent $\nu$ characterizing the divergence of the correlation length as the critical point is approached.
These exponents describe the scaling behavior of the correlation and response functions at or near criticality as
\begin{subequations}\label{scaling functions}
\begin{equation}
C_i(\mathbf{q},\omega,\{r_j\}) \propto |\mathbf{q}|^{-2 + \eta - z} \hat{C}_i\left(\frac{\omega}{|\mathbf{q}|^z},\left\{\frac{r_j}{|\mathbf{q}|^{1/\nu_j}}\right\}\right),
\end{equation}
\begin{equation}
\chi_i(\mathbf{q},\omega,\{r_j\}) \propto |\mathbf{q}|^{-2 + \eta'}\hat{\chi}_i\left(\frac{\omega}{|\mathbf{q}|^z},\left\{\frac{r_j}{|\mathbf{q}|^{1/\nu_j}}\right\} \right),
\end{equation}
\end{subequations}
where $\hat{C}_i,\hat{\chi}_i$ are general scaling 
functions. We have dropped the subscript $i$ from $\eta, \eta', z$ due to the strong dynamic scaling and in anticipation of the same spatial scaling dimensions for the two fields; however, we have kept the subscript in $r_j$ for $j=1,2$ since the RG equations couple them in a nontrivial way. 

The exponents at the fixed point can be extracted via what is known as the method of characteristics (see Appendix \ref{characteristic} for details). Noting that, for fixed bare (microscopic) parameters, the correlation and response functions are not affected by changing the RG momentum scale $\mu$, we can relate these critical exponents to the flow functions as  \begin{equation}
\begin{aligned}
\eta &= \gamma_T - \gamma_D, & \eta' &= - \gamma_{D}, & z &= 2 + \gamma_{D} - \gamma_\zeta.
\end{aligned}\label{exponents from gamma}
\end{equation}
The renormalization of the parameters $r_j$ and the corresponding exponent $\nu_j$ requires a special treatment and will be discussed later in this section. At the non-equilibrium critical point, we find [cf.~Eqs.~(\ref{gammaflow},\ref{fixedvals}) together with the $Z$ factors in Appendix \ref{RG}]
\begin{equation}
    \gamma_{\zeta} = - \frac{\epsilon^2}{6} \log(4/3),\quad \gamma_D = -\frac{\epsilon^2}{36},\quad \gamma_T = - \frac{\epsilon^2}{3} \log (4/3).
\end{equation}
Interestingly, we see that in contrast to an equilibrium fixed point where the temperature becomes scale invariant, the effective temperature at the NEFPs changes with the scale. In particular, the system becomes ``hotter'' at longer length scales since $\gamma_T < 0$. 
Using Eq.~(\ref{exponents from gamma}), the critical exponents at the NEFPs are given by
\begin{subequations}
\begin{equation}
\eta = \frac{\epsilon^2}{36}\left(1 - 12 \log\left(4/3\right)\right), 
\end{equation}
\begin{equation}
\eta' = \frac{\epsilon^2}{36},
\end{equation}
\begin{equation}
z = 2 + \eta' \left(6 \log\left(4/3\right)-1\right).
\end{equation}
\end{subequations}
While, in equilibrium, $\eta = \eta'$ as a consequence of the fluctuation-dissipation theorem,
we have $\eta \ne \eta'$ since the temperature itself is scale dependent, $\gamma_T\ne0$, at the NEFP. 
Note also that the critical exponents $z,\eta,\eta'$ are the same for both fields.
While strong dynamic scaling already guarantees the same dynamical critical exponent, the anomalous dimensions are also identical owing to the emergent symmetry of the fixed point where
$u_{1_R} = u_{2_R}$ and $v_R = w_R = 1$. However, the latter do not reflect any actual symmetry of the model and could be modified at higher orders in the epsilon expansion.

An interesting feature of the NEFPs is that $\eta < 0$ to the first nontrivial order in the epsilon expansion. This is in contrast with equilibrium where $\eta>0$, a fact that can eve ben proved on general grounds (e.g., unitarity in a related quantum field theory) \cite{Zinn96Book}.
If this feature ($\eta<0$) extends beyond perturbation theory to, say, 
two dimensions, it would indicate that the correlation function ($C(\mathbf{r}) \propto |\mathbf{r}|^{-d+2-\eta}$) diverges at large distances. However, this would invalidate the starting point of our field-theoretical treatment based on an expansion in field powers 
since large-scale fluctuations grow without bound. However, it might also indicate the absence of ordering in low dimensions. This possibility seems particularly natural in light of the effective temperature increasing at larger scales, which in turn tends to disallow ordering in low dimensions. While this may be an artifact of perturbative RG, it indicates that the behavior of the NEFPs in low dimensions is governed by different principles than  their equilibrium counterparts.
Finally, we note that, to the lowest nontrivial order considered, the dynamical critical exponent $z$ is related to $\eta'$ in an identical fashion as in equilibrium.

Next we consider the renormalization of the mass terms. This requires special care as their renormalization is intertwined. Similar to our redefinition of $u$, we should instead consider the renormalization of $r_i/D_i$ so that we only need to consider two flow equations, which is consistent with the scaling analysis in Appendix ~\ref{characteristic}. In a slight abuse of notation, we simply replace $r_i \to D_i r_i$. Defining $\tilde{\mu}(l) = \mu l$ and the flowing parameters $\tilde{r}_i(l)$ with $\tilde{r}_i(1) = r_{i_R}$, we find the flow equations [cf.~ Eqs.~(\ref{gammaflow},\ref{fixedvals}) together with the $Z$ factors in Appendix \ref{RG}]
\begin{subequations}
\begin{equation}
l \frac{d \tilde{r}_1(l)}{d l} = \gamma_{r_1} \tilde{r}_1 = \left(-2 + \frac{\epsilon}{2} \right) \tilde{r}_1(l) \pm  \frac{\epsilon}{2 \sqrt{3}} \tilde{r}_2(l),
\end{equation}
\begin{equation}
l \frac{d \tilde{r}_2(l)}{d l} = \gamma_{r_2} \tilde{r}_{2} = \left(-2 + \frac{\epsilon}{2}\right) \tilde{r}_2(l) \mp   \frac{\epsilon}{2 \sqrt{3}} \tilde{r}_1(l),
\end{equation}
\end{subequations}
where the $\pm$ refer to the two NEFPs with opposite signs of $\tilde{u}_{12_R}$. The flow equations can be solved as 
\begin{subequations}
\begin{equation}
\tilde{r}_1(l) = l^{-1/\nu'} \left[ r_{1_R} \cos \frac{\log l}{\nu''} + r_{2_R} \sin\frac{\log l}{\nu''} \right],
\end{equation}
\begin{equation}
\tilde{r}_2(l) = l^{-1/\nu'} \left[ r_{2_R} \cos\frac{\log l}{\nu''} - r_{1_R} \sin\frac{\log l}{\nu''} \right],
\end{equation}
\end{subequations}
where 
\begin{equation}
\begin{aligned}
{\nu'}^{-1} = 2 - \frac{\epsilon}{2},& & {\nu''}^{-1} = \pm \frac{\epsilon}{2 \sqrt{3}}.
\end{aligned}
\end{equation}
These equations can be cast in a more compact notation as
\begin{equation}
\tilde{r}_1(l) + i \tilde{r}_2(l) = l^{-1/\nu' - i /\nu''}(r_{1_R} + i r_{2_R}).
\end{equation}
Defining $r\equiv r_1+ir_2$, we can recast this equation as $\tilde r (l)=l^{-1/\nu} r_R$ where
the critical exponent $\nu$ emerges as 
\begin{equation}
    \nu^{-1} =  \nu'^{-1} + i\, {\nu''}^{-1}= 2 - \left(\frac{1}{2}\pm \frac{i}{2 \sqrt{3}}\right)\epsilon.
\end{equation}
Interestingly, the exponent $\nu$ becomes complex-valued at the NEFP. 
We can then express the scaling functions in Eq.~(\ref{scaling functions}) as
\begin{subequations}\label{scaling functions 2}
\begin{equation}
\hat{C}_i = \tilde{C}_i\left(\frac{\omega}{|\mathbf{q}|^z},\frac{|r_R|}{|\mathbf{q}|^{1/\nu'}}, P\Big(\frac{\log|\mathbf{q}|}{\nu''}-\angle r_R \Big)\right),
\end{equation}
\begin{equation}
\hat{\chi}_i = \tilde{\chi}_i\left(\frac{\omega}{|\mathbf{q}|^z},\frac{|r_R|}{|\mathbf{q}|^{1/\nu'}}, P\Big(\frac{\log|\mathbf{q}|}{\nu''}-\angle r_R \Big)\right),
\end{equation}
\end{subequations}
where $|r_R|=|r_{1_R}+ ir_{2_R}|=  \sqrt{r_{1_R}^2+r_{2_R}^2}$ while $\angle r_R$ denotes the polar angle in the $r_{1_R}$-$r_{2_R}$ plane. Additionally, $P$ is a $2 \pi$-periodic function. To obtain these equations, we have used the fact that $r/l^{1/\nu'+i/\nu''}$ can instead be written as a function of $|r|/l^{1/\nu'}$ and $e^{i(\log l)/\nu''-i \angle r}$. 
The former expression often appears in scaling functions of this type and characterizes the scaling of the correlation length; however, the latter gives rise to a log-periodic function as a change of $\log l \to \log l+2\pi\nu''$ leaves the exponential invariant.  

\begin{table*}
\centering
\def\arraystretch{1.5}
\setlength\tabcolsep{5.1mm}
\begin{tabular}{|c|c|c|c|c|c|c|c|}
\hline 
Fixed Point & $u_1$ & $u_2$ & $u_{12}$ & $\nu^{-1}$ & $\eta$ & $\eta'$  & $z-2$ \\ 
\hline
NEFP & $\frac{\epsilon}{6}$ &$\frac{\epsilon}{6}$ &$\frac{\pm \epsilon}{2\sqrt{3}}$ & $2- \big( \frac{1}{2} \pm \frac{i}{2 \sqrt{3}} \big) \epsilon$ & $\frac{\epsilon^2}{36}\big(1-12 \log\frac{4}{3}\big)$& $\frac{\epsilon^2}{36}$ & \multirow{4}{*}{$\eta' \big(6 \log\frac{4}{3} -1\big)$} \\ 
\cline{1-7} 
$O(2)$ &$\frac{\epsilon}{10}$ &$\frac{\epsilon}{10}$ &$\frac{\epsilon}{10}$ & $2 - \frac{2}{5} \epsilon$ & \multicolumn{2}{c|}{$\frac{\epsilon^2}{50}$} & \\ 
\cline{1-7} 
Biconical &$\frac{\epsilon}{18}$ &$\frac{\epsilon}{18}$ &$\frac{\epsilon}{6}$ & $2 - \frac{1}{3}\epsilon$ & \multicolumn{2}{c|}{$\frac{\epsilon^2}{54}$} & \\ 
\cline{1-7}  
$\mathbb{Z}_2 + \mathbb{Z}_2$ &$\frac{\epsilon}{9}$ &$\frac{\epsilon}{9}$ &$0$ & $2 - \frac{1}{3}\epsilon$ & \multicolumn{2}{c|}{$\frac{\epsilon^2}{54}$}  & \\ 
\hline 
\end{tabular} 
\caption{Fixed-point values of the coupling coefficients and critical exponents to the lowest order. In all cases, $v_R = w_R = 1$. At the NEFP, $\sigma=-1$. The decoupled $\mathbb{Z}_2 + \mathbb{Z}_2$ fixed point and the biconical fixed point in this case are unstable to the order ${\cal O}(\epsilon)$, while the other two fixed points are stable to the same order. The $\mathbb{Z}_2 + \mathbb{Z}_2$ and biconical fixed points exhibit the same critical behavior since they can be mapped to each other through a $\pi/4$ rotation in the $\phi_1$-$\phi_2$ plane. Fixed points involving the Gaussian fixed point are not included. \label{exponents}}
\end{table*}

The appearance of log-periodic functions has important consequences for the critical nature of the fixed points. They lead to a
discrete scale invariance rather than the characteristic continuous scale invariance at a typical critical point \cite{Sornette1998}. Rather than a self-similar behavior at all length scales, a preferred scaling factor emerges as 
\begin{equation}
    b_*=e^{2 \pi  \nu''},
\end{equation}
rescaling by which, or any multiple integer thereof, leaves the system scale invariant. In this sense, discrete scale invariance mimics a fractal-like structure, in which rescaling the system by a particular factor leaves the system self-similar. Note, however, that the discrete scale invariance and the fractal-like structure only emerges at long length scales (in the continuum) as opposed to the microscopic structure of a fractal (in discrete space). Additionally, if we were to consider, e.g., the effect of a physical momentum cutoff $\Lambda$, this would enter the periodic function as a phase shift, thus determining the phase of the oscillations.

Similar phenomena appear to arise in stock markets \cite{Feigenbaum1996}, earthquakes \cite{Sornette1995}, equilibrium models on fractals \cite{Karevski1996}, and several other systems \cite{Sornette1998}. 
Log-periodic functions and the emergence of a preferred scale have been identified in the early developments of renormalization group theory \cite{Jona-Lasinio75,Nauenberg75,Niemeijer:1976bs}, but they have been rejected as artifacts of position-space RG. On the other hand, their recent surge in diverse contexts from earthquakes to stock markets has instead relied on simple dynamical systems (with one or few variables) where the dynamics involves a discrete map itself
\cite{Sornette1998}. 
This phenomenon has also emerged in recent works in the context of driven-dissipative quantum criticality \cite{Marino2016,Marino16-2} as well as the dynamics of strongly interacting non-equilibrium systems \cite{Maki2018}.
A particularly well-known example of RG limit cycles is the behavior of Efimov states \cite{Braaten2006,Efimov1970}, whose binding energies form a geometric progression similar to discrete scale invariance. These quantum RG limit cycles have been noted to be closely related to Berezinskii-Kosterlitz-Thouless phase transitions \cite{LeClair2003,Kaplan2009}. 
Disordered systems provide another context where complex-valued exponents and discrete scale invariance have been noted in both classical \cite{Aharony75,Chen77,Khmelnitskii78,Weinrib83,Boyanovsky82} and quantum \cite{Hartnoll-16} settings.
The discrete scale invariance reported in this work, however, appears to be unique as it has emerged in an effectively classical yet non-equilibrium model in the absence of disorder.

The discrete scale invariance approaches a continuous one as the upper critical dimension, $d_c=4$, is approached. In three dimensions, perturbative values at the NEFP (with $\epsilon = 1$) yield a very large scaling factor 
($b_* \sim 10^9$);
however, with the exponential dependence on the critical exponents, the scaling factor is sensitive even to small corrections of the exponent beyond the lowest-order perturbation theory. Nevertheless, our results should be viewed as 
a proof of principle for the emergence of discrete scale invariance in macroscopic non-equilibrium systems.  
Additionally, higher harmonics in the periodic function $P$ can be significant, which then should be observed over smaller variations in the physical scale.

Finally, we elaborate on a possible connection between the log-periodic behavior and limit cycles. 
Indeed, the microscopic mean-field phase diagram near the multicritical point also includes a limit-cycle phase that displays persistent oscillations.
For a rapidly oscillating limit cycle, the corresponding phase transition can be described from the viewpoint of a rotating frame (defined by the oscillation frequency) 
and by making the rotating-wave approximation.
With this mapping, a limit-cycle phase transition is no different from a dissipative phase transition with an emergent $U(1)$  symmetry  \cite{Chan2015}. 
Near our multicritical point, however, the frequency of oscillations becomes small and thus no such mapping is possible. On the other hand, the discrete scale invariance discussed above also leads to an oscillatory behavior (in both space and time), albeit one that is log-periodic. 
Nevertheless, a natural possibility is that, at some intermediate regime away from the multicritical point, the discrete scale invariance merges into a limit-cycle solution. 
Moreover, we find that in the doubly-ordered phase, the Liouvillian gap becomes complex-valued (see Sec. \ref{gap closure}). This furthers the possible connection to the limit-cycle phase as a nonzero imaginary part implies that the system undergoes oscillations---which, however, decay---as the steady state is approached.

In Table \ref{exponents}, we summarize all of the fixed points (aside from those involving the trivial Gaussian fixed point) and their critical exponents to the lowest order. 

\subsection{Phase diagram}

The phase diagram itself is distinct in the vicinity of the NEFPs. In contrast to their equilibrium counterparts, these fixed points only give rise to a 
tetracritical point. With the effective magnetic field set to zero, $h=0$, four different phases emerge: A disordered phase with $\phi_1 = \phi_2 = 0$; two phases with either $\phi_1\ne 0$ corresponding to bistability or $\phi_2\ne0$ leading to antiferromagnetic ordering; and, finally, a doubly-ordered phase where both fields become ordered, $\phi_1\ne 0\ne \phi_2$. 
While the first three phases also emerge in the mean field theory of the microscopic model, the doubly-ordered phase only arises in the course of RG when the $A_{20}$ term is generated.

The phase boundaries are governed by the scaling behavior of $r_i$. Let us set the effective magnetic field to zero, $h=0$, and consider the scaling functions characterizing the correlation and response functions in Eq. (\ref{scaling functions}).
To determine the phase boundary, it suffices to take the limit $\omega,\bq  \to 0$. In this case, the scaling functions are solely determined as a $2\pi$-periodic function of $\frac{\nu'}{\nu''} \log\left(|r_R|\right)-\angle r_R$; this is achieved by eliminating the momentum scale in Eq.~(\ref{scaling functions 2}) in favor of $r_R$.
Since the correlation functions only depend on the mass terms through the above combination, the phase boundary itself---characterized by the divergence of correlations---arises at a fixed value of this quantity (up to integer multiples of $2\pi$). 
Therefore, the shape of the phase boundary is given by 
\begin{equation}
\frac{\nu'}{\nu''} \log(|r_R|)-\angle r_R = \mbox{const},
\end{equation}
which is a spiral, leading to the phase diagram illustrated in Fig.~\ref{phasediags}.
Similar to our discussion of the discrete scale invariance, the perturbative values of the exponents at $\epsilon = 1$ require very large scales to observe a full spiral.
But again these scales are highly sensitive to corrections to perturbative RG. Moreover, partial spirals can still be observed for reasonable scales. 
Since the spiraling boundaries all spiral in the same direction, distinguishing them from equilibrium critical points, the effects of this may be seen even for very weak spiraling. Additionally, the two NEFPs are distinguished from each other by the direction of spiraling, since each has a different sign of $\nu''$.

\begin{figure}
\includegraphics[scale=.18]{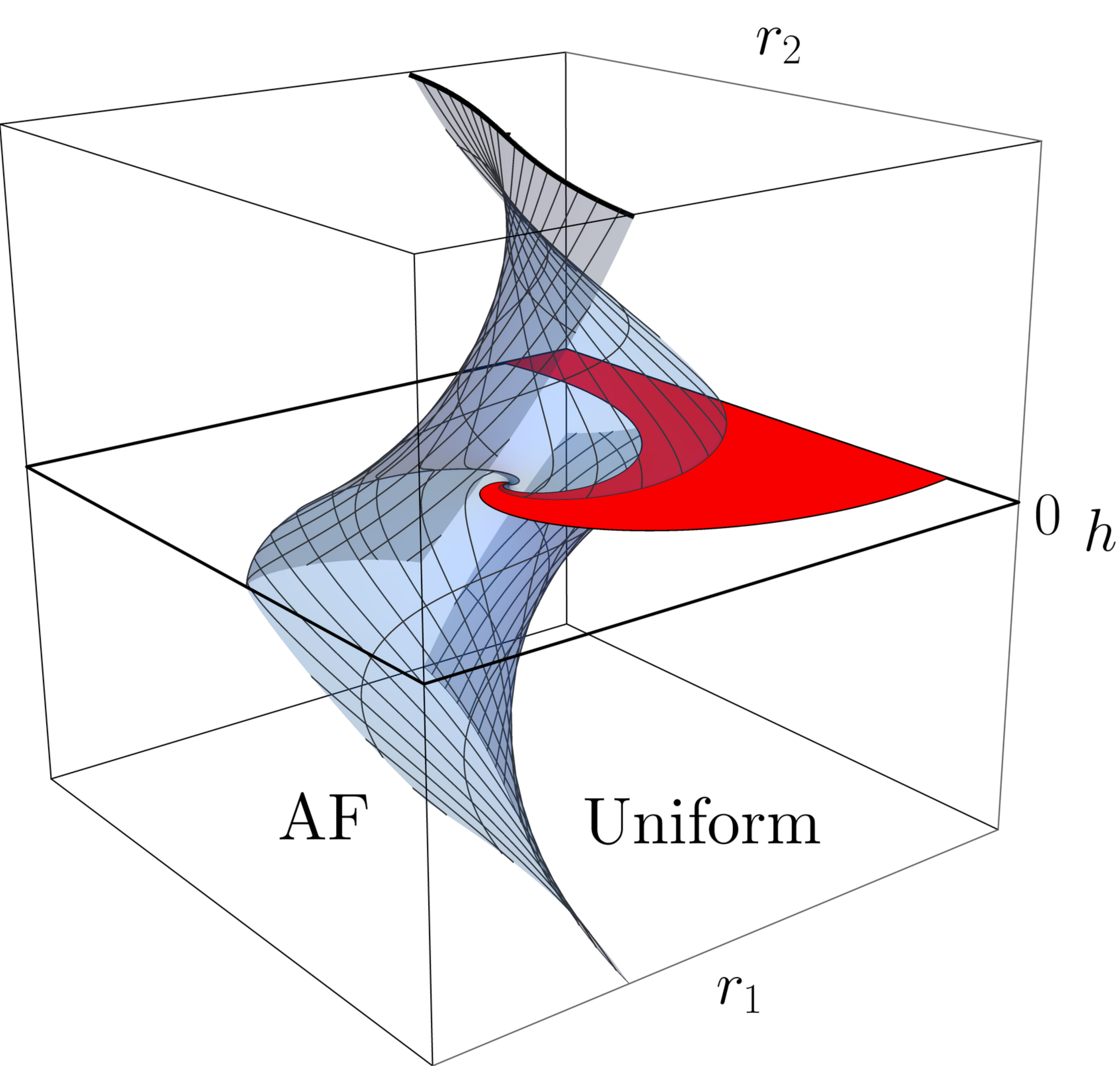}
\caption{Phase diagram for (one of the two) the non-equilibrium tetracritical points as a function of effective mass terms and  magnetic field. The transparent boundary indicates the location of an antiferromagnetic phase transition from a uniform phase. The horizontal (red) surface denotes a surface of first-order phase transitions. In both AF and uniform phases, the latter indicates a transition from a low-  to a high-population phase; the sublattice population difference changes continuously across this surface. The (black) square boundary in the middle indicates the plane of $h=0$. The diagram for the second non-equilibrium tetracritical point will spiral in the opposite direction.\label{tetradiag}}
\end{figure}

When the effective magnetic field is nonzero, there is no such symmetry as $\phi_1 \to -\phi_1$, leading to a surface of first-order phase transitions where $\phi_1$ undergoes spontaneous symmetry breaking. This first-order transition occurs in both the uniform phase (defined by the B phase at $h=0$) and the antiferromagnetic phase (defined by the B+AF phase at $h=0$). In both cases, the average population changes discontinuously while the sublattice population difference changes continuously.
Finally, we take into account the magnetic field in determining the phase boundaries by considering an effective mass as $r_R + |h_R|^{1/\beta \delta}$, where $\beta$ and $1/\delta$ describe the scaling behavior of the order parameter with $r$ and $h$, respectively, within the ordered phase.
The effect of the magnetic field $h$ is to ``unravel'' the spirals for small $r_i$. The corresponding phase diagram for nonzero effective magnetic field is illustrated in Fig.~\ref{tetradiag}.

\subsection{Hyperscaling relations}

	As $\eta \neq \eta'$ and the exponent $\nu$ is complex, the standard hyperscaling relations will be modified. For example, consider the exponents characterizing the order parameter and magnetic susceptibility, 
	\begin{equation}
	    \label{hyperdef}
	    \langle \phi \rangle \propto |r|^\beta, \hspace{1cm}  \frac{\partial \langle \phi \rangle}{\partial h}  \propto |r|^{-\gamma}.
	\end{equation}
	In the standard equilibrium setting, these exponents are  related to $\nu,\eta$ via
	\begin{equation}
	    \beta = \nu(d-2+\eta)/2, \hspace{1cm} \gamma = \nu(2-\eta).
	\end{equation}
	To see how the above hyperscaling relations are modified near the NEFPs, first note
	\begin{subequations}
	\begin{equation}
	    \langle \phi^2 \rangle = \lim_{|\bx|,t \to \infty} C(\mathbf{x},t,\{r_j\}),
    \end{equation} 
    \begin{equation}
	    \frac{\partial \langle \phi \rangle}{\partial h} = \lim_{\mathbf{q},\omega \to 0} \chi(\mathbf{q}, \omega,\{r_j\}),
	\end{equation}
	\end{subequations}
	and that in the ordered phases, $\langle \phi^2 \rangle$ and $\langle \phi \rangle^2$ will have the same scaling behavior. Similar to our phase-diagram analysis, we take the limit $\omega, \mathbf{q} \to 0$ or $t, |\mathbf{x}| \to \infty$ and allow $|r_R|$ to define the momentum scale in place of $|\mathbf{q}|$. This amounts to replacing the $|\mathbf{q}|$ prefactors in Eq.~(\ref{scaling functions}) with $|r_R|^{-\nu'}$. We then find the scaling of the order parameter and magnetic susceptibility to be
	\begin{subequations}
	\label{periodicbg}
	\begin{equation}
	    \langle \phi^2 \rangle \propto |r_R|^{\nu'(d-2+\eta)} P_C\left(\frac{\nu'}{\nu''} \log(|r_R|)-\angle r_R\right),
	\end{equation}
	\begin{equation}
	    \frac{\partial \langle \phi \rangle}{\partial h} \propto |r_R|^{\nu'(2-\eta')}P_\chi\left(\frac{\nu'}{\nu''} \log(|r_R|)-\angle r_R \right),
	\end{equation}
	\end{subequations}
where $P_C, P_\chi$ are $2\pi$-periodic functions. Given this scaling behavior, there are two approaches to identifying the exponents $\beta, \gamma$. The first is that, like $\nu$, the exponents $\beta$ and $\gamma$ assume complex values, corresponding to the log-periodic behavior. However, this is simply a reflection of the structure of the phase diagram: approaching the critical point directly, the system crosses in and out of all four phases, giving rise to discrete scale invariance. The second and perhaps more natural approach is to tune the system to the critical point along paths which fix the argument of the periodic functions. In doing so, the angle relative to the phase boundaries is fixed. According to this second approach, which confines the discrete scale invariance to $\nu$, we have the new hyperscaling relations
	\begin{equation}
	    \beta = \nu'(d-2+\eta)/2, \hspace{1cm} \gamma = \nu'(2-\eta'),
	\end{equation}
	where only the real part $\nu'$ enters.

\subsection{Liouvillian gap closure}\label{gap closure}

In this section, we investigate how the Liouvillian gap closes upon approaching the multicritical point.  
In contrast to the equilibrium fixed points where the gap always closes along the real axis (hence, purely dissipative or relaxational dynamics), the NEFPs exhibit a qualitatively different behavior with the gap closing along a complex path, indicating an interplay between reversible and irreversible relaxation in this phase.

We consider the system in the doubly-ordered phase where both fields take nonzero expectation values $M_i$; for notational convenience, we make the change of variables  $\phi_i \to \phi_i + M_i$ where the fields $\phi_i$ now represent the fluctuations around the order parameter. In addition to the original action, this transformation introduces new quadratic and linear terms as (including the original $r_1$ and $r_2$ terms too)
\begin{widetext}
\begin{equation}
\begin{aligned}
\int_{\mathbf{x},t}&(r_1 + 3 u_1 M_1^2 + u_{12} M_2^2) \phi_1 \tilde{\phi}_1 + (r_2 + 3 u_2 M_2^2 + \sigma u_{12} M_1^2) \phi_2 \tilde{\phi}_2 + 2 u_{12} M_1 M_2 \phi_2 \tilde{\phi}_1 + 2 \sigma u_{12} M_1 M_2 \phi_1 \tilde{\phi}_2\\
&+ M_1(r_1 + u_1 M_1^2 + u_{12} M_2^2) \tilde{\phi}_1 + M_2(r_2 + u_2 M_2^2 + \sigma u_{12} M_1^2) \tilde{\phi}_2.
\end{aligned}
\end{equation}
\end{widetext}
In addition, several cubic terms are also introduced, which are not reported for simplicity. 
We set the vertices $\tilde{\phi}_1$ and $\tilde{\phi}_2$ to zero since, by definition, $\phi_i$ solely represent the fluctuations. This in turns sets $r_1 = - u_1 M_1^2 -u_{12} M_2^2$ and $r_2 = -u_2 M_2^2 - \sigma u_{12} M_1^2$. Upon including the effect of fluctuations to $\mathcal{O}(u)$, the remaining quadratic vertices are then given by
\begin{equation}
2 u_{1_R} M_1^2 \phi_1 \tilde{\phi}_1 + 2 u_{2_R} M_2^2 \phi_2 \tilde{\phi}_2 + 2 u_{12_R} M_1 M_2 (\phi_2 \tilde{\phi}_1 + \sigma \phi_1 \tilde{\phi}_2),
\end{equation}
where the coupling terms have been replaced with their renormalized values due to the inclusion of fluctuations in the form of counterterms. The other parameters are not renormalized at this order, but if we were to include higher-order fluctuations, they too would be replaced with their renormalized values since the ordered phases do not give rise to any new $Z$ factors. 

Putting these terms together with the quadratic part of the action, we find
\begin{equation}
\begin{aligned}
S_0 = \int_{\mathbf{x},t}& \sum_i \tilde{\phi}_i ( \zeta \partial_t - D \nabla^2 + R_i)\phi_i - \zeta T \tilde{\phi}_i^2 \\
&+ R_{12}( \phi_2 \tilde{\phi}_1 + \sigma \phi_1 \tilde{\phi}_2),
\end{aligned}
\end{equation}
where $R_i = 2 u_{i_R} M_i^2$ and $R_{12} = 2 u_{12_R} M_1 M_2$. The poles of the corresponding propagators are then obtained as
\begin{equation}
0 = \sigma R_{12}^2 - (D \mathbf{k}^2 + R_1 + i \zeta \omega)(D \mathbf{k}^2 + R_2 + i \zeta \omega),
\end{equation}
or, more explicitly, as
\begin{equation}
-i \zeta \omega = D \mathbf{k}^2 + \frac{R_1 + R_2}{2} \pm \sqrt{\sigma R_{12}^2 + \left(\frac{R_1-R_2}{2}\right)^2}.
\end{equation}
From this equation, we can find a simple condition for when the poles do not lie along the imaginary axis (corresponding to the negative real eigenvalues of the Liouvillian) as
\begin{equation}
- \sigma R_{12}^2 > \left(\frac{R_1-R_2}{2}\right)^2.
\end{equation}
Indeed, in equilibrium, where $\sigma = 1$, this condition cannot be satisfied. This implies that the relaxation is purely relaxational in equilibrium as expected (in this case, for model A). 
However, at the NEFPs where $\sigma=-1$, the above condition can be satisfied. 
To see this, let us cast the above condition for $\sigma=-1$ in terms of $u$ and $M_i$ as
\begin{equation}\label{condition}
4 u_{12_R}^2 M_1^2 M_2^2 > (u_{1_R} M_1^2-u_{2_R} M_2^2)^2.
\end{equation}
Recalling that $u_{1_R} = u_{2_R}$ at the NEFPs, at least to the lowest order in the epsilon expansion, the above condition is trivially satisfied whenever $|M_1| = |M_2|$.
In this case, the pole with the lowest nonzero decay rate takes the form (with $|M_1|=|M_2|\equiv M$ and $\mathbf{k}\to 0$) 
\begin{equation}
-i\zeta \omega = 2M^2 (u_{1_R}\pm i u_{12_R} ).
\end{equation}
In fact, with $|M_1|=|M_2|$, the Liouvillian gap achieves its largest imaginary value relative to its real part. We can identify the ratio of the imaginary part to the real part of the gap as $\frac{u_{12_R}}{u_{1_R}} = \sqrt{3}$, which corresponds to the Liouvillian gap closing at the angle $\pi/3$ with respect to the real axis. Again, higher orders in the epsilon expansion may modify the value of this angle.
A generalization of the model considered here to the $O(N) \times O(N)$ model of $N$-component vector-like order parameters leads to similar behavior. In fact, we find that, for $N=2$ and $N=3$, the corresponding Liouvillian gap closes at the angles $\pi/4$ and $\pi/6$, respectively. We should then conclude that different non-equilibrium universality classes of our model and its generalization give rise to different angles of the Liouvillian gap closure in the complex plane. 
Further details on these generalized models will be presented in follow-up papers. 

The scaling of (the magnitude of) the gap  itself as a function of the distance from the critical point can be directly obtained by observing that the gap defines an inverse time scale which itself is associated with the exponent $z$. Thus the gap scales as $|r|^{z \nu'}$, where the exponent $\nu'$ is due to the scaling of $r$ itself.
With the order parameters $M_1$ and $M_2$ scaling similarly, the angle that defines the gap closure in the complex plane only depends on (the absolute value of) their ratio. As remarked earlier, this angle achieves its maximum when $|M_1|=|M_2|$.
We further note that the gap is purely real (relaxational) near phase boundaries where only one of the order parameters undergoes a transition
since the lhs of Eq.~(\ref{condition}) would be suppressed compared to the rhs. 

Finally, much like the discrete scale invariance in the previous section, the complex Liouvillian gap is somewhat reminiscent of limit cycles,
although a true limit-cycle phase is characterized by purely imaginary eigenvalues that characterize the steady state itself. 

\section{Experimental realization}

\label{expt}

An ideal avenue for realizing these multicritical points is via the use of cavity or circuit quantum electrodynamics (QED). Individual cavities and circuits have been studied experimentally in great depth due to their potential applications in quantum computation \cite{Blais2007,Houck2012,Devoret2013}. Furthermore, both cavity QED and circuit QED have been proposed as platforms for realizing many-body states of light via nearest-neighbor coupling arrays of cavities or circuits \cite{Hartmann2008,Tomadin2010,Carusotto2013,Peropadre2013}.
Generally, these cavities and circuits have non-negligible loss due to dissipation. While dissipation is detrimental when it comes to realizing the quantum ground state of a given system, it is a crucial ingredient in realizing driven-dissipative phase transitions. There have been a variety of theoretical proposals to realize different driven-dissipative models in cavity- and circuit-QED systems \cite{Torre2013,Jin2013,Jin2014,Wilson2016,Biondi2017,Iemini2018}. Recent experiments have even identified a driven-dissipative many-body phase transition \cite{Fitzpatrick2017}. 

For the model considered in this work (see Sec.~\ref{sum}), 
many-body experimental platforms already exist that include drive, hopping, as well as dissipation. The remaining ingredient is then the nearest-neighbor interaction [the quartic term in Eq.~(\ref{modeq1})] to be contrasted with a Hubbard term that characterizes on-site interaction. Both types of interaction are generally known as Kerr nonlinearities; we are interested in what is known as a cross-Kerr nonlinearity, which has been utilized experimentally in several few-mode systems \cite{Kumar2010,Hoi2013,Holland2015}. A more general version of our model has been considered in Refs.~\cite{Jin2013,Jin2014}, along with a discussion on how the nonlinear interaction terms can be tuned experimentally via Josephson nanocircuits. In a recent theoretical proposal,  a setting consisting of a capacitor in parallel with a superconducting quantum interference device is put forth as an alternative means of achieving tunable Kerr nonlinearities \cite{Kounalakis2018}.

While generic experimental settings 
introduce other nonlinear terms (e.g., Hubbard interactions and correlated hopping) in addition to the density-density interactions, we do not expect them  to dramatically affect the results of this paper. While such terms can change the location of the multicritical point \cite{Jin2013,Jin2014}, the universal properties of the latter should not be affected by the details of the microscopic model.

We close this section with a discussion of 
the sign of various terms (e.g., the negative cross-Kerr nonlinearity) arising in the proposals of Refs.~\cite{Jin2013,Jin2014,Kounalakis2018}. While a negative interaction term could lead to unbounded energy spectra, it would not pose a problem in the context of driven-dissipative systems where the steady state is not concerned with a minimum-energy ground state. 
Furthermore, one can change the sign of various terms in the Hamiltonian of a driven-dissipative system with a proper mapping \cite{Li2017}. For example, by sending $\Omega \to - \Omega$ and $a \to -a$ on one of the two sublattices, the sign of $J$ can be changed while leaving the remaining terms fixed. Similarly, one can also map $H \to -H$ by taking the complex conjugate of the master equation, which, together with the previous mapping, allows an appropriate choice for the sign of  $J, V$. Finally, the overall phase of $\Omega$ is unimportant while the parameter $\Delta$ can be easily tuned to a desired sign.

\section{Conclusion and Outlook}

\label{out}

In this work, we have considered an experimentally relevant driven-dissipative system where two distinct order parameters emerge that characterize a liquid-gas type transition (associated with the average density) as well as an antiferromagnetic transition (associated with the difference in the sublattice density). The two phase transitions coalesce and form a multicritical point where both transitions occur at the same time. We have investigated the  nontrivial interplay of two order parameters at the multicritical point. Using a field-theoretical approach---appropriate in the vicinity of the phase transition---we have shown that the critical behavior at this point can be mapped to a non-equilibrium stochastic model described by a $\mathbb{Z}_2 \times \mathbb{Z}_2$ symmetry. Using perturbative renormalization group techniques, we have determined the RG flow equations of the model and identified a pair of new classical non-equilibrium fixed points that exhibit several exotic properties. 
First, we obtain two different exponents for the critical scaling of fluctuations and dissipation at the critical point, underscoring the violation of the fluctuation-dissipation relation at all scales and resulting in a behavior where the system becomes hotter and hotter at larger and larger scales. Furthermore, these NEFPs are distinguished by the emergence of discrete scale invariance and a complex Liouvillian gap even close to the critical point. 
Additionally, the phase diagram near these multicritical points displays spiraling phase boundaries. The latter properties could be particularly useful in identifying these NEFPs in experiments. 

While generic driven-dissipative phase transitions tend to have effective equilibrium dynamics, we have shown that the interplay between several order parameters (in this case, two) 
could very well lead to exotic non-equilibrium behavior. This perspective opens a new avenue to investigate and experimentally realize non-equilibrium phases and phase transitions in the context of driven-dissipative systems without relying on the engineering of complicated nonlocal or non-Markovian dissipation.

Future experimental and numerical studies into the NEFPs discussed in this work are crucial to develop a more complete understanding of their properties. Characterizing the discrete scale invariance, either in the dynamics or the form of the phase boundary, defines a particularly important direction. Investigating the possible emergence and the critical behavior of such non-equilibrium phase transitions in low dimensions is worthwhile. It would be particularly interesting to identify low-dimensional ordering and phase transitions that are not otherwise possible in equilibrium settings.
Another question that remains open is the fate of the subspace $g_{12} g_{21} = 0$, namely if it contains new NEFPs. 
Beyond these non-equilibrium generalizations of model A systems, one can further consider similar non-equilibrium versions of other equilibrium universality classes.
While we focused on the particular case of an experimentally relevant model with only two scalar order parameters, our analysis indicates that a large class of new non-equilibrium multicritical points are yet to be discovered. A natural extension of our work is to 
identify possibly new NEFPs
in $O(N_\parallel) \times O( N_\perp)$ models involving vector-like order parameters \cite{Fisher1965,Nelson1974,Bruce1975,Kosterlitz1976,Folk2008a,Folk2008b,Eichhorn2013}.
While a driven-dissipative condensate of polaritons has been investigated theoretically in detail \cite{Sieberer2013,Tauber2014a}, recent experimental studies into condensate supersolids  \cite{Baumann2010,Mottl2012,Landig2016,Leonard2017,Leonard2017a} can provide excellent platforms for probing any emergent NEFPs. In addition to the $U(1)$ symmetry of the condensate, the two coupled optical cavities can provide either an additional $\mathbb{Z}_2 \times \mathbb{Z}_2$  symmetry (corresponding to a lattice supersolid) or an approximate $O(2)$ symmetry (corresponding to a continuous supersolid).

\begin{acknowledgments}
We thank M.~Kardar, S.~Diehl, R.~Fazio, J.~Marino, I.~Boettcher, S.~Mathey, R.~Matthew, S.-K.~Chu, J.~Curtis, and N.~Grabon for helpful discussions. J.~T.~Y. and A.~V.~G. acknowledge support from the NSF
PFC at JQI, the DOE BES QIS program (Award No.~DESC0019449), NSF PFCQC program, AFOSR, ARL CDQI,
ARO MURI, and the DOE ASCR Quantum Testbed
Pathfinder program (Award No.~DE-SC0019040). M. M.
acknowledges support from NSF under Grant No.~DMR1912799 and start-up funding from Michigan State
University.
\end{acknowledgments}

\appendix

\section{LANGEVIN EQUATIONS NEAR THE MULTICRITICAL POINTS}

\label{CH}

In this appendix, we present the details of the derivation of the Langevin equations in the main text. To this end, we follow the procedure detailed in Ref.~\cite{FossFeig2017}. We begin by constructing the Keldysh path integral, then identify a semi-classical limit, and derive a pair of complex Langevin equations that describe the dynamics near the steady state. Finally, we identify a pair of two massless real fields (i.e., soft modes) and two massive real fields (i.e., fast modes). We adiabatically eliminate the massive fields to obtain a pair of Langevin equations presented in the main text.

We first ignore the sublattice symmetry for simplicity; this would not affect the analysis presented here. We shall return to the latter symmetry once we identify the semi-classical limit and corresponding Langevin equations.
We cast our model in terms of a Keldysh path integral as
\begin{equation}
Z = \int \mathcal{D}[\psi_q,\psi_{cl}]e^{i S_K},
\end{equation}
where the action $S_K$ is defined as 
\begin{equation}
\begin{aligned}
S_K =& \int_{\mathbf{x},t} \psi_q^* \partial_t \psi_{cl} + \psi_q \partial_t \psi_{cl}^* \\
- &\int_{\mathbf{x},t}(H_n(\psi_{cl}+\psi_q)-H_n(\psi_{cl} - \psi_q))\\
+ &i \Gamma \int_{\mathbf{x},t} (|\psi_q|^2 - \psi_{cl} \psi_q^*/2 - \psi_{cl}^* \psi_q/2), 
\end{aligned}
\end{equation}
with $\psi_{cl/q}$ the classical/quantum fields and $H_n(\psi)$ the normal-ordered form of the Hamiltonian. The third line corresponds to the particular case of the Lindblad operator $\sqrt{\Gamma} a$, although this approach may easily be extended to more general Lindbladians \cite{Sieberer2016}. With the Hamiltonian in Eq.~(\ref{modeq1}), the action in the continuum (with the nearest-neighbor interactions expanded in powers of the gradient) is given by
\begin{equation}
\begin{aligned}
S_K &= \int_{\mathbf{x},t} \psi_q^*\left(i \partial_t  + J \nabla^2 + \Delta + \mathfrak{z} J + i \frac{\Gamma}{2}\right)\psi_{cl} + c.c.\\ 
&-\int_{\mathbf{x},t} V |\psi_{cl}|^2 (\nabla^2 + \mathfrak{z}) \psi_{cl} \psi_q^* - \sqrt{2} \Omega \psi_{q}^*  + c.c. \\
&+\int_{\mathbf{x},t} i \Gamma |\psi_q|^2- V |\psi_q|^2 (\nabla^2 + \mathfrak{z}) \psi_{cl} \psi_q^* + c.c.,
\end{aligned}
\end{equation}
where $\mathfrak{z}=2d$ is the coordination number. 
This expression bears a close resemblance to the action of Eq.~(12) in Ref.~\cite{FossFeig2017}, but they differ in the form of their interactions (which involve gradient terms here) and due to our use of normal ordering rather than the Weyl ordering of Ref.~\cite{FossFeig2017}. This motivates a similar rescaling of the parameters as
\begin{equation}
\begin{aligned}
\Psi_{cl} = \psi_{cl}/\sqrt{\mathcal{N}}, && \Psi_q = \psi_q \sqrt{\mathcal{N}},\\
\tilde{\Omega} =\Omega/\sqrt{\mathcal{N}}, && v = \mathfrak{z} V \mathcal{N}.
\end{aligned}
\end{equation}
The parameter $\mathcal{N}$ effectively describes a density scale for the microscopic model via $|\psi_{cl}|^2 = \mathcal{N} |\Psi_{cl}|^2$, where $|\Psi_{cl}|^2$ is $\mathcal{O}(1)$. Since varying the density scale also modifies the interaction energy per particle,
the interaction strength should be reduced correspondingly such that 
$V |\psi_{cl}|^2 = v |\Psi_{cl}|^2$; similarly, the drive should be increased so that $\Omega \psi_{q} = \tilde{\Omega} \Psi_q$. We can then rewrite the action as
\begin{equation}
\begin{aligned}
S_K &= \int_{\mathbf{x},t} \Psi_q^*\left(i \partial_t  + J \nabla^2 + \Delta + \mathfrak{z} J + i \frac{\Gamma}{2}\right)\Psi_{cl} + c.c.\\ 
&- \int_{\mathbf{x},t} v |\Psi_{cl}|^2 (\nabla^2/\mathfrak{z}+1) \Psi_{cl} \Psi_q^* - \sqrt{2} \tilde{\Omega} \Psi_{q}^*  + c.c. \\
&+ \int_{\mathbf{x},t} i \frac{\Gamma}{\mathcal{N}} |\Psi_q|^2- \frac{v}{\mathcal{N}^2} |\Psi_q|^2 (\nabla^2/ \mathfrak{z} + 1) \Psi_{cl} \Psi_q^* + c.c.
\end{aligned}
\end{equation}
In the limit of large $\mathcal{N}$, the last term (the second term in the last line) can be dropped, leading to an action that is at most quadratic in $\Psi_q$. This is simply because a large population $\mathcal{N}$ corresponds to the semi-classical limit represented by a large classical field $\psi_{cl}$ and small fluctuations due to the quantum field $\psi_q$.
Using this fact, we can map the action to a Langevin equation as \cite{Tauber2014} 
\begin{subequations}
\begin{equation}
i \partial_t \Psi = - (J \nabla^2 + \Delta + \mathfrak{z} J + i \Gamma/2 -  v(1+\nabla^2/\mathfrak{z}) |\Psi|^2) \Psi + \tilde{\Omega} + \xi,
\end{equation}
\begin{equation}
\langle \xi^*(\mathbf{x},t) \xi(\mathbf{x}',t') \rangle = \frac{\Gamma}{\mathcal{N}} \delta(\mathbf{x}-\mathbf{x}') \delta(t-t'), \ \langle \xi \rangle = 0.
\end{equation}
\end{subequations}
Note that the noise level is further suppressed at larger $\mathcal{N}$ as should be expected from our semi-classical treatment.

\begin{table*}
\def\arraystretch{1.8}
\setlength\tabcolsep{1mm}
\begin{tabular}{ |c|c|c|c|c|c|c|c|c|c|c|c|c|c|c| } 
 \hline
 & $\kappa_1$ & $\kappa_2$ & $\mathfrak{z} D_1$ & $\mathfrak{z} D_2$ & $h$ & $r_1$ & $A_{20}$ & $A_{02}$ & $A_{30}$ & $A_{12}$& $r_2$ & $B_{11}$ & $B_{03}$ & $B_{21}$ \\ \hline
 $\Delta_c = 1/3$ & $\frac{2 v \Gamma}{3}$&$2 v \Gamma$& $\frac{4 \sqrt{3}}{9}$& $\frac{4 \sqrt{3}}{9}$& $\frac{\delta_{\Gamma}}{\sqrt{6}} - \frac{2}{\sqrt{3}} \delta_{\tilde{\Omega}}$ & $\frac{\delta_{\Gamma}}{2}$ & $\frac{\sqrt{3}}{2} \delta_{\tilde{\Omega}} - \frac{\sqrt{6}}{4} \delta_{\Gamma} $ & $\frac{2 \sqrt{2}}{9}$ & $-\frac{1}{\sqrt{3}}$ & $-\frac{7}{9 \sqrt{3}}$ & $\frac{2 \sqrt{3}}{3} \delta_{\Delta}  + \frac{5}{6}\delta_{\Gamma} -\frac{\sqrt{2}}{3}\delta_{\tilde{\Omega}}$ & $-\frac{2 \sqrt{2}}{3}$ & $-\frac{5}{9 \sqrt{3}}$ & $-\frac{1}{3 \sqrt{3}}$\\ \hline
 $\Delta_c = 2/3$ &$\frac{2 v \Gamma}{3}$&$2 v \Gamma$&$\frac{\sqrt{3}}{3}$&$\frac{\sqrt{3}}{3}$& $\frac{\delta_{\Gamma}}{\sqrt{6}} - \frac{2}{\sqrt{3}} \delta_{\tilde{\Omega}}$ & $\frac{\delta_{\Gamma}}{2}$ & $\frac{\sqrt{3}}{2} \delta_{\tilde{\Omega}} - \frac{\sqrt{6}}{4} \delta_{\Gamma} $ & $\frac{\sqrt{2}}{9}$ & $-\frac{1}{\sqrt{3}}$ & $-\frac{5}{9 \sqrt{3}}$ & $\frac{2 \sqrt{3}}{3} \delta_{\Delta}  + \frac{5}{6}\delta_{\Gamma} -\frac{\sqrt{2}}{3}\delta_{\tilde{\Omega}}$ & $-\frac{2 \sqrt{2}}{3}$ & $-\frac{1}{9 \sqrt{3}}$ & $-\frac{1}{3 \sqrt{3}}$ \\ \hline
\end{tabular}
\caption{Langevin equation parameters for soft modes $\phi_1, \phi_2$ in Eq.~\eqref{Eq. Langevin eqs}. \label{langtab}}
\end{table*}

Next we include the sublattice symmetry by defining $\Psi_1$ as the sublattice average and $\Psi_2$ as the sublattice difference of the field $\Psi$; see Eq.~(\ref{BAF}), which differs by a factor of $\sqrt{\mathcal{N}}$ due to our semi-classical limit. Our new Langevin equations are now
\begin{subequations}
\begin{equation}
i \partial_t \Psi_1 = - (\Delta + J + i \Gamma/2) \Psi_1 + v(\Psi_1^2-\Psi_2^2)\Psi_1^* + \tilde{\Omega} + \xi_1,
\end{equation}
\begin{equation}
i \partial_t \Psi_2 = - (\Delta -  J + i \Gamma/2) \Psi_2 + v(\Psi_2^2-\Psi_1^2)\Psi_2^* + \xi_2,
\end{equation}
\begin{equation}
\langle \xi^*_i(\mathbf{x},t) \xi_j(\mathbf{x}',t') \rangle =  \frac{\Gamma}{\mathcal{N}}\delta_{ij} \delta(\mathbf{x}-\mathbf{x}') \delta(t-t'), \ \langle \xi_i \rangle = 0.
\end{equation}
\end{subequations}
Notice that we have dropped all the gradient terms as they do not play a role in identifying the massive fields and their adiabatic elimination. We also follow our convention in the main text to set $\mathfrak{z} J \to J$.

Our model exhibits two multicritical points where two modes (each a component of one of the two fields $\Psi_i$) become critical. Due to the sublattice symmetry, $\Psi_2=0$ at the multicritical points up to fluctuations.
Working in units where $\Delta + J = 1$, the two multicritical points occur at
\begin{equation}
\begin{gathered}
(\Delta_c, J_c) = \left(\frac{1}{3},\frac{2}{3}\right), \left(\frac{2}{3},\frac{1}{3}\right), \\
\Gamma_c = \sqrt{4/3}, \ \ \tilde{\Omega}_c = (2/3)^{3/2}/\sqrt{v},
\end{gathered}
\end{equation}
with $\Psi_1 = \Psi_c = \sqrt{2/3v} e^{-i \pi/3}$.

Next, we expand the Langevin equations in the vicinity of the two multicritical points as 
\begin{equation}
\begin{aligned}
\Delta = \Delta_c + \delta_\Delta,&&J_c = J_c - \delta_\Delta,\\
\Gamma = \Gamma_c + \delta_\Gamma,&&\tilde{\Omega} = \tilde{\Omega}_c + \delta_{\tilde{\Omega}} /\sqrt{v},\\
\Psi_1 = \Psi_c + \psi_1/ \sqrt{v},&&\Psi_2 = \psi_2/\sqrt{v}.
\end{aligned}
\end{equation}
The soft and gapped modes can be determined as linear combinations of the real and imaginary parts of the two fields $\psi_1, \psi_2$. As in the case of bistability, the first pair is identified as \cite{FossFeig2017}
\begin{equation}
\psi_1 = \phi_1' + e^{i \pi/3} \phi_1,
\end{equation}
where $\phi_1'$ is massive and relaxes quickly while $\phi_1$ defines the slow field. Identifying the massive and massless components of the field $\psi_2$ depends on the corresponding multicritical point as 
\begin{equation}
\begin{aligned}
\Delta_c=1/3:&& \psi_2 = \frac{1}{\sqrt{3}}(\phi_2 e^{- i \pi/6} + \phi_2' e^{i \pi/6}),\\
\Delta_c = 2/3:&&\psi_2 = \frac{1}{\sqrt{3}}(\phi_2 + \phi_2' e^{i \pi/3}),
\end{aligned}
\end{equation}
where again the primed (unprimed) field indicates the massive (massless) field. Note that these differ from the main text by a factor of $\sqrt{V}$, which is done to simplify the resulting parameters in the Langevin equations by moving all the $V$ dependence to the noise term.

Upon adiabatically eliminating the massive fields and restoring the gradient terms, we arrive at the Langevin equations
\begin{subequations}\label{Eq. Langevin eqs}
\begin{equation}
\begin{aligned}
\dot{\phi_1} =& \ h - r_1 \phi_1 + D_1 \nabla^2 \phi_1 +\xi_1  \\
&+ A_{20} \phi_1^2 + A_{02} \phi_2^2 + A_{12} \phi_1 \phi_2^2+ A_{30} \phi_1^3  ,
\end{aligned}
\end{equation}
\begin{equation}
\begin{aligned}
\dot{\phi_2} = &-r_2 \phi_2 + D_2 \nabla^2 \phi_2  + \xi_2\\
&+ B_{11} \phi_1 \phi_2 + B_{21} \phi_1^2 \phi_2 + B_{03} \phi_2^3,
\end{aligned}
\end{equation}
\end{subequations}
with Gaussian noise 
\begin{equation}
\langle \xi_i(\mathbf{x},t) \xi_j(\mathbf{x}',t') \rangle = 2 \frac{ \kappa_i}{\mathcal{N}} \delta_{ij} \delta(\mathbf{x}-\mathbf{x}') \delta(t-t').
\end{equation}
The various numerical factors are summarized in Table \ref{langtab} for the two multicritical points under consideration. Note that $\zeta_i = 1$ and $T_i = \kappa_i/\mathcal{N}$. 
We can see at this point that the opposite signs of $A_{02}$ and $B_{11}$ indicate that no Hamiltonian description is possible. Indeed, as will be discussed in the following section, this will carry over to the signs of $g_{12}$ and $g_{21}$, leading to the critical behavior defined by the NEFPs.

\section{REDUNDANT OPERATORS}

\label{Redundant}

In this appendix, we identify the redundant operators in the Langevin equations~\eqref{Eq. Langevin eqs}. In general, this can be done at the level of the Schwinger-Keldysh action; however, we shall focus on the equivalent description in terms of the Langevin equations.
This perspective is particularly suitable in dealing with the (It\^{o}) regularization that is required to properly define the stochastic equations.

Consider a pair of Langevin equations that define an \textit{It\^{o} process} in the differential form \cite{Gardiner2009}
\begin{subequations}\label{Ito eqs}
\begin{equation}
d \phi_1 = f_1(\phi_1, \phi_2) dt + \sqrt{\kappa_1} dW_1,
\end{equation}
\begin{equation}
d \phi_2 = f_2(\phi_1, \phi_2) dt + \sqrt{\kappa_2} dW_2,
\end{equation}
\end{subequations}
where $dW_i$ is the stochastic noise that obeys 
the It\^{o} rules:
\begin{subequations}
\begin{equation}
dW_i dW_j = \delta_{ij} dt,
\end{equation}
\begin{equation}
dW_i dt = dt dW_i = 0,
\end{equation}
\begin{equation}
dt^2 = 0.
\end{equation}
\end{subequations}
At the multicritical point, where the effective masses and the magnetic field are set to zero, the Langevin equations~\eqref{Eq. Langevin eqs} can be written in the form of Eq.~\eqref{Ito eqs} with
\begin{subequations}
\begin{equation}
f_1(\phi_1,\phi_2) = D_1 \nabla^2 \phi_1 + A_{20} \phi_1^2 + A_{02} \phi_2^2 + A_{12} \phi_1 \phi_2^2+ A_{30} \phi_1^3,
\end{equation}
\begin{equation}
f_2(\phi_1, \phi_2) =  D_2 \nabla^2 \phi_2 + B_{11} \phi_1 \phi_2 + B_{21} \phi_1^2 \phi_2 + B_{03} \phi_2^3.
\end{equation}
\end{subequations}

In order to identify the redundant operators, we should examine the Langevin equations under a general change of the field variables. 
We should then find the dynamics in terms of new variables defined as $\Phi_1 = g_1(\phi_1, \phi_2)$ and $\Phi_2 = g_2(\phi_1, \phi_2)$; the functions $g_i$ are general (but local) nonlinear maps that are invertible in a neighborhood around the multicritical point and preserve the sublattice symmetry $\phi_2 \to -\phi_2$. The equations governing the dynamics of the new variables take the form 
\begin{widetext}
\begin{align}
\label{nonlin}
\begin{split}
d \Phi_1 &= \frac{\partial g_1}{\partial \phi_1} d\phi_1 + \frac{\partial g_1}{\partial \phi_2} d\phi_2 + \frac{1}{2} \frac{\partial^2 g_1}{\partial \phi_1^2} d\phi_1^2 +\frac{1}{2} \frac{\partial^2 g_1}{\partial \phi_2^2} d \phi_2^2 + \frac{\partial g_1}{\partial \phi_1 \partial \phi_2} d \phi_1 d \phi_2 \\
&= \left(f_1 \frac{\partial g_1}{\partial \phi_1} + f_2 \frac{\partial g_1}{\partial \phi_2} + \frac{\kappa_1}{2} \frac{\partial^2 g_1}{\partial \phi_1^2} + \frac{\kappa_2}{2} \frac{\partial^2 g_1}{\partial \phi_2^2}\right) dt + \sqrt{\kappa_1} \frac{\partial g_1}{\partial \phi_1} dW_1 + \sqrt{\kappa_2} \frac{\partial g_1}{ \partial \phi_2} dW_2.
\end{split}
\end{align}
\end{widetext}
We have used the It\^{o} rules to derive the above equation, which is known as 
It\^{o}'s formula or It\^{o}'s lemma \cite{Gardiner2009}. A similar stochastic equation can be derived for $\Phi_2$ by switching $1\leftrightarrow 2$. Note that the terms on the rhs should be expressed in terms of $\Phi_i$ through 
the inverse functions $g_i^{-1}$. 
One notices that there are new contributions to the deterministic dynamics due to the noise. However, since we are working under the assumption that the noise is parametrically small compared to the deterministic terms (with a strength proportional to $1/\cal N$), we can ignore such terms.
Additionally, the noise terms are no longer additive but are instead multiplicative, which introduces new terms in the action. However, since $\kappa_i \neq 0$, the latter are irrelevant in the sense of RG and can be neglected as well. 
The same holds for nonlinear terms (beyond quadratic terms) that involve gradients. 

We shall assume without loss of generality that
\begin{equation}
\left. \frac{\partial g_i}{\partial \phi_j} \right|_{\phi_i,\phi_j = 0} = \delta_{ij};
\end{equation}
rescaling the fields by a constant factor does not allow us any additional freedom while rotations obscure the symmetry $\phi_2 \to - \phi_2$. Additionally, we do not consider a constant shift in the field $\phi_1$ (a shift in $\phi_2$ is disallowed due to symmetry) for now, but discuss it separately later in this appendix. 
Based on the structure of Eq.~\eqref{nonlin}, we notice that the quadratic terms in $f_i$ and $g_i$ result in additional cubic terms. All other new terms in the deterministic part of the dynamics involve fourth- or higher-order terms that are irrelevant under RG. 
Expressing a general nonlinear transformation as
\begin{subequations}\label{eq. g fns}
\begin{equation}
g_1 (\phi_1,\phi_2) = \phi_1 + c_{20} \phi_1^2 + c_{02} \phi_2^2,
\end{equation}
\begin{equation}
g_2 (\phi_1,\phi_2) = \phi_2 + c_{11} \phi_1 \phi_2,
\end{equation}
\end{subequations}
the modification of the cubic terms to lowest order in the coefficients and interaction terms is given by
\begin{equation}
\left(
\begin{array}{c}
A_{12} \\
B_{21} \\
B_{03}
\end{array}
\right) \to \left(
\begin{array}{c}
A_{12} \\
B_{21} \\
B_{03}
\end{array}
\right) + 
\mathbf{M}
\left(
\begin{array}{c}
c_{20} \\
c_{02} \\
c_{11}
\end{array}
\right),
\end{equation}
where the matrix $\mathbf{M}$ is given by
\begin{equation}
\mathbf{M} = \left(
\begin{array}{ccc}
2 A_{02} & 2 B_{11}- 2 A_{20} & -2 A_{02} \\
-B_{11} & 0 & A_{20} \\
0 & -B_{11} & A_{02} 
\end{array}
\right).
\end{equation}
Notice that the coefficient $A_{30}$ is left unchanged. The rank of matrix $\mathbf{M}$ is 2, which then determines the number of corresponding redundant operators. 

In addition to the two redundant operators above, a third one emerges due to a constant shift $\phi_1 \to \phi_1 + c_{00}$. Under this transformation, the quadratic terms transform as
\begin{equation}
\label{const}
\left(
\begin{array}{c}
A_{20} \\
A_{02} \\
B_{11}
\end{array}
\right) \to \left(
\begin{array}{c}
A_{20} \\
A_{02} \\
B_{11}
\end{array}
\right) + 
\left(
\begin{array}{c}
3 A_{30} \\
A_{12} \\
4 A_{21}
\end{array}
\right)
c_{00}.
\end{equation}
The effective mass and magnetic field terms also change, but this simply shifts the location of the critical point.

The three redundant operators derived here can be used to always set the terms $A_{20},A_{02},B_{11}$ to zero. This can be understood by noting that the transformation corresponding to $\mathbf{M}$ allows one to adjust the ratios of $A_{12}$ and $A_{21}$ relative to $A_{30}$ without changing the quadratic terms. By properly using this redundancy, these ratios can be tuned until the constant shift in Eq.~\eqref{const} shifts all three quadratic terms to zero. At the same time, the cubic terms transform as
\begin{subequations}
\begin{equation}
A_{30} \to A_{30},
\end{equation}
\begin{equation}
B_{03} \to B_{03} - \frac{A_{12} A_{20} B_{11} - 6 A_{02} A_{30} B_{11} + 2 A_{02} A_{20} B_{21}}{2 A_{20} (A_{20}-B_{11})},
\end{equation}
\begin{equation}
A_{12} \to 2 A_{02} \frac{ 3 A_{30}}{ 2 A_{20}},
\end{equation}
\begin{equation}
B_{21} \to B_{11} \frac{ 3 A_{30}}{ 2 A_{20}}.
\end{equation}
\end{subequations}
Having exhausted the three redundant operators to remove the three quadratic terms, there is no further freedom in tuning other terms and, specifically, all the cubic terms are fixed. While we could in principle include cubic or higher-order terms in the nonlinear transformation [Eq.~\eqref{eq. g fns}], these would only modify the fourth- or higher-order terms, which are irrelevant under RG due to the presence of the cubic terms. We also see that the relative sign of $A_{12}$ and $B_{21}$ is indeed directly determined by the relative sign of $A_{02}$ and $B_{11}$, leading to criticality described by the NEFPs. Finally, we note that the coefficient $A_{20}$ appears in the denominator of the above transformations. We assumed that this term is generated under coarse graining and thus should pose no problem in making the above transformations. 
However, if there is a mechanism where this coefficient could be tuned to zero, the above transformations are no longer valid and the two nonzero cubic terms should be kept.

\section{Perturbative RG}

In this appendix, we discuss the details of the calculations in our perturbative RG analysis. In the first part, we introduce the diagrammatic techniques we have used in the main text. In the second part, we compute the one-loop diagrams, while, in the third part, we compute the two-loop diagrams for terms that are unrenormalized at one loop. 

\label{RG}

\subsection{Diagrammatic techniques}

\begin{figure}[b]
\subfloat[]
{
\includegraphics[scale=1]{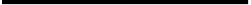}
}
\hspace{1cm}
\subfloat[]{
\centering
\includegraphics[scale=1]{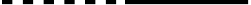}
}

\caption{Diagrammatic representation of Gaussian propagators. Solid (dotted) lines correspond to classical (response) fields. The two fields are later distinguished by thickness and color. In this figure, we have shown (a) the response propagator, and (b) the correlation propagator. \label{gauss}}

\end{figure}

To define the Gaussian propagators, we start with the Gaussian model with the corresponding action
\begin{equation}
\mathcal{A}_0[\tilde{\phi}_i,\phi_i] = \sum_i \int_{t,\bx}  \tilde{\phi}_i ( \zeta_i \partial_t - D_i \nabla^2 + r_i) \phi_i - \zeta_i T_i \tilde{\phi_i}^2.
\end{equation}
The Gaussian response and correlation functions are then given by 
\begin{subequations}
\label{gaussprop}%
\begin{equation}
\chi_0^i(\mathbf{q},\omega) = \mathcal{F}\langle \tilde{\phi}_i (\mathbf{0},0) \phi_i (\mathbf{r},t) \rangle = \frac{1}{- i \zeta_i \omega + D_i \mathbf{q}^2 + r_i},
\end{equation}
\begin{equation}
C_0^i(\mathbf{q},\omega) = \mathcal{F}\langle \phi_i (\mathbf{0},0) \phi_i (\mathbf{r},t) \rangle= \frac{2 \zeta_i T_i}{\zeta_i^2 \omega^2 + (D_i \mathbf{q}^2 + r_i)^2},
\end{equation}
\end{subequations}
where $\mathcal{F}$ denotes the Fourier transform in both space and time.
These propagators can be expressed in a diagrammatic representation as shown in Fig.~\ref{gauss}.

\begin{figure}[t!]
\centering
\subfloat[]
{
\centering
\includegraphics[scale=1]{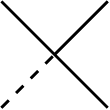}
}
\hspace{.7cm}
\subfloat[]
{
\centering
\includegraphics[scale=1]{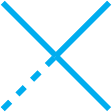}
}
\hspace{.7cm}
\subfloat[]
{
\centering
\includegraphics[scale=1]{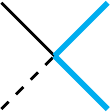}
}
\hspace{.7cm}
\subfloat[]
{
\centering
\includegraphics[scale=1]{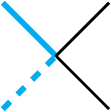}
}
\caption{Interaction vertices. Thin black (thick cyan) lines correspond to the first (second) field and solid (dashed) lines correspond to the classical (response) field. (a) $u_1 \phi_1^3 \tilde{\phi}_1$. (b) $u_2 \phi_2^3 \tilde{\phi}_2$. (c) $u_{12} \phi_2^2 \phi_1 \tilde{\phi}_1$. (d) $\sigma u_{12} \phi_1^2 \phi_2 \tilde{\phi}_2$. \label{ints}}
\end{figure}

The four interaction vertices from Eq.~(\ref{Aint}) are illustrated in Fig.~\ref{ints}. Due to the structure of the action, one can find the corresponding $Z$ factors for $\phi_2$ by switching the subscripts $1\leftrightarrow2$ and multiplying $u_{12}$ by a factor of $\sigma$.

Finally, in order to determine the $Z$ factors, we employ dimensional regularization using the minimal subtraction procedure. This means that only the ultraviolet divergences, in the form of powers of $1/\epsilon$, 
are incorporated into the $Z$ factors. For simplicity, we only present these divergences in the evaluation of integrals in the following sections; non-divergent terms are dropped. Technical details for evaluating integrals of the type presented here may be found in, e.g.,  Ref.~\cite{Tauber2014}.

\subsection{One-loop diagrams}

\subsubsection{Mass terms}

In this section, we consider corrections to $r_1$. The corrections to $r_2$ can be obtained by switching the roles of $\phi_1$ and $\phi_2$. There are two diagrams that provide corrections, which are illustrated in Fig.~\ref{oneloop}(a,b). The combinatorial and interaction factors are (a) $3 u_1$ and (b) $ u_{12}$. 
Thus the one-loop contribution to the renormalization of $r_1$ is
\begin{equation}
\int \frac{d^d\mathbf{p}}{(2 \pi)^d} \int \frac{d \omega}{2 \pi} \left[ 3 u_1 C_0^1(\mathbf{p},\omega) + u_{12} C_0^2(\mathbf{p},\omega)\right],
\end{equation}
so we should evaluate the integral of $C_0^i$. The latter can be written as
\begin{align}
2\zeta_i& T_i \int \frac{d^d\mathbf{p}}{(2 \pi)^d} \int \frac{d \omega}{2 \pi} \frac{1}{\zeta_i^2 \omega^2 + (D_i \mathbf{p}^2 + r_i)^2} \nonumber \\
= \ & T_i \int \frac{d^d\mathbf{p}}{(2 \pi)^d}\frac{1}{ D_i \mathbf{p}^2 + r_i},
\end{align}
where the last line follows once we integrate over frequency. 
The phase transition occurs where the renormalized mass term vanishes. The critical value of the mass parameter $r_{i_c}$ is then determined by
\begin{equation}
r_{1_c} = -3 u_1  T_1 \int \frac{d^d\mathbf{p}}{(2 \pi)^d}\frac{1}{ D_1 \mathbf{p}^2} - u_{12} T_2 \int \frac{d^d\mathbf{p}}{(2 \pi)^d}\frac{1}{ D_2 \mathbf{p}^2},
\end{equation}
and similarly for $r_{2_c}$; note that the factors of $r$ in the denominator have been dropped as they introduce $O(u^2)$ corrections. Defining an additive renormalized mass term $\overline{r}_i = r_i - r_{i_c}$, we can determine the $Z$ factors for $\overline{r}_i$.
To this end, we need to compute integrals of the form 
\begin{equation}
\int \frac{d^d\mathbf{p}}{(2 \pi)^d}\frac{\overline{r}_i}{D_i \mathbf{p}^2( D_i \mathbf{p}^2 + \overline{r}_i)} = \frac{A_d \mu^{-\epsilon}}{\epsilon} \frac{\overline{r}_i}{D_i^2},
\end{equation}
where we have included the geometrical factor $A_d$.
We can then compute the corresponding $Z$ factors as 
\begin{subequations}
\begin{equation}
Z_{\overline{r}_1} = 1 - 3 u_1 \frac{A_d \mu^{-\epsilon}}{\epsilon} \frac{T_1}{D_1^2} - u_{12} \frac{A_d \mu^{-\epsilon}}{\epsilon} \frac{T_2}{D_2^2} \frac{\overline{r}_2}{\overline{r}_1},
\end{equation}
\begin{equation}
Z_{\overline{r}_2} = 1 - 3 u_2 \frac{A_d \mu^{-\epsilon}}{\epsilon} \frac{T_2}{D_2^2} - \sigma u_{12} \frac{A_d \mu^{-\epsilon}}{\epsilon} \frac{T_1}{D_1^2} \frac{\overline{r}_1}{\overline{r}_2}.
\end{equation}
\end{subequations}
From this point on, we simply write $\overline{r}_i$ as $r_i$. 

\subsubsection{Coupling terms}

\begin{figure*}
\centering
\subfloat[]{
\centering
\includegraphics[scale=1]{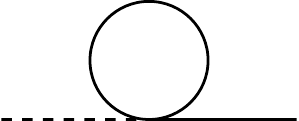}
}
\hspace{1cm}
\subfloat[]{
\centering
\includegraphics[scale=1]{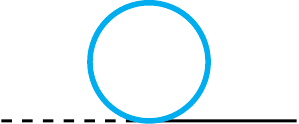}
}
\hspace{1cm}
\subfloat[]{
\centering
\includegraphics[scale=1]{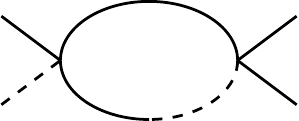}
}
\hspace{1cm}
\subfloat[]{
\centering
\includegraphics[scale=1]{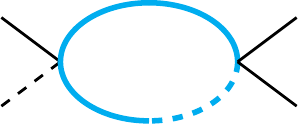}
}

\subfloat[]{
\includegraphics[scale=1]{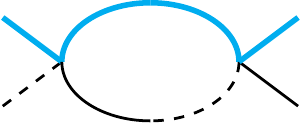}
}
\hspace{1cm}
\subfloat[]{
\centering
\includegraphics[scale=1]{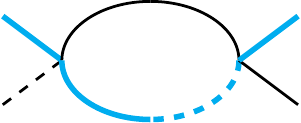}
}
\hspace{1cm}
\subfloat[]{
\centering
\includegraphics[scale=1]{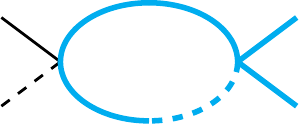}
}
\hspace{1cm}
\subfloat[]{
\centering
\includegraphics[scale=1]{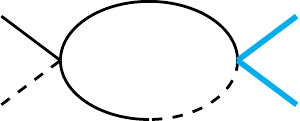}
}

\caption{One-loop corrections to (a,b) $r_1 \phi_1 \tilde{\phi}_1$, (c,d) $u_1 \phi_1^3 \tilde{\phi}_1$, and (e-h) $u_{12} \phi_2^2 \phi_1 \tilde{\phi}_1$. Analogous diagrams for $r_2 \phi_2 \tilde{\phi}_2$, $u_2\phi_2^3 \tilde{\phi}_2$ and $\sigma u_{12} \phi_1^2 \phi_2 \tilde{\phi}_2$ can be obtained by switching thin black and thick cyan lines. \label{oneloop}}

\end{figure*}

We first consider one-loop corrections to $u_1$. We need to consider two diagrams as illustrated in Fig.~\ref{oneloop}(c,d). The combinatorial and interaction factors are (c) $-18 u_1^2$ and (d) $-2 \sigma u_{12}^2$.
Thus the one-loop contribution to
the renormalization of $u_1$ is
\begin{equation}
\begin{aligned}
-\int \frac{d\mathbf p}{(2 \pi)^d} \int \frac{d \omega}{2 \pi} &\left[  18 u_1^2 \chi_0^1 (\mathbf{p},\omega) C_0^1(-\mathbf{p},-\omega)\right. \\ 
& \left. + 2 \sigma u_{12}^2 \chi_0^2 (\mathbf{p},\omega) C_0^2(-\mathbf{p},-\omega) \right],
\end{aligned}
\end{equation}
whose evaluation we put aside for the moment.

The renormalization of $u_{12}$ involves eight diagrams, four of which are illustrated in Fig.~\ref{oneloop}(e-h) and renormalize $u_{12} \phi_2^2 \phi_1 \tilde{\phi}_1$. The combinatorial and interaction factors are (e) $-4 u_{12}^2$, (f) $-4 \sigma u_{12}^2$, (g) $-6 u_{12} u_2 $, and (h) $-6  u_{12} u_1 $.
The remaining four diagrams, corresponding to the renormalization of $\sigma u_{12} \phi_1^2 \phi_2 \tilde{\phi}_2$,
can be simply obtained by interchanging the two fields. The resulting set of internal diagrams and combinatorial factors are the same up to a factor of $\sigma$, so $\phi_2^2 \phi_1 \tilde{\phi}_1$ and $\phi_1^2 \phi_2 \tilde{\phi}_2$ are renormalized in the same way at this order. Indeed, this reflects the fact fact that $g_{12}/g_{21}$ is not renormalized at the one-loop order. 

Thus the one-loop contribution to the renormalization of $u_{12}$ is
\begin{widetext}
\begin{multline}
-\int \frac{d\mathbf p}{(2 \pi)^d} \int \frac{d \omega}{2 \pi} \left[4 u_{12}^2 \chi_0^1 (\mathbf{p},\omega) C_0^2(-\mathbf{p},-\omega) + 4 \sigma u_{12}^2 \chi_0^2 (\mathbf{p},\omega) C_0^1(-\mathbf{p},-\omega) \right. \\ \left. + 6 u_{12} u_{2} \chi_0^2 (\mathbf{p},\omega) C_0^2(-\mathbf{p},-\omega) + 6 u_{12} u_1  \chi_0^1 (\mathbf{p},\omega) C_0^1(-\mathbf{p},-\omega) \right].
\end{multline}
\end{widetext} 
This expression involves a nontrivial integral of the form 
\begin{equation}
U_{ij} = - \int \frac{d\mathbf p}{(2 \pi)^d} \int \frac{d \omega}{2 \pi} \chi_0^i (\mathbf{p},\omega) C_0^j(-\mathbf{p},-\omega) ,
\end{equation}
which can be reduced to
\begin{equation}
- \frac{T_j}{\zeta_i} \frac{\Gamma(2-d/2)}{(4 \pi)^{d/2}} \int_0^1 dx \frac{( \tilde{D}_j + \tilde{D}_i x)^{-d/2}}{(\tilde{r}_j +  \tilde{r}_i x)^{2-d/2}},
\end{equation}
where $D_i = \zeta_i \tilde{D}_i$ and $r_i = \zeta_i \tilde{r}_i$.
Noting that $\tilde{r}_i = \mu^2 \tilde{r}_{i_R} + \mathcal{O}(u)$ and $\tilde{r}_{i_R}$ is a finite constant, then according to the minimal subtraction procedure, $(\tilde{r}_j + \tilde{r}_i x)^{2-d/2} \approx \mu^{-\epsilon} (\tilde{r}_{j_R}+\tilde{r}_{i_R} x)^0 = \mu^{-\epsilon}$, where $\epsilon$ has been set to 0 in the non-divergent part to extract the residue of the pole. 
Similarly, we expand $\Gamma(2-d/2) = 2/\epsilon + \mathcal{O}(1)$ to extract the pole.
Thus the above integral becomes
\begin{equation}
U_{ij} = - \frac{T_j}{\zeta_i \zeta_j}\frac{1}{\tilde{D}_j(\tilde{D}_i+ \tilde{D}_j)} \frac{A_d \mu^{- \epsilon}}{\epsilon},
\end{equation}
where we have included the geometrical factor $A_d$.
This result in combination with the diagrams considered above results in the following $Z$ factors
\begin{subequations}
\begin{equation}
Z_{u_1} = 1 + 18 U_{11} u_1 + 2 U_{22} \sigma u_{12}^2/u_1,
\end{equation}
\begin{equation}
Z_{u_2} = 1 + 18 U_{22} u_2 + 2 U_{11} \sigma u_{12}^2/u_2,
\end{equation}
\begin{equation}
Z_{u_{12}} = 1 + 4 U_{21} u_{12} + 4 U_{12} \sigma u_{12} + 6 U_{11} u_1 + 6 U_{22} u_2.
\end{equation}
\end{subequations}

\begin{figure}[b]
\subfloat[]{
\centering
\includegraphics[scale=1]{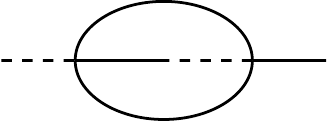}
}
\hspace{.7cm}
\subfloat[]{
\centering
\includegraphics[scale=1]{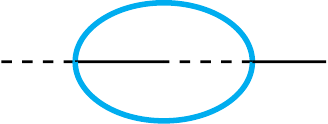}
}

\subfloat[]{
\centering
\includegraphics[scale=1]{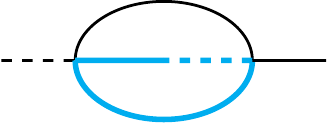}
}

\subfloat[]{
\centering
\includegraphics[scale=1]{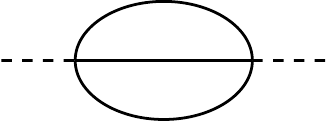}
}
\hspace{.7cm}
\subfloat[]{
\centering
\includegraphics[scale=1]{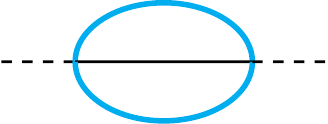}
}

\caption{Two-loop corrections to (a-c) $\zeta_1$ and $D_1$ as well as (d,e) $\zeta_1 T_1$. Analogous diagrams for $\zeta_2,D_2$, and $\zeta_2 T_2$ can be obtained by switching thin black and thick cyan lines. \label{twoloop}}

\end{figure}

\subsection{Two-loop diagrams}

First, we consider the two-loop corrections that arise from the $\tilde{\phi}_i \phi_i$ terms. There are three two-loop diagrams that renormalize $\zeta_1$ and $D_1$ as shown in Fig.~\ref{twoloop}(a-c). The corresponding combinatorial and interaction factors are (a) $-18 u_1^2$, (b) $ -2 u_{12}^2$, and (c) $-4 \sigma u_{12}^2$.

As in the case of the coupling terms, the internal diagrams are all of the same form, so we consider the generic internal diagram, which corresponds to the integral
\begin{widetext}
\begin{equation}
I_{ij}^k = \int \frac{d^d\mathbf{p} d^d\mathbf{q}}{(2 \pi)^{2d}} \int \frac{d \omega_1 d \omega_2}{(2 \pi)^2} C_0^i(\mathbf{p},\omega_1) C_0^j(\mathbf{q},\omega_2)\chi_0^k(\mathbf{k}-\mathbf{p}-\mathbf{q},\omega-\omega_1-\omega_2),
\end{equation}
where $C_0^i,C_0^j,\chi_0^k$ correspond to the top, bottom, and middle propagators, respectively, in the diagrams. This may be reduced to the form
\begin{subequations}
\begin{equation}
I_{ij}^k = \frac{T_i T_j}{\zeta_i \zeta_j \zeta_k} \frac{\Gamma(3-d)}{(4 \pi)^{d}}\int_0^1 \int_0^1 dx dy \frac{y^{1-d/2} (\alpha_2 \beta_2)^{-d/2} }{(\alpha_0-\pmb{\alpha}_1^2/\alpha_2)^{3-d}},
\end{equation}
\begin{equation}
\begin{gathered}
\alpha_0 = (1-y) \tilde{r}_i + y(\tilde{r}_j + x \tilde{D}_k )\mathbf{k^2}+\tilde{r}_k x + \tilde{r}_i x - i x \omega- \frac{x^2 \tilde{D}_k^2 \mathbf{k}^2}{\tilde{D}_j + x \tilde{D}_k},\\
\pmb{\alpha}_1 = \frac{-x y \tilde{D}_k  \tilde{D}_j}{\tilde{D}_j + x \tilde{D}_k}\mathbf{k}, \hspace{1cm}
\alpha_2 = (1-y+x y) \tilde{D}_i +  \frac{\tilde{D}_k \tilde{D}_j }{\tilde{D}_j + \tilde{D}_k x}x y, \hspace{1cm}
    \beta_2 = \tilde{D}_i +x \tilde{D}_k.
    \end{gathered}
\end{equation}
\end{subequations}
In order to determine corrections to $\omega$ and $\mathbf{k}^2$, we consider $W_{ij}^k \equiv \frac{\partial I_{ij}^k}{\partial (-i \omega)}$ and $K_{ij}^k \equiv \frac{\partial I_{ij}^k}{\partial (\mathbf{k}^2)}$, respectively, in the limit $\omega \to 0, \mathbf{k} \to 0$. Additionally, noting $\Gamma(3-d) = -\frac{1}{\epsilon} + \mathcal{O}(1)$ and taking $d\to 4$ while extracting a factor $\mu^{-2 \epsilon}$, we have
\begin{subequations}
\begin{equation}
W_{ij}^k = -\frac{T_i T_j}{4 \zeta_i \zeta_j \zeta_k} \frac{A_d^2 \mu^{-2 \epsilon}}{\epsilon}\int_0^1 \int_0^1 \frac{x}{(\tilde{D}_k \tilde{D}_j x y + \tilde{D}_i(\tilde{D}_j + \tilde{D}_k x)(1-y+xy))^2}dx dy,
\end{equation}
\begin{equation}
K_{ij}^k = -\frac{T_i T_j}{4 \zeta_i \zeta_j \zeta_k} \frac{A_d^2 \mu^{-2\epsilon}}{\epsilon}\int_0^1 \int_0^1 \frac{\tilde{D}_i \tilde{D}_j \tilde{D}_k x(1-y+xy)}{(\tilde{D}_k \tilde{D}_j x y + \tilde{D}_i(\tilde{D}_j +\tilde{D}_k x)(1-y+xy))^3}dx dy,
\end{equation}
\end{subequations}
where we have included the geometrical factor $A_d^2$. We evaluate both of these integrals exactly to find the resulting corrections
\begin{subequations}
\begin{equation}
W_{ij}^k = -\frac{ T_i T_j}{4 \zeta_i \zeta_j \zeta_k} \frac{A_d^2 \mu^{-2\epsilon}}{\epsilon} \frac{1}{\tilde{D}_k^2 \tilde{D}_i \tilde{D}_j} \log \left( \frac{(\tilde{D}_k+\tilde{D}_i)(\tilde{D}_k+\tilde{D}_j)}{\tilde{D}_i \tilde{D}_j + \tilde{D}_k(\tilde{D}_i + \tilde{D}_j)}\right),
\end{equation}
\begin{equation}
K_{ij}^k = -\frac{ T_i T_j}{4 \zeta_i \zeta_j \zeta_k} \frac{A_d^2 \mu^{-2\epsilon}}{\epsilon} \frac{(\tilde{D}_i+\tilde{D}_j)\tilde{D}_k^2+2 \tilde{D}_i \tilde{D}_j \tilde{D}_k}{2 \tilde{D}_i \tilde{D}_j(\tilde{D}_i+\tilde{D}_k)(\tilde{D}_j+\tilde{D}_k)(\tilde{D}_i \tilde{D}_j + \tilde{D}_i \tilde{D}_k + \tilde{D}_j \tilde{D}_k)},
\end{equation}
\end{subequations}
\end{widetext}
from which we identify the following $Z$ factors
\begin{subequations}
\begin{equation}
Z_{\zeta_1} = 1 - (18 u_1^2 W_{11}^1 + 2 u_{12}^2 W_{22}^1 + 4 \sigma u_{12}^2 W_{12}^2)/\zeta_1,
\end{equation}
\begin{equation}
Z_{\zeta_2} = 1 - (18 u_2^2 W_{22}^2 + 2 u_{12}^2 W_{11}^2 + 4 \sigma u_{12}^2 W_{21}^1)/\zeta_2,
\end{equation}
\begin{equation}
Z_{D_1} = 1 -(18 u_1^2 K_{11}^1 + 2 u_{12}^2 K_{22}^1 + 4 \sigma u_{12}^2 K_{12}^2)/D_1,
\end{equation}
\begin{equation}
Z_{D_2} = 1 -(18 u_2^2 K_{22}^2 + 2 u_{12}^2 K_{11}^2 + 4 \sigma u_{12}^2 K_{21}^1)/D_2,
\end{equation}
\end{subequations}
where the three corrections correspond to diagrams (a), (b), and (c) in Fig.~\ref{twoloop}, respectively.

Next, we consider the two-loop corrections to the $\tilde{\phi}_i^2$ terms. There are two such diagrams, which are illustrated in Fig.~\ref{twoloop}(d,e). The combinatorial and interactions factors are (d) $6 u_1^2$ and (e) $ 2 u_{12}^2$. Note the lack of minus sign due to the sign difference in $\mathcal{A}[\tilde{\phi}_i,\phi_i]$.
Again, the internal diagrams are all of the same form, so we instead consider the generic internal diagram, which corresponds to the integral
\begin{widetext}
\begin{equation}
S_{ijk} = \int \frac{d^d\mathbf{p} d^d \mathbf{q}}{(2 \pi)^{2d}} \int \frac{d \omega_1 d \omega_2}{(2 \pi)^2} C_0^i(\mathbf{p},\omega_1) C_0^j(\mathbf{q},\omega_2) C_0^k(\mathbf{p}+\mathbf{q},\omega_1+\omega_2),
\end{equation}
where $C_0^i ,C_0^j, C_0^k$ correspond to the the three propagators in the diagrams. This integral can be reduced to the form
\begin{subequations}
\begin{equation}
2 \frac{T_i T_j T_k}{\zeta_i \zeta_j \zeta_k}  \frac{\Gamma(4-d)}{(4 \pi)^d}\int_0^1 \int_0^1 \int_0^1 dx dy dz \frac{y z^{2-d/2}(\alpha_2 \beta_2)^{-d/2}}{\alpha_0^{4-d}},
\end{equation}
\begin{equation}
\begin{gathered}
\alpha_0 = (1 - yz) \tilde{r}_i + (1-xy)z \tilde{r}_j + (1-y+xy) z \tilde{r}_k, \\ 
\alpha_2 = (1-y z)\tilde{D}_i + \frac{\tilde{D}_j \tilde{D}_k (1-y+xy)(1-x y) z}{\tilde{D}_k(1-y+xy)+\tilde{D}_j (1-xy)}, \hspace{1cm}
\beta_2 = (1-xy) \tilde{D}_j + (1-y+xy) \tilde{D}_k.
\end{gathered}
\end{equation}
\end{subequations}
We note that $\Gamma(4-d) = 1/\epsilon + \mathcal{O}(1)$. Taking the limit $d \to 4$ and extracting a factor $\mu^{-2\epsilon}$, we rewrite the latter integral as
\begin{subequations}
\begin{equation}
C \int_0^1 \int_0^1 \int_0^1 dx dy dz \frac{y}{(\tilde{D}_j \tilde{D}_k (-1 +y - x y^2 + x^2 y^2)z + \tilde{D}_i(\tilde{D}_j(1-xy) + \tilde{D}_k(1-y+xy))(-1+yz))^2},
\end{equation}
\begin{equation}
C = \frac{T_i T_j T_k}{2 \zeta_i \zeta_j \zeta_k } \frac{A_d^2 \mu^{-2\epsilon}}{\epsilon} ,
\end{equation}
\end{subequations}
where we have included the geometrical factor $A_d^2$. Taking advantage of the fact that at least two of the $\tilde{D}$ must be the same, this integral can be evaluated analytically as
\begin{equation}
S_{ijj} = S_j^i = \frac{T_i T_j^2}{2 \zeta_i \zeta_j^2} \frac{A_d^2 \mu^{-2\epsilon}}{\epsilon} \frac{1}{\tilde{D}_i^2 \tilde{D}_j^2} \log\left(2^{2 \tilde{D}_i/\tilde{D}_j} \frac{\tilde{D}_i+\tilde{D}_j}{\tilde{D}_j} \left( \frac{\tilde{D}_i+\tilde{D}_j}{2\tilde{D}_i+\tilde{D}_j} \right)^{1+2\tilde{D}_i/\tilde{D}_j} \right),
\end{equation}
where $i$ corresponds to the field with one propagator and $j$ to the field with two.
\end{widetext}

Thus we identify the corresponding $Z$ factors 
\begin{subequations}
\begin{equation}
Z_{\zeta_1} Z_{T_1} = 1 + (3 u_1^2 S^1_1 +  u_{12}^2 S^1_2)/(\zeta_1 T_1),
\end{equation}
\begin{equation}
Z_{\zeta_2} Z_{T_2} = 1 + (3 u_2^2 S^2_2 +  u_{12}^2 S^2_1)/(\zeta_2 T_2),
\end{equation}
\end{subequations}
where the two corrections correspond to diagrams (d) and (e) in Fig.~\ref{twoloop}, respectively.
Note that the factors are half of their combinatorial factors. This is because the zeroth-order vertex is $2 \zeta_i T_i$ rather than $\zeta_i T_i$.

\section{METHOD OF CHARACTERISTICS}\label{characteristic}
In this section, we employ the method of characteristics in order to derive the scaling behavior of the correlation and response functions at or near a given fixed point. Since the correlation and response functions do not depend on the renormalized parameters, they are independent of the momentum scale of renormalization $\mu$. Thus for the response functions we have
\begin{subequations}
\begin{equation}
\mu \frac{d}{d \mu} \chi_i(\mathbf{q}, \omega, \{p_R\}, \{u_R\})=0,
\end{equation}
\end{subequations}
where $\{p\} = \{r_i,\zeta_{i}, D_{i},T_{i}\}$ and 
$\{u\}=\{u_1,u_2,u_{12}\}$ are the interaction strengths. 
Additionally, we define a
scaling function via $\chi_i = \mu^{-2} \hat{\chi}_i$, where the scaling factors are due to the scaling dimensions of the fields as well as the delta functions---which are factored out---corresponding to momentum and energy conservation. We can rewrite the total derivative $\mu \frac{d}{d \mu}$ in terms of the partial derivatives with respect to other parameters as
\begin{equation}
\mu\frac{d}{d\mu}\chi_i=\mu^{-2}\left(\sum_p \gamma_p p_R \partial_{p_R} + \sum_s \beta_u \partial_{u_R} -2\right)\hat \chi_i.
\end{equation}
Next we employ the method of characteristics and define $\tilde{\mu}(l) = \mu l$. 
We then introduce the flowing dimensionless parameters $\tilde{p}(l)$, $\tilde{u}(l)$ via 
\begin{align}
l \frac{d \tilde{p}(l)}{dl} = \gamma_p (l) \tilde{p}(l), && \tilde{p}(1) = p_R, \\
l \frac{d \tilde{u} (l)}{dl} = \beta_u(l), &&\tilde{u}(1) = u_R.
\end{align}
At the fixed point, $\beta_u(l) = 0$ for all $u$ and the parameters and flow functions assume their fixed-point values $\tilde{u}(l) = u^*$ and $\gamma_p(l) = \gamma_p^*$. This allows us to easily solve the flowing parameters as $\tilde{p}(l) = p_R l^{\gamma_p^*}$. Additionally, since all the terms in the perturbation series involve integrals of Gaussian propagators, we can reduce the scaling functions to [cf.~Eq.~(\ref{gaussprop})],
\begin{widetext}
\begin{equation}
\chi_i(\mathbf{q}, \omega, \{p_R\}, \{u_R\}) \propto 
\mu^{-2} D_R^{-1} l^{-2 - \gamma_D^*} \hat{\chi}_i \left(\frac{|\mathbf{q}|}{\mu l},\frac{\zeta_{i_R}\omega}{\mu^2 D_{i_R}^2} l^{-2- \gamma_D^*+\gamma_{\zeta}^*},\left\{\frac{r_{j_R}}{D_{i_R}} l^{\gamma_{r_j}^*-\gamma_D^*}\right\}\right),
\end{equation}
where we have utilized the fact that the scaling behavior of $\zeta,D,T$ is the same for both fields.
Upon applying the matching condition $|\mathbf{q}| = \mu l$ and carrying out a similar procedure for the correlation function, we can 
express the response and correlation functions as
\begin{subequations}
\begin{equation}
\chi_i(\mathbf{q},\omega,\{r_j\}) \propto |\mathbf{q}|^{-2 - \gamma_D^*} \hat{\chi}_i\left( \frac{\omega}{|\mathbf{q}|^{2+ \gamma_D^*-\gamma_{\zeta}^*}}, \left\{ \frac{r_{j_R}}{|\mathbf{q}|^{-\gamma_{r_j}^*+\gamma_D^*}} \right\} \right),
\end{equation}
\begin{equation}
C_i(\mathbf{q},\omega,\{r_j\}) \propto |\mathbf{q}|^{-4+\gamma_{\zeta}^* + \gamma_T^* -2 \gamma_D^*} \hat{C}_i\left( \frac{\omega}{|\mathbf{q}|^{2+ \gamma_D^*-\gamma_{\zeta}^*}}, \left\{ \frac{r_{j_R}}{|\mathbf{q}|^{-\gamma_{r_j}^*+\gamma_D^*}} \right\} \right),
\end{equation}
\end{subequations}
\end{widetext}
where we have further simplified the arguments of the scaling functions by dropping factors of $\mu$ and $p_R$ and excluding the argument $|\mathbf{q}|/\mu l = 1$ in a slight abuse of notation. Comparing these scaling functions against those in Eq.~(\ref{scaling fns}), we identify the critical exponents
\begin{equation}
\begin{aligned}
\eta &= \gamma_T^* - \gamma_D^*, & \eta' &= - \gamma_{D}^*, & z &= 2 + \gamma_{D}^* - \gamma_\zeta^*.
\end{aligned}
\end{equation}
Similarly, we can identify $\nu_j^{-1} = -\gamma_{r_j}^* + \gamma_D^*$, although the subtleties of a complex-valued exponent $\nu$ at the NEFPs are discussed in detail in the main text. 

\bibliography{NonEqMulti,article-driven-dissipative}

\end{document}